\documentclass{elsarticle}

\usepackage{amssymb}

\usepackage{lineno}
\usepackage{mathtools}
\usepackage{amsmath} 
\usepackage{amsfonts}
\usepackage{latexsym}
\usepackage{siunitx}
\usepackage{isomath}
\usepackage{bm}
\usepackage{tikz}
\usepackage{xfrac}
\usepackage{xcolor, colortbl}
\usepackage{booktabs}
\usepackage[T1]{fontenc}
\usepackage[utf8]{inputenc}
\usepackage{algorithm, algpseudocode}
\usepackage{setspace}
\usepackage{xspace}
\usepackage{geometry}
\usepackage{tabularx}
\usepackage{comment}
\usepackage[font=small, labelfont=bf, labelformat=parens, position=top]{subcaption}

\usepackage[textsize=footnotesize]{todonotes}
\definecolor{Gray}{gray}{0.90}
\newcolumntype{a}{>{\columncolor{Gray}}r}
\newcolumntype{L}{>{\raggedright\let\newline\\\arraybackslash\hspace{0pt}}X}
\newcolumntype{R}{>{\raggedleft\let\newline\\\arraybackslash\hspace{0pt}}X}

\usepackage{hyperref}
\usepackage[nameinlink]{cleveref}

\Crefname{equation}{Eq.}{Eqs.}
\Crefname{figure}{Fig.}{Figs.}
\Crefname{table}{Tab.}{Tabs.}
\hypersetup{colorlinks}
\definecolor{darkgreen}{rgb}{0.05, 0.5, 0.06}

\newcommand{\reviewerOne}[1]{{\textcolor{black}{#1}}}
\newcommand{\reviewerTwo}[1]{{\textcolor{black}{#1}}}
\newcommand{\generalChange}[1]{{\textcolor{black}{#1}}}

\newcommand{\tensor}[1]{\boldsymbol{\mathbf{#1}}}             
\newcommand{\isotensor}[1]{\overline{\tensor{#1}}}             
\newcommand{\iso}[1]{\overline{#1}}                         
\renewcommand{\vec}[1]{{\mathbf{#1}}}  
\newcommand{\Rvec}[1]{\underline{#1}}                
\newcommand{\Xvec}[1]{\boldsymbol{\mathbf{#1}}} 
\newcommand{\Grad}{\operatorname{Grad}}      
\newcommand{\Div}{\operatorname{Div}}        

\newcommand{\pcfem}[1][]{%
  \ifthenelse{\equal{#1}{}}{\Rvec{p}_\mathcal{C}}{\Rvec{p}_{\mathcal{C},{#1}}}%
}
\newcommand{\dd}{\,\mathrm{d}}

\newcommand{\dX}{\,\mathrm{d}\Xvec{X}}

\newcommand{\dsX}{\,\mathrm{d}s_{\Xvec{X}}}

\newcommand{\normalout}{{\mathbf{n}^{\mathrm{out}}_0}}
\newcommand\aug{\fboxsep=-\fboxrule\!\!\!\vrule\!\!\!}

\newcommand{\Ea}{\ensuremath{E_\mathrm{a}}}
\newcommand{\Ees}{\ensuremath{E_\mathrm{es}}}
\newcommand{\Vd}{\ensuremath{V_\mathrm{d}}}
\newcommand{\Ved}{\ensuremath{V_\mathrm{ed}}}
\newcommand{\Ves}{\ensuremath{V_\mathrm{es}}}
\newcommand{\ped}{\ensuremath{p_\mathrm{ed}}}
\newcommand{\pes}{\ensuremath{p_\mathrm{es}}}
\newcommand{\ppeak}{\ensuremath{\hat{p}}}
\newcommand{\Rsys}{\ensuremath{R_\mathrm{sys}}}

\newcommand{\Sa}{\ensuremath{S_\mathrm{a}}}

\newcommand{\ca}{\emph{CircAdapt}\xspace}  
\newcommand{\AO}{AO}  
\newcommand{\AP}{AP}  
\newcommand{\VC}{VC}  
\newcommand{\VP}{VP}  
\newcommand{\LV}{LV}  
\newcommand{\RV}{RV}  
\newcommand{\SV}{Sep}  
\newcommand{\LA}{LA}  
\newcommand{\RA}{RA}  
\newcommand{\AV}{AV}  
\newcommand{\PV}{PV}  
\newcommand{\MV}{MV}  
\newcommand{\TV}{TV}  
\newcommand{\PO}{PO}  
\newcommand{\SO}{SO}  

\journal{Computer Methods in Applied Mechanics and Engineering}

\begin{document}

\begin{frontmatter}



\title{A computationally efficient physiologically comprehensive 3D--0D closed-loop model of the heart and circulation
}

\author[1]{Christoph M. Augustin}
\author[1]{Matthias A. F. Gsell}
\author[1]{Elias Karabelas}
\author[3]{Erik Willemen}
\author[3]{Frits W. Prinzen}
\author[3]{Joost Lumens}
\author[4]{Edward J. Vigmond}
\author[1,2]{Gernot Plank\corref{cor1}}
\address[1]{%
  Gottfried Schatz Research Center: Division of Biophysics,
  Medical University of Graz, Graz, Austria}
\address[2]{%
  BioTechMed-Graz, Graz, Austria}
\address[3]{%
  Department of Biomedical Engineering, CARIM School for Cardiovascular Diseases,
  Maastricht University, Maastricht, Netherlands}
\address[4]{%
	IHU Liryc, Electrophysiology and Heart Modeling Institute, fondation Bordeaux Université, Pessac-Bordeaux, France}
\cortext[cor1]{Address correspondence to Gernot Plank,
    Gottfried Schatz Research Center: Division of Biophysics,
    Medical University of Graz, Neue Stiftingtalstrasse 6/IV, Graz 8010, Austria.
    Email: gernot.plank@medunigraz.at}

\begin{abstract}
  Computer models of cardiac electro-mechanics (EM) show promise as an effective means
  for the quantitative analysis of clinical data
  and, potentially, for predicting therapeutic responses.
  To realize such advanced applications
  methodological key challenges must be addressed.
  \reviewerOne{Enhanced computational efficiency and robustness is crucial
      to facilitate, within tractable time frames, model personalization,
      the simulation of prolonged observation periods
      under a broad range of conditions,
      and physiological completeness encompassing therapy-relevant mechanisms is needed
      to endow models with predictive capabilities beyond the mere replication
      of observations. }


  Here, \reviewerOne{we introduce a universal feature-complete cardiac EM modeling framework
  that builds on} a flexible method for coupling
  a 3D model of bi-ventricular EM to the physiologically comprehensive 0D \ca model
  representing atrial mechanics and closed-loop circulation.
  A detailed mathematical description is given
  and efficiency, robustness, and accuracy of numerical scheme and solver implementation
  are evaluated.
  After parameterization and stabilization of the coupled 3D-0D model
  to a limit cycle under baseline conditions,
  the model's ability to replicate physiological behaviors is demonstrated,
  by simulating the transient response to alterations in loading conditions and contractility,
  as induced by experimental protocols used for assessing systolic and diastolic ventricular properties.
  Mechanistic completeness and computational efficiency of this novel model
  render advanced applications geared towards predicting acute outcomes of EM therapies feasible.

\end{abstract}


\end{frontmatter}


\section{Introduction}%
\label{sec:introduction}
%
Cardiovascular diseases (CVDs) are the primary cause of mortality and morbidity in industrialized nations,
posing a significant burden on health care systems worldwide~\cite{laslett2012worldwide,timmis2020european,wilkins2017european}.
Despite continuous diagnostic and therapeutic advances,
their optimal treatment remains a challenge~\cite{smith2011euHeart}.
In no small part, this is due to the complex multiphysics nature of cardiovascular function
-- the heart is an electrically controlled mechanical pump driving blood through the circulatory system.
Advanced clinical modalities provide a wealth of disparate data,
but effective tools allowing their comprehensive quantitative analysis are lacking.
Computer models able to capture mechanistic relations between clinical observations quantitatively
show promise to fill this void.
In recent single physics cardiac electrophysiology (EP) studies,
the added value of models
in improving therapy stratification \cite{arevalo2016arrhythmia}
and planning \cite{prakosa2018personalized,strocchi2020simulating} has been demonstrated already.

Multiphysics models of cardiovascular EM are even more challenging to apply in a clinical context.
Their utility depends on the ability
to comprehensively represent mechanisms underlying a broader range of physiological function,
and to tailor these to approximate -- with acceptable fidelity --
anatomy and cardiovascular function of a given patient.
Such models are complex
as all major mechanisms governing a heart beat \reviewerOne{bidirectionally interact with each other and, thus,} must be taken into account.
These comprise models of cardiac EP
producing electrical activation and repolarization patterns
that drive EM coupling to models of contractile function,
cardiac mechanics describing deformation and stresses
under given mechanical boundary and hemodynamic loading conditions
imposed by the intra-thoracic embedding of the heart and the circulatory system, respectively.

Pumping function is regulated
through a bidirectional interaction between the heart and both the systemic and pulmonary vascular systems.
The circulatory system as an extracardiac factor imposes a pressure and volume load upon the heart
and, \emph{vice versa}, pressure and flow in the circulatory system are determined
by the mechanical state of cardiac cavities.
Optimal function depends on matching the coupling between
these two systems \cite{toorop1988matching}.
From a physics point of view, coupling poses a fluid-structure interaction (FSI) problem,
with pressure and blood flow velocity fields as coupling variables
~\cite{nordsletten2011fluid,karabelas2019towards}.
These are relevant for investigating flow patterns or wall shear stresses,
but are less suitable for systems level investigations.
Simpler, computationally less costly 0D and 1D lumped models have been preferred
to provide appropriate hemodynamic loading conditions to the
heart~\cite{Heusinkveld2019,Shi2011}.

Most EM modeling studies consider ventricular afterload only
represented by lumped 0D Windkessel type models
comprising 2-, 3- or 4-elements~\cite{elzinga1973pressure,liu2015multi,segers2008three,stergiopulos1999total,wang2003time,westerhof1991normalized,westerhof2000models},
or, less common, by 1D models derived from Navier--Stokes equations~\cite{Alastruey2012a,Blanco2014,Formaggia2003b,Mynard2008,Muller2013}.
The latter also account for pulse wave transmission and reflection,
but identifying parameters
is more challenging than for 0D models~\cite{marx2020personalization}.
A fundamental limitation of isolated models of pre- and afterload
is the lack of regulatory loops
which respond to altered loading or contractility in one chamber
by balancing preload conditions in all chambers
until a new stable limit cycle with common compatible stroke volumes is reached.
Isolated afterload models are thus best suited
for approximating the immediate responses in a single beat \cite{niederer2011:_length},
but less so for predicting transient behaviors over multiple beats.
Closed-loop circulatory  systems~\cite{arts2005adaptation,Blanco2010computational,guidoboni2019cardiovascular,neal2007subject,paeme2011mathematical}
take into account these feedback mechanisms
and ascertain the conservation of blood volume throughout the cardiovascular system.

Achieving a flexible, robust, and efficient coupling of 3D EM models of cardiac chambers
to a 0D closed-loop model of the circulatory system remains challenging.
Hydrostatic pressure, $p$, in cavities
and blood flow, $q$, between cavities and circulatory system serve as coupling variables
that act as pressure boundary condition and impose volume constraints on the 3D cavity models.
Previous studies addressed 3D solid-0D fluid coupling problems
using simpler
partitioned~\cite{eriksson2013influence,kerckhoffs2007coupling,usyk2002compuational}
or more advanced strongly coupled monolithic approaches~\cite{fritz2014simulation,Gurev2011a,Gurev2015high,Hirschvogel2017monolithic,marx2020personalization,Sainte-Marie2006modeling}.
Yet, reports on coupling of a closed-loop 0D to 3D solid models are sparse.
Mostly simplified circuit
models~\cite{Gurev2015high,kerckhoffs2007coupling,Hirschvogel2017monolithic,sack2018construction},
were used for simulating a single heart beat,
\reviewerOne{where fixed compliances and 0D chambers
based on time-varying elastance models are used are used
that do not account for pressure-volume relations or the Frank-Starling effect, respectively.}
Thus, attempts to demonstrate agreement with known physiological principles -- fundamental to cardiac pump function -- under experimental protocols requiring multibeat simulations have been limited.


Based on previous work on cardiac EM models~\cite{Augustin2016anatomically,Augustin2016patient,crozier2016image}
we report on the development of a monolithic 3D solid - 0D fluid coupling approach.
Feasibility is demonstrated by building a 3D canine bi-ventricular EM PDE model
coupled to the state-of-the art \ca model \cite{arts2005adaptation, walmsley2015fast}
-- a non-linear 0D closed-loop ODE model of the cardiovascular system
that implements dynamic adaptation processes based on physiological principles --
to represent \reviewerOne{physiologically realistic} atrial EM
as well as systemic and pulmonary circulation.
A detailed description of numerical underpinnings is given,
including a complete mathematical description of the \ca model in a single manuscript
that has been lacking so far.
Efficiency, robustness, and accuracy of numerical scheme and solver implementation are evaluated.
The coupled model is first parameterized and stabilized to a limit cycle
representing baseline conditions,
and then rigorously tested by demonstrating its ability
to predict physiological behaviors under experimental standard protocols
altering loading conditions and contractility
that are used for the experimental assessment of systolic and diastolic ventricular properties.
Transient responses under these protocols are simulated over prolonged observation periods,
covering up to 25 beats.
\reviewerOne{The presented framework can be considered a first feature-complete realization
of an universal cardiac EM simulator that can be applied,
given appropriate parameterization and initialization,
under a much broader range of protocols and conditions as any previously reported model.}

%
\section{Methodology}%
\label{sec:methods}

\subsection{Experimental data acquisition}

In a previous study, see~\cite{prinzen2009}, mongrel dog data were acquired to
investigate the influence of different pacing protocols on cardiac mechanics,
pump function and efficiency.
The animals were handled according to the Dutch Law on Animal
Experimentation (WOD) and the European Directive for the Protection
of Vertebrate Animals Used for Experimental and Other Scientific
Purposes (86/609/EU). The protocol was approved by the Maastricht
University Experimental Animal Committee.
Anatomical Magnetic Resonance Images (MRI) were acquired on a Philips Gyroscan $\SI{1.5}{\tesla}$
(NT, Philips Medical Systems, Best, the Netherlands)
using a standard synergy receiver coil for thorax examinations.
Images of seven short-axis cross-sections of slice thickness $\SI{8}{\milli \meter}$
with $\SI{0}{\milli \meter}$ inter-slice distance were obtained
to capture the whole heart.
LV pressure and volume were determined
using the conductance catheter technique (CD-Leycom, The Netherlands), see
\cite{verbeek2003}, and the signals were digitized at $\SI{1}{\kilo \hertz}$.


\subsection{Biventricular Finite Element Models} \label{sec:fe-model}

\begin{figure}[h!]
  \centering
  \includegraphics[width=\textwidth,keepaspectratio]{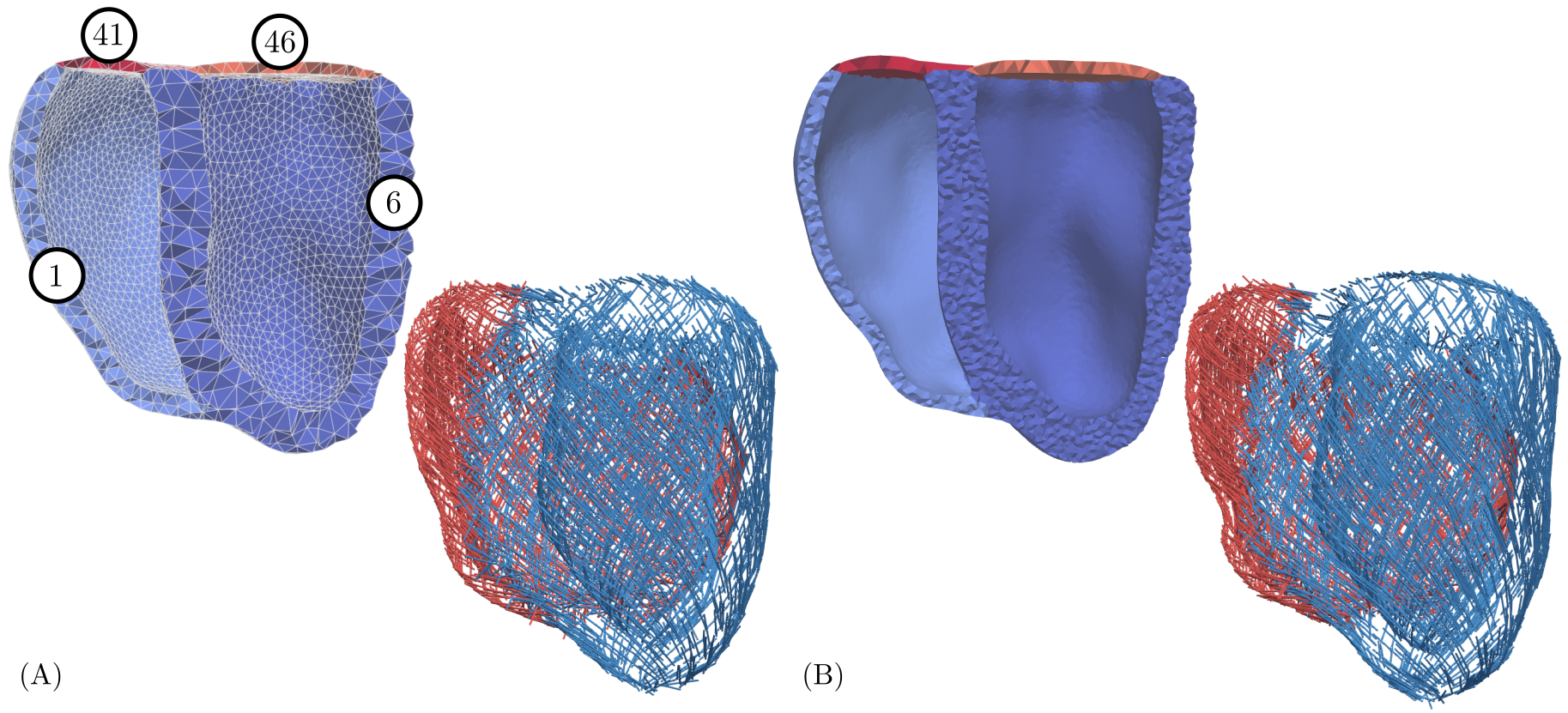}
  \caption{coarse (A) and fine (B) resolution meshes
  with domain labels and corresponding fiber fields.
  Note the difference in fiber angles due to spatial resolution. }
  \label{fig:model}
\end{figure}

Multilabel segmentations of right ventricular (RV) (tag $36$) and LV blood pool (tag $31$)
and of the LV myocardium (tag $1$), see~\Cref{fig:model}, were generated
from seven MRI short axis slices using Seg3D \cite{sci:seg3D}.
Each slice was first segmented semi-automatically
using thresholding techniques with manual correction.
Segmentations were upsampled to isotropic resolution,
followed by an automated iterative erosion and dilation smoothing scheme
implemented in Meshtool \cite{neic2020:meshtool}.
The RV wall (tag $6$) and lids representing the atrio-ventricular valves
were automatically generated by dilation of the adjacent blood pool (tags $41$ and $46$).
Biventricular multilabel meshes were created then from labeled segmentations \cite{crozier2016image}
using the Computational Geometry Algorithms Library, CGAL
(\href{https://www.cgal.org}{www.cgal.org}) and subsequently smoothed with Meshtool \cite{neic2020:meshtool}.
A rule-based method according to \cite{bayer2012a:fibers} was applied
to define fiber and sheet architecture,
with fiber angles changing linearly from $-60^{\circ}$ at the epicardium
to $+60^{\circ}$ at the endocardium \cite{streeter1969fiber}.
Universal ventricular coordinates were computed \cite{bayer2018universal}
to support the flexible definition of stimulation sites and \reviewerOne{mechanical} boundary conditions.
Two meshes of different resolution were generated,
a coarse mesh to reduce computational expenses
and to facilitate the fast exploration of experimental protocols over prolonged observation periods,
and a higher resolution mesh for investigating potential inaccuracies
introduced by the coarser spatial resolution.
For the coarse mesh, average edge lengths of $\sim\!\SI{3.4}{\milli \meter}$ and $\sim\!\SI{2.4}{\milli \meter}$,
were chosen in LV and RV, respectively, to ascertain
that at least two elements were generated transmurally across the myocardial walls,
as illustrated in \Cref{fig:model}.
For the finer mesh, average edge lengths of
$\sim\!\SI{1.3}{\milli\meter}$ and
$\sim\!\SI{1.2}{\milli\meter}$,
were chosen in LV and RV, respectively.

\subsection{Electromechanical PDE Model}
\paragraph{Tissue Mechanics}
Cardiac tissue is mechanically characterized as a hyperelastic,
nearly incompressible, orthotropic material with a nonlinear stress-strain
relationship.
The deformation gradient $\tensor{F}$ describes the deformation $\vec{u}$ of a
body from the reference configuration $\Omega_0(\vec{X})$ to the current
configuration $\Omega_t(\vec{x})$,
\begin{equation}
  F_{ij}=\frac{\partial x_i}{\partial X_j}, \quad i,j=1,2,3.
  \label{eq:defGradient}
\end{equation}
By convention, we denote $J=\det \tensor{F}>0$ and introduce the
right Cauchy--Green tensor $\tensor{C}=\tensor{F}^\top \tensor{F}$.  
The nearly incompressible behavior is modeled by a multiplicative decomposition
of the deformation gradient~\cite{flory1961thermodynamic} of the form
\begin{equation}\label{eq:decompositionFlory}
 \tensor{F}=J^{1/3}\overline{\tensor{F}},\
 \tensor{C}=J^{2/3}\overline{\tensor{C}},\
 \text{with } \det\overline{\tensor{F}}=\det\overline{\tensor{C}}=1.
\end{equation}
Mechanical deformation is described by \reviewerOne{Cauchy's first equation of motion} given as
\begin{equation}\label{eq:current}
\rho_0 \ddot{\vec{u}}(t,\vec X)
  -\Div\left[ \tensor{F}\tensor{S}(\vec{u},\vec{X})\right] = \vec{0}
  \quad \mbox{for} \; \vec{X} \in \Omega_0\times (0,T),
\end{equation}
with initial conditions
\begin{align*}
  \vec u(\vec X, 0) = \vec 0,\quad
  \dot{\vec u}(\vec X, 0) = \vec 0.
\end{align*}
Here,
$\rho_0$ is the density in reference configuration;
$\ddot{\vec{u}}$ are nodal accelerations;
$\dot{\vec{u}}$ are nodal velocities;
$\tensor{S} (\vec{u},\vec{X})$ is the second Piola--Kirchhoff stress tensor;  
and $\Div$ denotes the divergence operator in the reference configuration.

The boundary of the bi-ventricular models was decomposed in three parts, $\partial\Omega_0=\overline{\Gamma}_{\mathrm{endo},0} \cup
\overline{\Gamma}_{\mathrm{epi},0}\cup \overline{\Gamma}_{\mathrm{base},0}$,
with $\overline{\Gamma}_{\mathrm{endo},0}$ the endocardium, $\overline{\Gamma}_{\mathrm{epi},0}$ the epicardium, and
$\overline{\Gamma}_{\mathrm{base},0}$ the base of the ventricles.

Normal stress boundary conditions were imposed on the endocardium
\begin{equation}\label{eq:neumann_bc}
  \tensor{F}\tensor{S}(\vec{u},\vec{X})\,\normalout(\vec{X})
  = -p(t) J\tensor{F}^{-\top}\normalout(\vec{X})
  \quad\text{on}\quad {\Gamma}_{\mathrm{endo},0}\times (0,T)
\end{equation}
with $p(t)$ the pressure and $\normalout$ the outer normal vector; omni-directional spring type boundary conditions
constrained the ventricles at the basal cut plane $\overline{\Gamma}_{\mathrm{base},0}$~\cite{land2018influence};
and to simulate the mechanical constrains imposed by
the pericardium spatially varying normal Robin boundary conditions
were applied at the epicardium $\overline{\Gamma}_{\mathrm{epi},0}$~\cite{strocchi2020simulating}.

Apart from external loads the deformation of cardiac tissue is in particular governed by
active stresses intrinsically generated during contraction. To simulate both the
active and passive properties of the tissue, the total stress $\tensor{S}$ is
 additively decomposed according to
\begin{equation} \label{eq:additiveSplit}
  \tensor{S}= \tensor{S}_\mathrm{p}+ \tensor{S}_\mathrm{a},
\end{equation}
where $\tensor{S}_\mathrm{p}$ and $\tensor{S}_\mathrm{a}$
refer to the passive and active stresses, respectively.

\paragraph{Passive Stress}
Passive stresses are modeled based on the constitutive equation
\begin{equation}
  \tensor{S}_p = 2 \frac{\partial\Psi(\tensor{C})}{\partial \tensor{C}},
\end{equation}
where $\Psi$ is a strain-energy function to model the
orthotropic behavior of cardiac tissue.
The prevailing orientation of myocytes, referred to as fiber orientation,
is denoted as $\vec{f}_0$.
Individual myocytes are surrounded and interconnected by collagen,
forming sheets, which is described by the sheet orientation $\vec{s}_0$,
perpendicular to $\vec{f}_0$.
Together with the sheet-normal axis $\vec{n}_0$,
orthogonal to the sheet and the fiber orientations, this forms
a right-handed orthonormal set of basis vectors.

Following~\citet{usyk2000effect} the orthotropic constitutive relation is defined as
\begin{equation} \label{eq:usykStrainEnergy}
  \Psi(\tensor{C}) = \frac{\kappa}{2} {\left( \log\,J \right)}^2 +
  \frac{a}{2}\left[\exp(\mathcal{Q})-1\right],
\end{equation}
where the first term is the volumetric energy with the bulk modulus $\kappa\gg\SI{0}{\kPa}$
which penalizes local volume changes to enforce near incompressible behavior of the tissue, \reviewerOne{parameter $a$ is a stress-like scaling parameter},
and the term in the exponent is
\begin{equation} \label{eq:Usyk2000Q}
    \mathcal{Q} =  b_\mathrm{ff} \iso{E}_\mathrm{ff}^2
          + b_\mathrm{ss} \iso{E}_\mathrm{ss}^2
          + b_\mathrm{nn} \iso{E}_\mathrm{nn}^2
          + b_\mathrm{fs} \left(\iso{E}_\mathrm{fs}^2+\iso{E}_\mathrm{sf}^2\right)
          + b_\mathrm{fn} \left(\iso{E}_\mathrm{fn}^2+\iso{E}_\mathrm{nf}^2\right)
          + b_\mathrm{ns} \left(\iso{E}_\mathrm{ns}^2+\iso{E}_\mathrm{sn}^2\right).
\end{equation}
Here, \generalChange{$b_\bullet$ are dimensionless parameters} and the
directional strains read
\begin{align*}
    \iso{E}_{\mathrm{ff}}&=\vec{f}_0\cdot\isotensor{E}\vec{f}_0,\
    \iso{E}_{\mathrm{ss}}=\vec{s}_0\cdot\isotensor{E}\vec{s}_0,\
    \iso{E}_{\mathrm{nn}}=\vec{n}_0\cdot\isotensor{E}\vec{n}_0,\
    \iso{E}_{\mathrm{fs}}=\vec{f}_0\cdot\isotensor{E}\vec{s}_0,\
    \iso{E}_{\mathrm{fn}}=\vec{f}_0\cdot\isotensor{E}\vec{n}_0,\\
    \iso{E}_{\mathrm{ns}}&=\vec{n}_0\cdot\isotensor{E}\vec{s}_0, \
    \iso{E}_{\mathrm{sf}}=\vec{s}_0\cdot\isotensor{E}\vec{f}_0,\
    \iso{E}_{\mathrm{nf}}=\vec{n}_0\cdot\isotensor{E}\vec{f}_0,\
    \iso{E}_{\mathrm{sn}}=\vec{s}_0\cdot\isotensor{E}\vec{n}_0,
\end{align*}
with $\isotensor{E}=\frac{1}{2}(\overline{\tensor C}-\tensor{I})$
the modified isochoric Green--Lagrange strain tensor.  
\generalChange{All passive material parameters are} given in~\Cref{tab:pde_input_params}.

\paragraph{Active Stress}
Stresses due to active contraction are assumed to be orthotropic with full
contractile force along the myocyte fiber orientation $\vec{f}_0$ and
\SI{40}{\%} contractile force along the sheet orientation $\vec{s}_0$~\cite{Genet2014, walker2005mri}.
Thus, the active stress tensor is defined as
\begin{equation} \label{eq:act}
  \tensor{S}_{\mathrm{a}}
  = S_{\mathrm{a}} {\left(\vec{f}_0\cdot\tensor{C}\vec{f}_0\right)}^{-1}
      \vec{f}_0 \otimes \vec{f}_0
    + \num{0.4}\,S_{\mathrm{a}} {\left(\vec{s}_0\cdot\tensor{C}\vec{s}_0\right)}^{-1}
      \vec{s}_0 \otimes \vec{s}_0,
\end{equation}
where $S_\mathrm{a}$ is the scalar active stress describing the contractile
force.
A simplified phenomenological contractile model was used to represent active
stress generation~\cite{niederer2011:_length}.
Owing to its small number of parameters and its
direct relation to clinically measurable quantities such as peak pressure,
and the maximum rate of rise of pressure this model is fairly easy to fit
and thus very suitable for being used in clinical EM modeling studies.
Briefly, the active stress transient is given by
\begin{equation} \label{eq:tanh_stress}
  S_\mathrm{a}(t,\lambda) = S_\mathrm{peak} \,
  \phi(\lambda) \,
  \tanh^2 \left( \frac{t_\mathrm{s}}{\tau_\mathrm{c}} \right) \,
  \tanh^2 \left( \frac{t_\mathrm{dur} - t_\mathrm{s}}{\tau_\mathrm{r}} \right),
  \qquad \text{for } 0 < t_\mathrm{s} < t_\mathrm{dur},
\end{equation}
with
\begin{equation} \label{eq:tanh_stress2}
  \phi = \tanh (\mathrm{ld} (\lambda - \lambda_0)),\quad
  \tau_\mathrm{c} = \tau_\mathrm{c_0} + \mathrm{ld}_\mathrm{up}(1-\phi),\quad
  t_\mathrm{s} = t - t_\mathrm{a} - t_\mathrm{emd}
\end{equation}
and $t_\mathrm{s}$ is the onset of contraction;
$\phi (\lambda)$ is a non-linear length-dependent function
in which $\lambda$ is the fiber stretch and
$\lambda_0$ is the lower limit of fiber stretch below which no further active
tension is generated;
$t_{\mathrm{a}}$ is the local activation time from~\Cref{eq:_eikonal},
defined when the local transmembrane potential passes the threshold voltage $V_\mathrm{m,thresh}$;
$t_{\mathrm{emd}}$ is the EM delay between the onsets of
electrical depolarization and active stress generation;
$S_{\mathrm{peak}}$ is the peak isometric tension;
$t_\mathrm{dur}$ is the duration of active stress transient;
$\tau_{\mathrm{c}}$ is time constant of contraction;
$\tau_{\mathrm{c_0}}$ is the baseline time constant of contraction;
$\mathrm{ld}_{\mathrm{up}}$ is the length-dependence of $\tau_{\mathrm{c}}$;
$\tau_{\mathrm{r}}$ is the time constant of relaxation;
and $\mathrm{ld}$ is the degree of length dependence.
For the parameter values used in the simulations see~\Cref{tab:pde_input_params}.
Note that active stresses in this simplified model are only length-dependent,
but dependence on fiber velocity, $\dot{\lambda}$, is ignored.
\paragraph{Electrophysiology}
A recently developed reaction-eikonal (R-E) model~\cite{neic17:_efficient} was
employed to generate electrical activation sequences
which serve as a trigger for active stress generation in cardiac tissue.
The hybrid R-E model combines a standard reaction-diffusion (R-D) model
based on the monodomain equation with an eikonal model.
Briefly, the eikonal equation is given as
\begin{equation} \label{eq:_eikonal}
  \left\{
    \begin{array}{rcll}
      \sqrt{\nabla_{\vec X} t_\mathrm{a}^\top \, \tensor{V} \,
      \nabla_{\vec X} t_\mathrm{a}} & =
      & 1 \qquad & \text{in }  \Omega_0, \\
      t_\mathrm{a} & = & t_0 & \text{on } \Gamma_0^{\ast},
    \end{array}
  \right.
\end{equation}
where $(\nabla_{\vec X})$ is the gradient with respect to the end-diastolic
reference configuration $\Omega_{0}$;
$t_\mathrm{a}$ is a positive function describing the wavefront arrival
time at location $\vec{X} \in \Omega_0$;
and $t_0$ are initial activations at locations
$\Gamma_0^\ast \subseteq \Gamma_{\mathrm{N},0}$.
The symmetric positive definite $3 \times 3$ tensor $\tensor{V}(\vec{X})$
holds the squared velocities
$\left(v_\mathrm{f}(\vec{X}),v_\mathrm{s}(\vec{X}),v_\mathrm{n}(\vec{X})\right)$
associated to the tissue's eigenaxes $\vec{f}_0$, $\vec{s}_0$, and $\vec{n}_0$.
The arrival time function $t_\mathrm{a}(\vec{X})$ was subsequently used in a
modified monodomain R-D model given as
\begin{equation} \label{eq:_monodomain_R_E}
  \beta C_\mathrm{m} \frac{\partial V_\mathrm{m}}{\partial t} =
    \nabla_{\vec X} \cdot \bm{\sigma}_\mathrm{m} \nabla_{\vec X} V_\mathrm{m}
     - \beta I_\mathrm{ion} + I_\mathrm{foot},
\end{equation}
with $\beta$ the membrane surface-to-volume ratio;
$C_\mathrm{m}$ the membrane capacitance;
$V_\mathrm{m}$ the unknown transmembrane voltage;
$\bm{\sigma}_\mathrm{m}$ the monodomain conductivity tensor which holds the
scalar conductivities ($g_\mathrm{f}(\bm{X}),\ g_\mathrm{s}(\bm{X}),\ g_\mathrm{n}(\bm{X})$)
and is coupled to $\tensor{V}(\vec{X})$ proportionally~\cite{costa13:_automatic_parameterization};
and $I_\mathrm{ion}$ the membrane ionic current density.
Additionally, an arrival time dependent foot current, $I_{\mathrm{foot}}(t_\mathrm{a})$, was added
which is designed to mimic subthreshold electrotonic currents
to produce a physiological foot of the action potential.
The key advantage of the R-E model is its ability
to compute activation sequences at coarser spatial resolutions
that are not afflicted by the spatial undersampling artifacts
leading to conduction slowing or even numerical conduction block,
as it is observed in standard R-D models \cite{niederer11:_nversion_ep}.
Ventricular EP was represented by the tenTusscher--Noble--Noble--Panfilov model  
of the human ventricular myocyte~\cite{tentusscher04:_TNNP}.

\paragraph{Computation of Volumes}%
\label{sec:computationVolumes}
To compute the flow across the interface between 3D cavities and the 0D cardiovascular system,
\reviewerOne{the cavitary volume of each chamber that is described as a
3D PDE model} has to be tracked as a function of time:
$V^\mathrm{PDE}(\vec{x},t)$.
A reduction in cavitary volume
\[\frac{\partial V^\mathrm{PDE}(\vec{x},t)}{\partial t} < 0, \]
drives a positive flow into the circulatory system.
In a pure EM simulation context where the fluid domain is
not modeled explicitly, the cavitary blood pool volume is not discretized,
only the surface $\Gamma$ enclosing the volume is known.
Assuming that the entire surface of the cavitary volume is available, that is,
also the faces representing the valves are explicitly discretized,
the enclosed volume $V^\mathrm{PDE}$ can be computed from this surface
using the divergence theorem
\begin{equation}  \label{eq:volumeOmega}
  V^\mathrm{PDE}(\vec{u},t)=V^\mathrm{PDE}(\vec{x},t)=\frac{1}{3}\int_{\Gamma_{t}}\vec{x}\cdot\vec{n}\dd \Gamma_{t}.
\end{equation}
Using this approach, the volume $V^\mathrm{PDE}(\vec{x}, t)$
can be computed for each state of deformation at time $t$ and the
flow \reviewerOne{can be derived by a numerical approximation using a difference quotient.}

\subsection{Lumped ODE model of the circulatory system: the \ca model}%
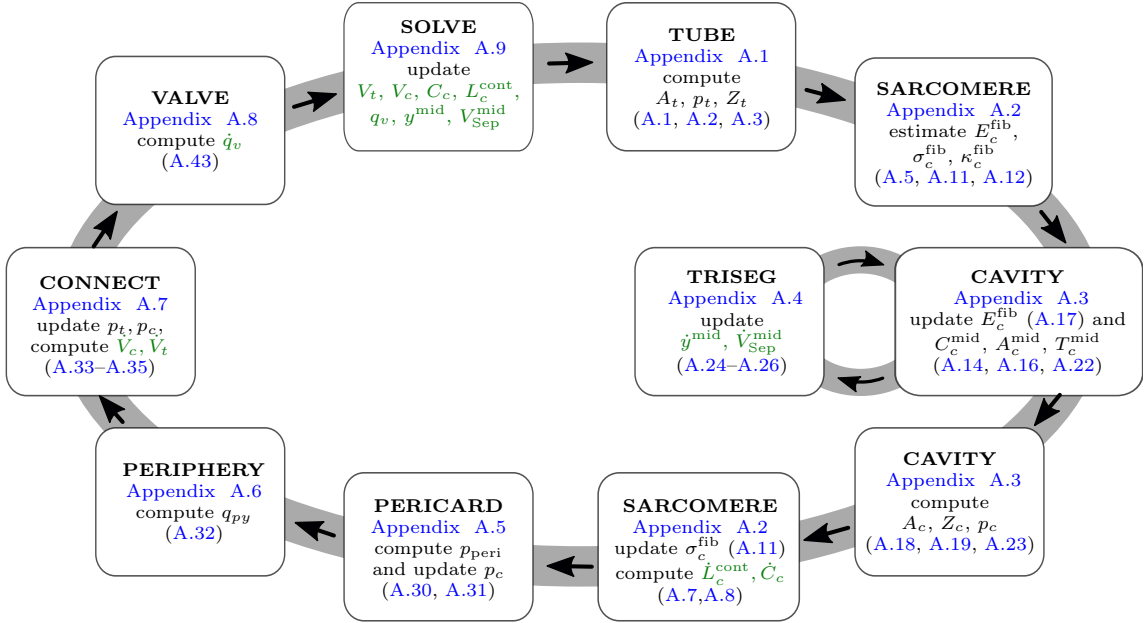
\begin{figure}\scriptsize
  \begin{tikzpicture}
  \node[inner sep=0pt] (structure) at (0,0)
    {\includegraphics[width=\textwidth]{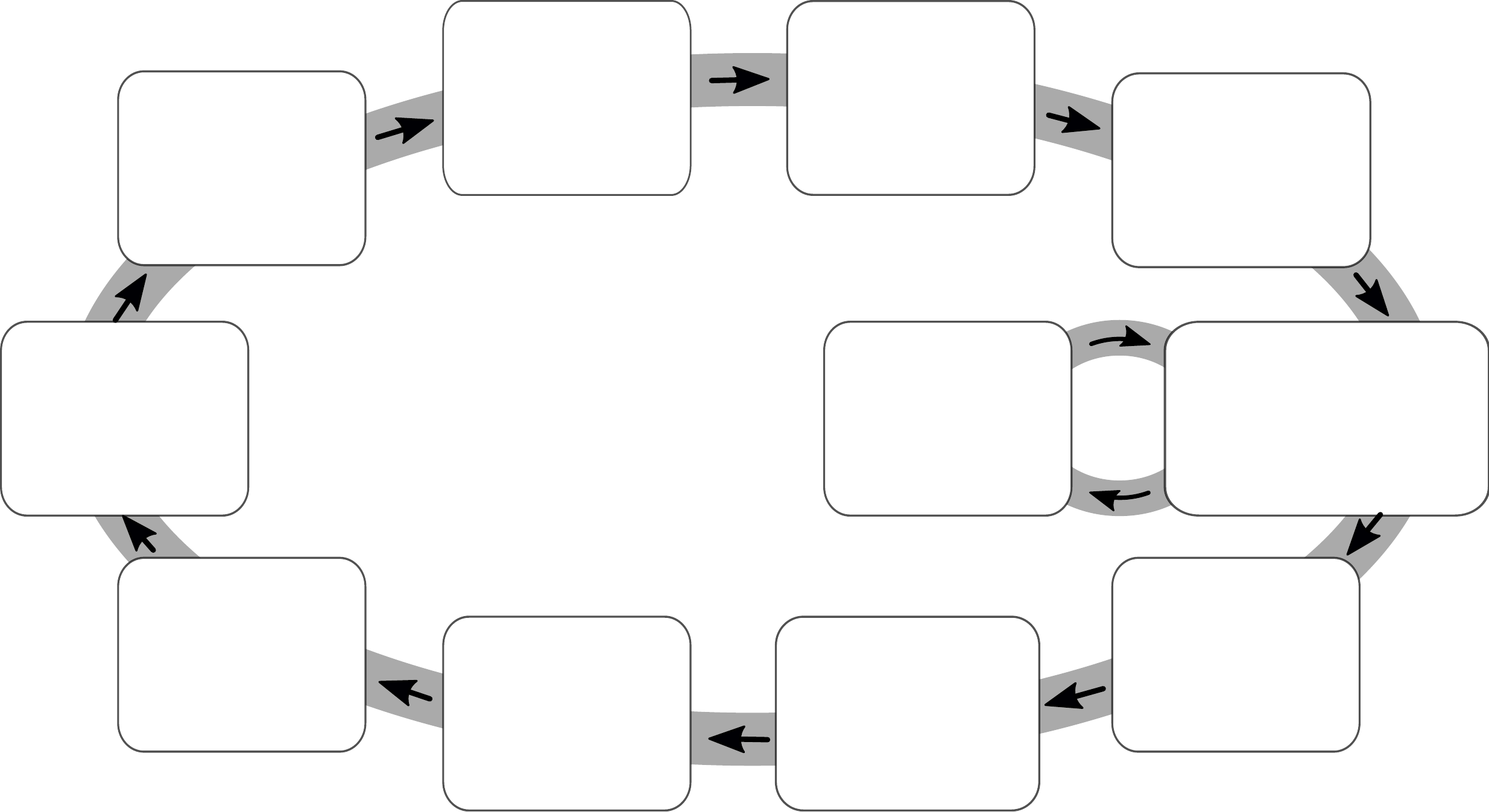}};

    \node[align=center, text width=8em] at (1.7,3.1) (tube)
    {\textbf{TUBE}\\\ref{sec:ca:tube}\\
      compute $A_t$, $p_t$, $Z_t$
      \\(\ref{eq:ca:tube_cross_section},~\ref{eq:ca:ptube},~\ref{eq:impedance})
    };

    \node[align=center, text width=8em] at (5.05,2.35) (sarc1)
    {\textbf{SARCOMERE}\\\ref{sec:ca:sarcomere} \\
      estimate $E_c^\mathrm{fib}$, $\sigma_c^\mathrm{fib}$, $\kappa_c^\mathrm{fib}$
      \\(\ref{eq:ca:sarc_strain},~\ref{eq:ca:sarc_total},~\ref{eq:ca:sarc_stiff})
    };

    \node[align=center, text width=11em] at (5.9,-0.15) (cav1)
    {\textbf{CAVITY}\\~\ref{sec:ca:chamber} \\
        update $E_c^\mathrm{fib}$ (\ref{eq:cav_fiber_stress}) and \\
       $C_c^\mathrm{mid}$, $A_c^\mathrm{mid}$, $T_c^\mathrm{mid}$
      \\ (\ref{eq:ca:Ccmid},~\ref{eq:ca:Acmid},~\ref{eq:midwall_tension})
    };

    \node[align=center, text width=8em] at (2.1,-0.15) (triseg)
    {\textbf{TRISEG}\\~\ref{sec:ca:triseg} \\
      update
      {\color{darkgreen}$\dot{y}^\mathrm{mid}$, $\dot{V}_\mathrm{\SV}^\mathrm{mid}$}
      \\(\ref{eq:tension_function}--\ref{eq:ca:ode_septum})
    };

    \node[align=center, text width=9em] at (5.0,-2.55) (cav2)
    {\textbf{CAVITY}\\~\ref{sec:ca:chamber} \\
      compute \\$A_c$, $Z_c$, $p_c$
      \\(\ref{eq:ca:cav_cross_section},~\ref{eq:ca:wave_impedance},~\ref{eq:ca:pc_trans})
    };

    \node[align=center] at (1.68,-3.20) (sarc2)
    {\textbf{SARCOMERE}\\\ref{sec:ca:sarcomere} \\
      update $\sigma_c^\mathrm{fib}$ (\ref{eq:ca:sarc_total})\\
      compute {\color{darkgreen}$\dot{L}_c^\mathrm{cont}, \dot{C}_c$}
      \\(\ref{eq:ca:sarcomerelength},\ref{eq:ca:contractility})
    };

    \node[align=center, text width=8em] at (-1.80,-3.17) (peri)
    {\textbf{PERICARD}\\\ref{sec:ca:pericardium} \\
      compute $p_\mathrm{peri}$ and update $p_c$
      \\ (\ref{eq:ca:pperi},~\ref{eq:ca:periupdate})
    };

    \node[align=center] at (-5.1,2.4) (valve)
    {\textbf{VALVE}\\\ref{sec:ca:valve}\\
      compute {\color{darkgreen}$\dot{q}_v$}
      \\(\ref{eq:ca:valve_flow})
    };

    \node[align=center] at (-5.05,-2.55) (periphery)
    {\textbf{PERIPHERY}\\\ref{sec:ca:periphery} \\
      compute $q_{py}$
      \\(\ref{eq:ca:qperiphery})
    };

    \node[align=center] at (-6.3,-0.2) (connect)
    {\textbf{CONNECT}\\\ref{sec:ca:connect} \\
      update $p_t, p_c$,\\ compute {\color{darkgreen}$\dot{V}_c, \dot{V}_t$}
      \\(\ref{eq:ca:py_vol_dot}--\ref{eq:ca:vlv_p_up})
    };

    \node[align=center, text width=8em] at (-1.8,3.15) (solve)
    {\textbf{SOLVE}\\\ref{sec:ca:solve}\\
      update \\{\color{darkgreen}$V_t$, $V_c$, $C_c$, $L_c^\mathrm{cont}$, $q_v$,
      $y^\mathrm{mid}$, $V_\mathrm{\SV}^\mathrm{mid}$}
    };

  \end{tikzpicture}
  \caption{\reviewerOne{Solution process of the lumped ODE model of the
    circulatory system. The \ca model}
    connects tubes ($t$), cavities ($c$), valves ($v$), and
    pulmonary and systemic periphery ($py$).
    In each timestep the ODE system is solved using a
    Runge--Kutta--Fehlberg method, see~\ref{sec:ca:solve}, to update the  
    ODE variables (in green), i.e., volumes of tubes ($V_t$) and cavities ($V_c$);
    sarcomere contractility ($C_c$) and sarcomere length ($L_c^\mathrm{cont}$) for
    each of the cavities and the septum;
    flow over valves ($q_v$);
    and septal midwall volume ($V_\mathrm{\SV}^\mathrm{mid}$)
    and radius ($y^\mathrm{mid}$), see~\Cref{fig:triseg}c.
    In the following steps the updated variables are used to compute current
    pressures ($p_{c},p_{t}$), cross sectional areas ($A_{c},A_{t}$), and
    impedances ($Z_{c},Z_{t}$) for tubes and cavities;
    fiber strain ($E_c^\mathrm{fib}$),
    fiber stiffness ($\kappa_c^\mathrm{fib}$),
    and fiber stress ($\sigma_c^\mathrm{fib}$) for the sarcomeres of each cavity and
    the septum; midwall curvature ($C_c^\mathrm{mid}$),
    midwall area ($A_c^\mathrm{mid}$), and midwall tension ($T_c^\mathrm{mid}$)
    for each cavity and the septum;
    pericardial pressure $p_\mathrm{peri}$;
    and flow over the systemic and pulmonary periphery $q_{py}$.
    Based on~\cite{walmsley2015fast}.
    }%
    \label{fig:circadapt}
\end{figure}

\label{sec:CircAdapt}
\ca~\cite{arts2005adaptation}, as shown schematically in~\Cref{fig:circadapt},
is a lumped 0D model of heart and circulation.
It enables real-time simulation of cardiovascular system dynamics
under a wide variety of physiological and pathophysiological situations.
The entire cardiovascular system is modeled as a concatenation of modules:
a tube module representing the systemic and pulmonary arteries and veins~(\ref{sec:ca:tube});
a chamber module modeling actively contracting chambers, i.e.,
left and right atria and ventricles~(\ref{sec:ca:chamber}), respectively,
where myofiber mechanics and contraction is described by a sarcomere module~(\ref{sec:ca:sarcomere});
following~\citet{lumens2009triseg} this also includes inter-ventricular mechanical interaction
through the inter-ventricular septum~(\ref{sec:ca:triseg});
a valve module representing the aortic, mitral, pulmonary, and tricuspid valves~(\ref{sec:ca:valve});
a module representing systemic and pulmonary peripheral microvasculatures~(\ref{sec:ca:periphery});
and a module accounting for effects of the pericardium~(\ref{sec:ca:pericardium}).
The modules are connected by flows over valves and venous-atrial inlets~(\ref{sec:ca:connect}).
The whole lumped model consists of 26 ordinary differential equations (ODEs)
which are solved using an adaptive Runge--Kutta--Fehlberg method (RKF45)~(\ref{sec:ca:solve}).

In~\ref{sec:CircAdaptequations} the mathematical underpinnings of the \ca model are outlined.
Briefly, cavity pressures and cavity volumes are interconnected as follows:
volumes regulate cavity wall areas, which in turn determine strain of the myofibers in the wall.
Strain is used to calculate myofiber stress,~(\ref{eq:ca:sarc_active}, \ref{eq:ca:sarc_passive}),
which drives wall tension in each cardiac wall~\eqref{eq:midwall_tension}.
Using Laplace's law, transmural pressure is calculated from wall tension and curvature for each wall~\eqref{eq:ca:pc_trans}.
Cavity pressures are found by adding the transmural pressures to the
intra-pericardial pressure surrounding the myocardial walls~\eqref{eq:ca:periupdate}.
Consecutively, cavity pressures are used to update flow over valves~\eqref{eq:ca:valve_flow}
and thus intra-cavitary volumes~\eqref{eq:ca:vlv_vol_dot}.

A significant advantage of the modular setup of the model is
that a simple 0D module can be straightforwardly replaced by the more complicated finite element (FE) model in~\Cref{sec:fe-model}.
In this setup \ca provides realistic boundary conditions to the FE problem, see~\Cref{sec:pde_ode_coupling}.

The version of the \ca model used for all simulations has been published previously~\cite{walmsley2015fast}
and can also be downloaded from the \ca website (\url{http://www.circadapt.org}).

\subsection{PDE-ODE Coupling} \label{sec:pde_ode_coupling}
\reviewerOne{We introduce the set of cavities
$\mathcal{C}=\{\mathrm{\LV, \RV, \LA, \RA}\}$,
with the left ventricle (\LV), the left atrium (\LA), the right ventricle (\RV),
and the right atrium (\RA);
the set of cavities $\mathcal{C}^\mathrm{PDE}\subseteq\mathcal{C}$ that
are modeled as a 3D PDE model;
and the set of cavities
$\mathcal{C}^\mathrm{ODE}=\mathcal{C}\setminus\mathcal{C}^\mathrm{PDE}$
that are modeled as a 0D ODE model.}
Coupling between PDE and ODE models can be achieved in various ways.
Fundamentally, the problem is to find the new state of deformation $\mathbf{u}_{n+1}$
as a function of the pressure $p_{n+1}$ in a given cavity at time $n+1$.
The pressure $p_{n+1}$ is applied as a Neumann boundary condition at the cavitary surface, see~\Cref{eq:neumann_bc}.
This pressure is not known and has to be determined in a way
which depends on the current state of the cavity.
Basically, two scenarios have to be considered:
(i) when all valves are closed, the cavity is in an isovolumetric state.
 That is, the muscle enclosing the cavity may deform,
 but the volume has to remain constant.
 Therefore if active stresses vary over time during an isovolumetric phase,
 the pressure $p_{n+1}$ in the cavity has to vary as well to keep the cavitary volume constant;
(ii) when at least one valve is open or regurging, the cavitary volume is changing.
 In this case the pressure $p_{n+1}$ is influenced by the state of the circulatory system
 or of a connected cavity.
 Thus $p_{n+1}$ has to be determined in a way that matches mechanical deformation
 and state of the system. Pressure $p_{n+1}$ in the cardiovascular system
 depends on flow and flow rate which are governed by cardiac deformation
 and as such the two models are tightly bidirectionally coupled.

The simplest approach for the PDE-ODE coupling is to
\reviewerOne{determine $p_{n+1}$ using a partitioned scheme}~\cite{eriksson2013influence,kerckhoffs2007coupling,usyk2002compuational}.
During ejection phases this is achieved by updating cavity volumes and flow
based on the current prediction on the change in the state of deformation
under the currently predicted pressure $p_{n+1}$.
In this scenario the pressure boundary condition in each non-linear solver step
is modified within each Newton iteration $k$.
The new prediction $p_{n+1}^{k+1}$ is then prescribed explicitly as a Neumann boundary condition.
While this \reviewerOne{partitioned} approach is easy to implement and may be
incorporated into an existing FE solver package without difficulty, it may introduce inaccuracies during ejection phases and
\reviewerOne{its convergence may deteriorate during isovolumetric
    phases~\cite{Hirschvogel2017monolithic}.
Instabilities are related to the so-called balloon dilemma~\cite{Kuttler2006} and}
stem from the problem of estimating the change in pressure
necessary to maintain the volume.
Inherently, this requires to know the pressure-volume ($pV$) relation of the cavity
at this given point in time.
However, this knowledge on chamber elastance is not available
and thus iterative estimates are necessary to gradually inflate or deflate
a cavity to its prescribed volume.
As the elastance properties of the cavities are highly non-linear,
an overestimation may induce oscillations
and an underestimation may lead to very slow convergence
and a punitively large numbers of Newton iterations.

\begin{figure}\scriptsize
   \includegraphics[width=\textwidth]{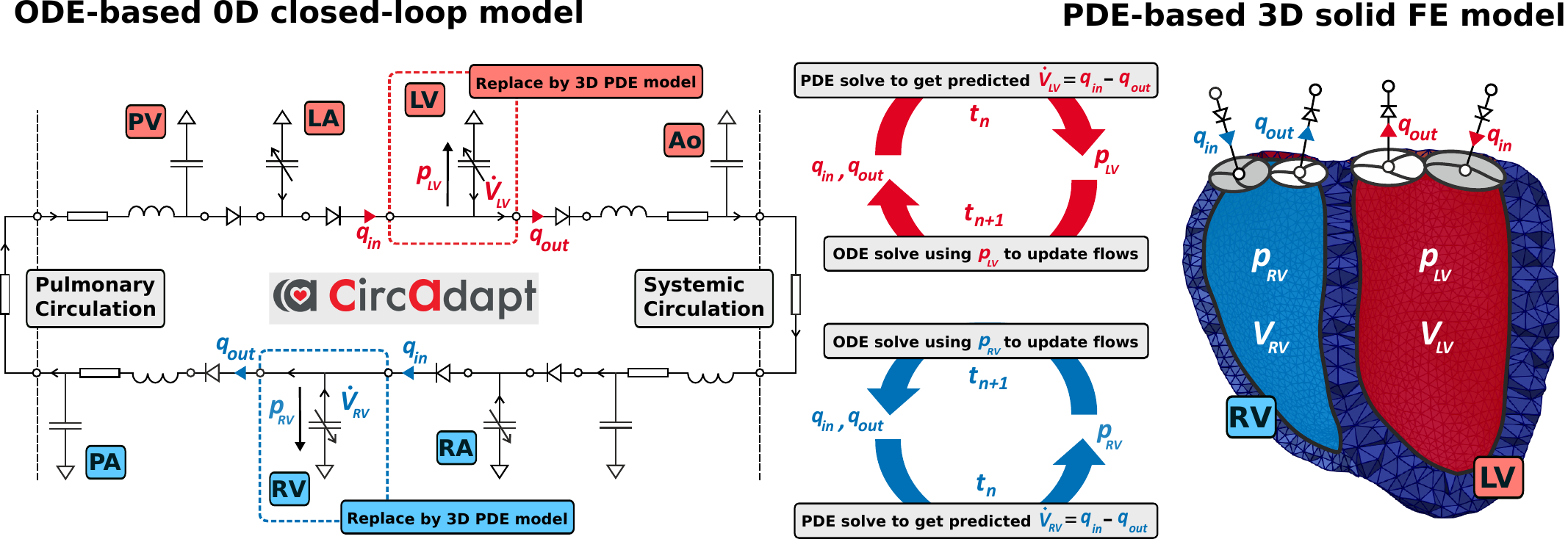}
   \caption{\reviewerOne{Schematic showing the coupling of the 0D ODE model,
    represented by the electrical equivalent circuit, to the 3D PDE model,
    represented by the FE mesh.
    In this case the ventricles (LV, RV) in the lumped model are replaced by
    3D PDEs, while the atria (LA, RA) are modeled as lumped cavities in the \ca model.
    Volume changes of the 3D cavities $\dot{V}_\mathrm{LV},\dot{V}_\mathrm{RV}$
    are driven by flow $q_\bullet$ of blood over valves and outlets computed by the
    0D model. In turn, updated pressures $p_\mathrm{LV}$ and $p_\mathrm{RV}$
    are used as an input to the lumped model in the next time step $t_\mathrm{n+1}$.
    The opening and closure of valves is only modeled in the lumped
    model and in the 3D model triangulated membranes are used to close
    the LV and RV cavities.
    Red colors indicate oxygenated and blue colors de-oxygenated blood.}}%
  \label{fig:3D0Dcoupling}
\end{figure}

A more elaborate approach is to treat $p_{n+1}$ as an
\reviewerOne{additional unknown in a monolithic scheme~\cite{kerckhoffs2007coupling,gurev2011models,Hirschvogel2017monolithic,rumpel2003volume}}.
In addition to the equilibrium equations~(\ref{eq:current}--\ref{eq:neumann_bc}) this requires one further
equation \reviewerOne{for each cavity $c\in\mathcal{C}^\mathrm{PDE}$.
Using this approach}, we get $N_\mathrm{cav}=\left|\mathcal{C}^\mathrm{PDE}\right|$, the number of PDE cavities, additional equations of the form
\begin{equation} \label{eq:VolumeEqu}
    V_c^\mathrm{PDE}(\vec{u},t)-V_c^\mathrm{ODE}(p_c, t)=0,\quad
    c\in\mathcal{C}^\mathrm{PDE}
\end{equation}
where $V_c^\mathrm{PDE}(\vec{u}, t)$ is the cavity volume computed
as the integral over the current surface $\Gamma_{c,t}$, see~\Cref{eq:volumeOmega},
and $V_c^\mathrm{ODE}(p_c, t)$ is the cavity volumes as predicted by the
\ca model for the intra-cavitary pressure $p_c$, see~\Cref{sec:CircAdapt} and~\ref{sec:CircAdaptequations}.

We write $\pcfem={[p_{c}]}_{c\in\mathcal{C}}$
for the vector of up to $1 \le N_\mathrm{cav} \le 4$ pressure unknowns.
Then, linearization of the variational problem, see~\ref{sec:linearization},
a Galerkin FE discretization, see~\ref{sec:assembling},
and a time integration using a generalized-$\alpha$ scheme, see~\ref{sec:generalized_alpha},
result in solving the
block system to find $\delta\Rvec{u}\in\mathbb{R}^{3N}$ and
$\delta \pcfem\in\mathbb{R}^{N_\mathrm{cav}}$ such that
\begin{equation} \label{eq:block_system}
  \tensor{K}'(\Rvec{u}^k, \pcfem^k)
  \begin{pmatrix} \delta\Rvec{u}\\ \delta \pcfem \end{pmatrix}
  = -\Rvec{K}(\Rvec{u}^k, \pcfem^k), \quad
  \Rvec{K}(\Rvec{u}^k, \pcfem^k) := \begin{pmatrix}\Rvec{R}_\alpha(\Rvec{u}^k, \pcfem^k) \\ \Rvec{R}_\mathrm{p}(\Rvec{u}^k, \pcfem^k)
      \end{pmatrix},
\end{equation}
with the updates
\begin{align}
    \Rvec{u}^{k+1} &= \Rvec{u}^{k} + \delta\Rvec{u},\\
    \pcfem^{k+1}   &= \pcfem^k + \delta\pcfem.
\end{align}
Here, $\Rvec{u}^k\in\mathbb{R}^{3N}$ and
$\pcfem^k\in\mathbb{R}^{N_\mathrm{cav}}$ are the solution vectors at the $k$-th Newton
step. The block tangent stiffness matrix $\tensor{K}'$ is assembled according to
\Crefrange{eq:lhs_assemblingv}{eq:Cp_assembling} and \Cref{eq:ga:stiffness}
and the right hand side vector $\Rvec{R}_\alpha$ according to~\Cref{eq:residual_K,eq:residual_B} and \Cref{eq:ga:residual}.
The residual $\Rvec{R}_\mathrm{p}$ which measures the accuracy of the current coupling is the discrete version
of~\eqref{eq:VolumeEqu}, i.e.,
\begin{equation} \label{eq:VolumeEquDiscrete}
  \Rvec{R}_\mathrm{p}(\Rvec{u}^k, \pcfem^k)
    := \Rvec{V}^\mathrm{PDE}(\Rvec{u}^k)-\Rvec{V}^\mathrm{ODE}(\pcfem^k).
\end{equation}
The whole procedure to perform the PDE-ODE coupling is
\reviewerOne{given in~\Cref{alg:iter_afterload}. In short,
volume changes of the 3D cavities are driven by flow of blood over
valves and outlets computed by the 0D model.
In turn, updated pressures in the 3D cavities are used as an input to the
lumped model in the next time step.}
\reviewerOne{Note that~\Cref{eq:block_system} is a block system with
  $\delta \pcfem$ holding at most four unknowns.
  Hence, we can apply a Schur complement approach for a small number of
  constraints, as described in~\ref{sec:SchurComplement}, to simplify the
  numerical solution of this linearized system, see~\Cref{sec:numerical_framework}.}
\reviewerOne{the cavitary volume of each chamber that is described as a
3D FE model has to be tracked as a function of time:
$V^\mathrm{PDE}_c(\vec{x},t)$ for $c\in\mathcal{C}$,}

\reviewerOne{While the described approach works for any
combination of 3D PDE chambers and 0D ODE chambers, we consider
biventricular FE models for our numerical examples in~\Cref{sec:results}.
This is, the ventricles are modeled as 3D PDEs as in \Cref{sec:fe-model}
and the atria are modeled as 0D ODEs described by the \ca
model as in~\ref{sec:ca:chamber}.
See~\Cref{fig:3D0Dcoupling} for a schematic of this 3D solid--0D fluid coupling.
}
\begin{algorithm}
\caption{Coupling of the lumped ODE model to the 3D PDE model}%
\label{alg:iter_afterload}
\begin{spacing}{1.2}
\begin{algorithmic}[1]
 \vspace{0.5em}
 \State{Initialize time $n=0$}
 \State{\reviewerOne{Initialize the set of cavities
        $\mathcal{C}=\{\mathrm{\LV, \RV, \LA, \RA}\}$,
        the set of PDE cavities
        $\mathcal{C}^\mathrm{PDE}\subseteq\mathcal{C}$,
        and the set of ODE cavities
        $\mathcal{C}^\mathrm{ODE}=\mathcal{C}\setminus\mathcal{C}^\mathrm{PDE}$}}
 \State{Initial displacement $\Rvec{u}_0 = \Rvec{0}$}
 \State{Initial cavity pressures $\Rvec{p}_{\mathcal{C},0} = {[p_c]}_{c\in\reviewerOne{\mathcal{C}^\mathrm{PDE}}}$ at time $n=0$}
 \State{Initialize final time point $n_{\max}$ and maximal number of Newton iterations $k_{\max}$}
 \State{Initialize Newton tolerance $\epsilon=10^{-6}$}
 \State{Compute initial cavity volumes $\Rvec{V}^\mathrm{PDE}(\Rvec{u}^{k}) = {[V_c^\mathrm{PDE}(\Rvec{u}^{k})]}_{c\in\reviewerOne{\mathcal{C}^\mathrm{PDE}}}$}
 \State{Run \ca ODE system, see~\Cref{fig:circadapt}, until steady-state is found and get
 $\Rvec{V}^\mathrm{ODE}(\Rvec{p}_{\mathcal{C},0}) = {[V_c^\mathrm{ODE}(p_c)]}_{c\in\reviewerOne{\mathcal{C}^\mathrm{PDE}}}$}
 \While{$n<n_{\max}$}
 \State{Initialize Newton iterator: $k=0$}
 \State{Initial guesses for Newton: $\Rvec{u}^0=\Rvec{u}_{n}$,
 $\pcfem^0=\pcfem[n]$}
 \While{$k<k_{\max}$}
 \State{Assemble block matrix}
 \Comment{\Crefrange{eq:lhs_assemblingv}{eq:Cp_assembling} and \Cref{eq:ga:stiffness}}
 \Statex{ \hskip\algorithmicindent\hskip\algorithmicindent and right hand side.}
 \Comment{\Cref{eq:residual_K,eq:residual_B,eq:VcPDE_assembling,eq:VcODE_assembling,eq:ga:residual,eq:ga:mass_damping,eq:ga:stiffness_damping}}
 \State{Solve linearized system for $\delta\Rvec{u}$ and $\delta \pcfem$}
 \Comment{\Cref{eq:blockCVSystem}}
 \State{Update displacement $\Rvec{u}^{k+1} = \Rvec{u}^{k} + \delta\Rvec{u}$ and cavity pressures $\pcfem^{k+1} = \pcfem^k + \delta\pcfem$}
 \State{Update cavity volumes $\Rvec{V}^\mathrm{PDE}(\Rvec{u}^{k+1})$}
 \Comment{\Cref{eq:volumeOmega}}
 \State{Update ODE system and get $\Rvec{V}^\mathrm{ODE}(\pcfem^{k+1})$}
 \Comment{\Cref{fig:circadapt}}
 \vspace*{2mm}
 \State{\textbf{Convergence test:}}
 \If{$\lVert\Rvec{R}_\alpha(\Rvec{u}^{k+1}, \pcfem^{k+1})\rVert_{L^2} < \epsilon$ \textbf{and} $\lVert\Rvec{R}_{\mathrm{p}}(\Rvec{u}^{k+1}, \pcfem^{k+1})\rVert_{\infty} < \epsilon$}
 \State{Solution at time $n+1$: $\Rvec{u}_{n+1} = \Rvec{u}^{k+1}$, $\pcfem[n+1]=\pcfem^{k+1}$}
 \State{\textbf{break}} \Comment{Newton converged}
 \Else%
 \State{$k=k+1$}  \Comment{Next Newton step}
 \EndIf%
\EndWhile%
 \State{$n=n+1$}\Comment{Next time step}
\EndWhile%
\end{algorithmic}
\end{spacing}
\end{algorithm}
\paragraph{Temporal synchronization of chamber contraction}
In the lumped \ca model contraction in individual chambers is controlled
by prescribed trigger events.
Based on the measured heart rate (HR) of 103 beats per minute
contraction of the RA
was triggered at intervals corresponding to a basic cycle length
of $1/\text{HR}=\SI{0.585}{s}$.
In all other chambers contraction was triggered by prescribed delays
relative to the instant of contraction of the RA.
In a hybrid coupled model contraction times
used in 3D EM~(\ref{eq:tanh_stress},\ref{eq:tanh_stress2})
and in the lumped \ca model~\eqref{eq:ca:contractility}
must be synchronized accordingly.
For this sake \reviewerOne{an interconnected event-driven} finite-state machine (FSM) was used
to control activation cycles in both 3D and 0D chamber models.
\reviewerOne{Two types of FSMs were used,
an autorhythmic FSM to generated triggers at a prescribed cycle length
independently of any input,
and a reactive excitable FSM of two possible states, excitable or non-excitable.
The excitable FSM reacts to external trigger input.
If the machine is in excitable state a transition is initiated to the non-excitable state,
otherwise, if in non-excitable state, the FSM does not accept the input
and remains in a non-excitable state.
The FSM returns to its excitable state automatically
after a prescribed effective refractory period.
A transition from an excitable to a non-excitable state sends out a trigger event
to all interconnected FSMs.
These interconnections are implemented as delays
representing the travel time needed for depolarization wave fronts to propagate
to all neighboring interconnected FSMs.}

\begin{figure}[ht!]
	\centering
	\includegraphics[width=0.75\linewidth]{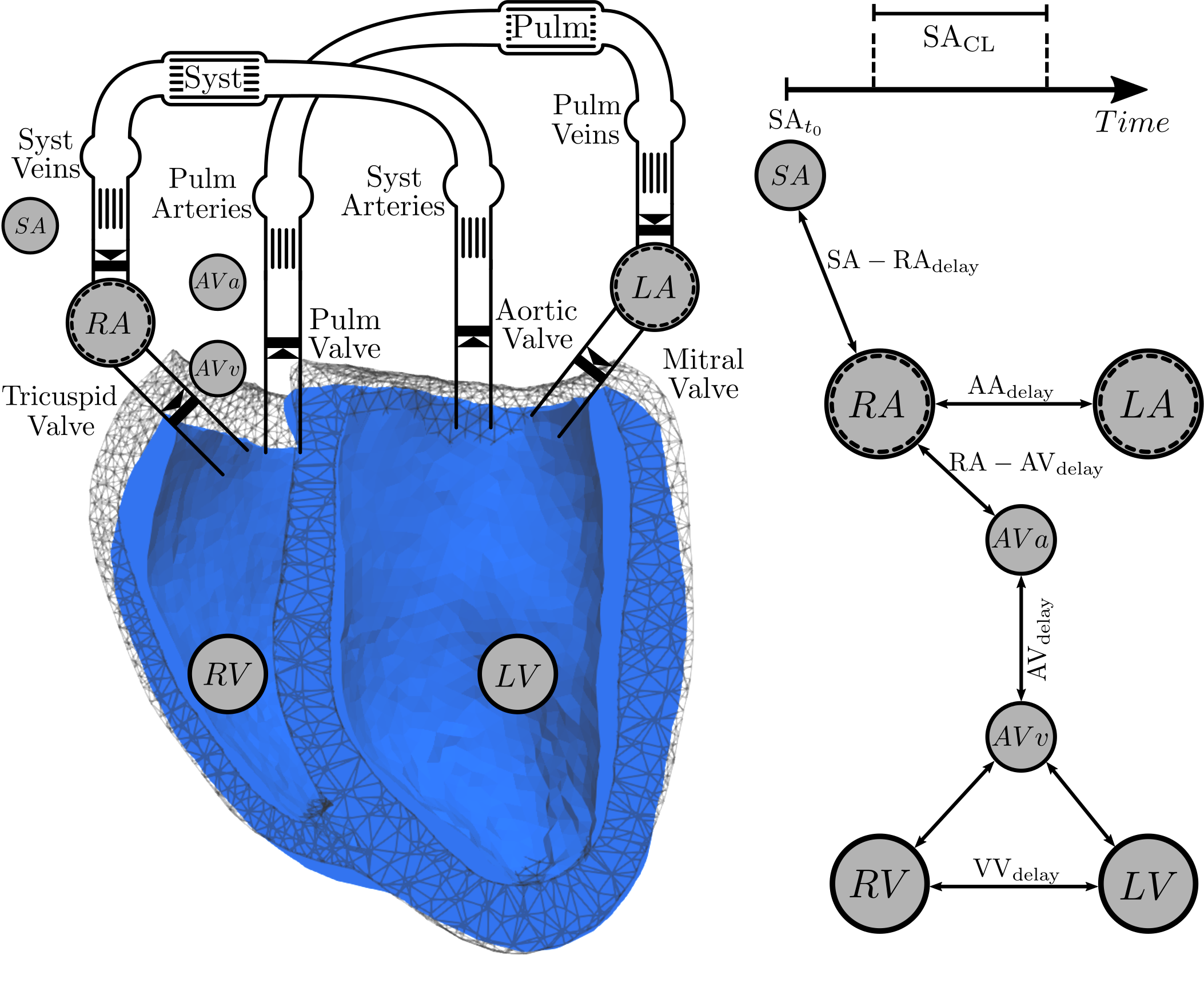}
	\caption{EM activation of the coupled 3D--0D model
	is steered by \reviewerOne{an event-driven interconnected} FSM
    that provides triggers for electrical activation
	of the 3D EM model and for mechanical activation of the 0D lumped atrial cavities.
	The sino-atrial node clock (SA) activates the RA at a prescribed cycle length, SA$_{\rm{CL}}$,
    starting at time SA$_{\rm {t_0}}$.
    \reviewerOne{The LA initiates contraction with a delay of AA$_{\rm {delay}}$ after the RA.
    The atrial entrance into the AV node activates at AV$_{\rm a}$
    which triggers, after the AV$_{\mathrm{delay}}$ elapsed,
    the ventricular exit of the AV node that is connected to the His bundle
    at AV$_{\rm v}$.}
    Fascicles in the LV are activated then relative to the $LV$ trigger
    to initiate electrical propagation in the EP model.
	Similarly, the RV is activated with an interventricular delay of VV$_{\rm d}$
	before (VV$_{\rm d}$<0) or after VV$_{\rm d}>0$ the LV.}%
	\label{fig:_fsm}
\end{figure}

The triggers provided by the FSM can be flexibly linked to
entities within both 3D and 0D model.
\reviewerOne{Specifically, the sino-atrial node is represented by an autorhythmic FSM
that is directly interconnected to the RA.
Both atrial cavities are implemented as a 0D model
that initiate contraction in RA and LA based on FSM trigger events.
The RA FSM connects to the atrial entrance to the atrio-ventricular node
that transduces excitation through the atrio-ventricular (AV) node with a given delay.
The ventricular exit of the AV node is connected to the left and right His bundle
that trigger electrical activation of LV and RV.
In the LV excitation is initiated by antero-septal, septal and posterior fascicle,
and in the RV by a septal and a moderator band fascicle in the RV.
As the timing of all fascicles was synchronous between all fascicles of a given chamber,
fascicular timings were lumped together under RV and LV (see~\Cref{fig:_fsm}).
RV and LV triggers prescribe fascicular activation times $t_{\rm a}$ to the Eikonal equation
that governs electrical activation of the ventricular cavities implemented as PDE model.}
Mechanical contraction of the ventricles was initiated then within a prescribed
EM delay, $t_{\rm {emd}}$.
The overall concept for synchronizing contraction in the coupled model is illustrated
in~\Cref{fig:_fsm} and FSM input parameters are given in \Cref{tab:fsm}.


\begin{table}[htbp]
	\centering
	\footnotesize
	\begin{tabularx}{0.65\textwidth}{ccccccc}
		\toprule
		HR  & CL    & RA  & AA delay  & AV delay & VV delay & ERP \\
		$[\text{beats}/\si{\min}]$ & $[\si{\s}]$ & $[\si{\s}]$ & $[\si{\s}]$ & $[\si{\s}]$  & $[\si{\s}]$ & $[\si{\s}]$\\
		\midrule
		103 & 0.585 & 0.0 &  0.02  & 0.1 & 0.0 & 0.35\\
		\bottomrule
	\end{tabularx}
	\caption{FSM input parameters used for synchronizing electro-mechanical activity
        \reviewerOne{comprise
    heart rate (HR) or cycle length (CL), right atrium (RA), left atrium (LA), intra-atrial (AA), atrio-ventricular (AV) and inter-ventricular (VV) delays and effective refractory period (ERP)}.}%
	\label{tab:fsm}
\end{table}

\subsection{Parameterization of the baseline model}\label{sec:parameterization}

\reviewerTwo{For the sake of physiological validation
the available experimental data were used to calibrate the model
in terms of stroke volume (SV) and peak systolic pressure in the LV ($\ppeak_{\mathrm{\LV}}$).}
Following~\cite{Augustin2016patient},
initial parameters of passive biomechanics,
characterized by the material model given in~\eqref{eq:Usyk2000Q},
were taken from~\cite{sugiura2012multiscale};
the model was unloaded using a backward displacement
algorithm~\cite{Sellier2011} and the material law's scaling
parameter $a$ was determined then by fitting the LV model to an
empirical Klotz relation~\cite{klotz2007computational},
using end-diastolic pressure ($\ped$) and volume ($\Ved$) as input.
Active stress model parameters $\tau_{\mathrm{c}}$, $S_\mathrm{peak}$,
$\tau_{\mathrm{r}}$ and duration of the force transient, $t_{\rm {dur}}$
were determined as described previously \cite{marx2020personalization}.
Initial values and parameters for the \ca model
were chosen following \cite{willemen2019:_crt}.
To replicate the observed SV and $\ppeak_{\mathrm{\LV}}$
in the left ventricle, input parameters of the active stress model
as well as \ca model parameters were iteratively adjusted.
The final parameterized model beating at 103 bpm
produced a cardiac output of $\approx$\SI{2.1}{\liter/\min} with a SV of \SI{21}{\milli \liter},
in keeping with the experimental data.

\subsection{Physiological testing}
The coupled 3D-0D model was subjected to thorough physiological testing
by evaluating its transient response to alterations in loading conditions and contractile state.
The model under baseline condition was used as a reference working point
relative to which the effect of perturbations in loading and contractility was compared.
Standard protocols for assessing of systolic and diastolic properties of the ventricles
based on $pV$ analysis \cite{burkhoff2005assessment} were implemented
to qualitatively gauge the model's ability to consistently predict known cardiovascular physiology.
For all perturbations in preload, afterload, or contractility,
two points in time were considered,
the immediate acute response after perturbing the system
and the new approximate limit cycle reached after 8 beats.
For baseline and each limit cycle,
the end-systolic pressure volume relation (ESPVR) of the LV
was interrogated by imposing additional step changes in afterload.
For this sake, end-systolic pressure ($p_{\rm {es}}$) and volume
($V_{\rm {es}}$) were determined in the $pV$ loops at  the instant of
end-systole, as determined by the cessation of flow out of the LV.
Linear regression was used then to determine end-systolic elastance, $\Ees$,
as the slope of the regression curve,
and the volume intercept, $\Vd$, of the ESPVR.
Step changes in preload, afterload, and contractility were implemented
by varying the cross-sectional area of the pulmonary veins,
the systemic vascular resistance, \Rsys,
and the active peak stress $S_{\rm {peak}}$ generated by the myofilament model, respectively.
Overall pump function was also assessed.
Following \cite{westerhof2019:_snapshots},
the heart as a pump can be described by the pump function graph (PFG),
the relation between mean ventricular pressure,
i.e., the ventricular pressure averaged over the entire cardiac cycle, and Cardiac Output.
A PFG comprehensively describes cardiac pump function
similar to the characterization of industrial pumps and ventricular assist devices.
To construct a PFG,
data were gathered under all protocols,
including additional afterload variations over a wider range,
between $\Ea \approx 0$ by setting the system resistance $\Rsys \approx 0$
and $\Ea \approx \infty$, by closing the aortic valve,
to obtain data points under extreme conditions
corresponding to the LV beating in absence of external loading
and under isovolumetric conditions, respectively.

\subsection{Numerical framework}\label{sec:numerical_framework}
After discretization, at each Newton--Raphson step the block system~\eqref{eq:block_system}
has to be solved.
For this sake, we applied a Schur complement approach,
see~\ref{sec:SchurComplement}, to cast the problem in a pure displacement formulation,
to be able to reuse previously established solver methods~\cite{Augustin2016anatomically}.
In brief, we used the generalized minimal residual method (GMRES) with an
relative error reduction of $\epsilon=10^{-8}$.
Efficient preconditioning was based on \emph{PETSc}~\cite{petsc-user-ref}
and the incorporated solver suite
\emph{hypre/BoomerAMG}~\cite{henson2002boomeramg}.

\reviewerOne{The dynamic version of the mechanics equations~\eqref{eq:current}
was also used in other recent studies on cardiac EM~\cite{Regazzoni2021,Pfaller2019,Hirschvogel2017monolithic} and
showed advantages in performance of the linear solver -- compared to the more common quasi-static approach -- due to a more diagonal-dominant tangent
stiffness matrix~\cite{Jacob1994}.}
For the time integration we used a generalized-$\alpha$ scheme, see~\ref{sec:generalized_alpha},
with spectral radius $\rho_\infty=0$ and damping parameters $\beta_\mathrm{mass}=\SI{0.1}{\ms^{-1}}$, $\beta_\mathrm{stiff}=\SI{0.1}{\milli \second}$.

We implemented the coupling scheme in the FE framework Cardiac Arrhythmia Research Package (CARPentry)~\cite{vigmond2008solvers,neic17:_efficient},
built upon extensions of the openCARP EP framework (\url{http://www.opencarp.org}).
Based on the MATLAB code presented in~\cite{walmsley2015fast}, which is available on the \ca website (\url{http://www.circadapt.org}),
a C\texttt{++} circulatory system module was implemented into CARPentry
to achieve a computationally efficient and strongly scalable numerical scheme
that allows fast simulation cycles.

\reviewerOne{
Execution of the 3D-0D model was sped up
by limiting the number of Newton steps to $k_{\max}=1$
for the initial series of heart beats
that were simulated to stabilize the coupled 3D-0D model to a limit cycle.
This corresponds to a semi-implicit (linearly-implicit) discretization method
\cite{deuflhard2011newton} which worked very well in combination with the generalized-$\alpha$ scheme.
Finally, after arriving at a stable limit cycle two further beats were simulated
using a fully converging Newton method with $k_{\max}=20$ and an
relative $\ell_2$ norm error reduction of the residual of $\epsilon=10^{-6}$.}

\section{Results}%
\label{sec:results}
\subsection{Parameterization of the baseline model}

\begin{figure}[htpb]
	\centering
	\includegraphics[width=0.99\linewidth]{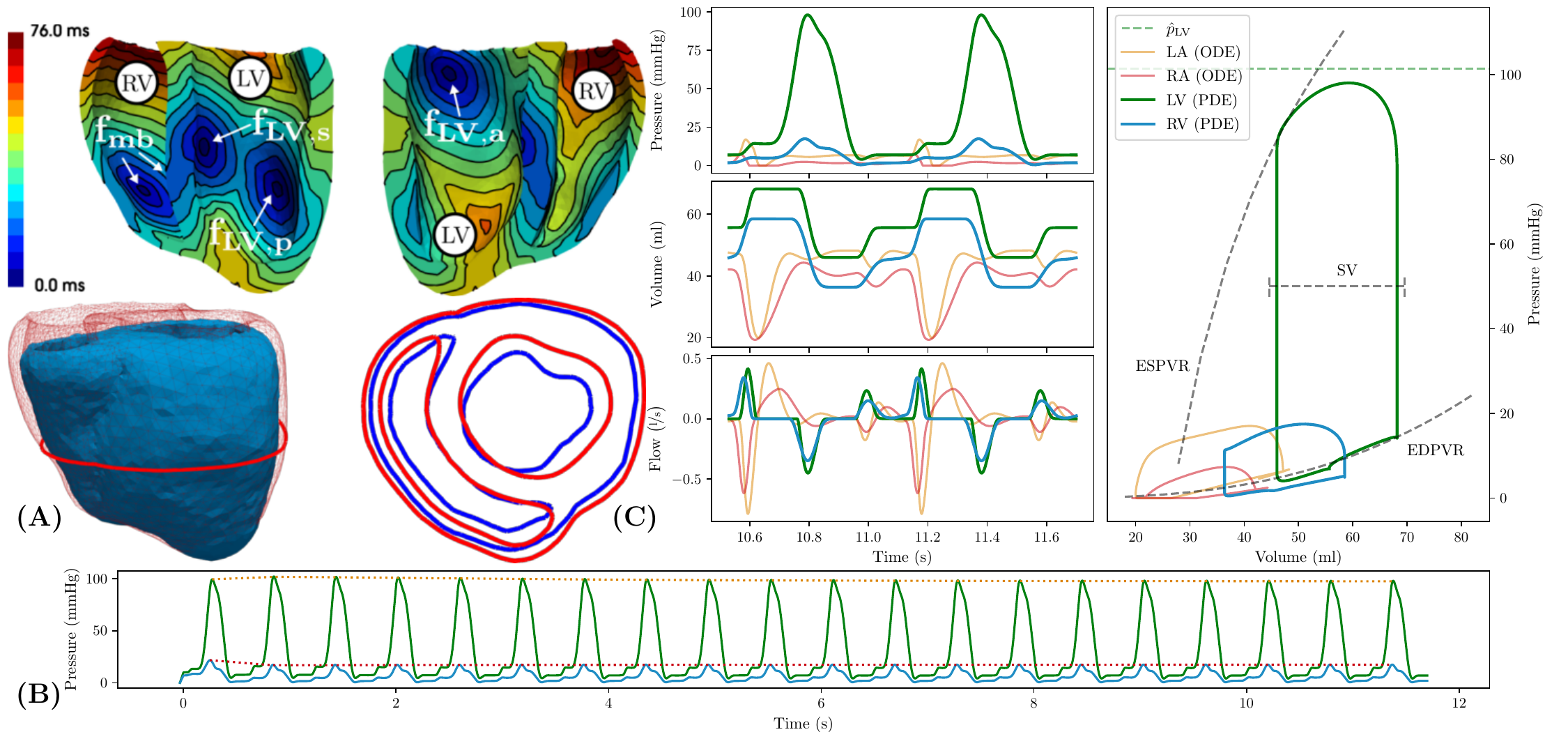}
	\caption{Model parameterization under baseline conditions.
        (A) The top panels show ventricular sinus activation sequence
          induced by three LV ($\mathrm{f}_\mathrm{\LV,a}$, $\mathrm{f}_\mathrm{\LV,s}$,
          $\mathrm{f}_\mathrm{\LV,p}$)
          and two RV fascicles ($\mathrm{f}_\mathrm{mb}$).
          The bottom panels show the mechanical end-diastolic (red) and end-systolic (blue) configuration.
          Note the minor change in epicardial shape due to the pericardial boundary conditions.
        (B) Simulated pressure traces in LV (green) and RV (blue) are shown
          for the entire pacing protocol using a train of 20 beats.
          Envelopes (dotted traces) indicate that an approximate limit cycle was reached
          after 3 beats.
        (C) Left panels show time traces of pressure $p$, flow $q$ and volume $V$
          in lumped 0D atrial cavities and PDE-based ventricular cavities
          for the last two beats of the limit cycle pacing protocol.
          variables traverse the state space along limit cycle trajectories.
          Right panel shows $pV$ loops in all four cavities.
          For PDE-based ventricular cavities EDPVR and ESPVR are indicated.
	  Experimental data on peak LV pressure $\ppeak_{\mathrm{\LV}}$ and stroke volume
	  used for fitting are indicated (dashed lines).
}
    \label{fig:_baseline}
\end{figure}

The coupled 3D-0D model was fit to approximate the experimental observed data
on peak pressure $\ppeak_{\mathrm{\LV}}$ and stroke volume in the LV under baseline conditions.
Electrical activation was driven by a tri- and bi-fascicular model
in LV and RV, respectively, see~\Cref{fig:_baseline}(A).
Conduction velocities were chosen for the given activation pattern
to obtain a total ventricular activation time of $\sim$\SI{75}{\milli \second},
compatible with the observed QRS duration of the ECG.
Mechanical boundary conditions were set to limit radial contraction of the model,
thus leading to a heart beat where ejection was mediated largely by atrio-ventricular plane displacement,
i.e.\ long-axis shortening of the ventricles, and myocardial wall thickening.
The resulting end-diastolic and end-systolic configuration of the model
is shown in~\Cref{fig:_baseline}\reviewerOne{(A)}.
Trains of 20 heart beats were simulated to arrive at an approximate stable limit cycle,
as verified by inspecting the slope of the envelope of key hemodynamic state variables, see~\Cref{fig:_baseline}(C).
Corresponding hemodynamic data on pressure, volume, and flows for all four chambers
along with the corresponding $pV$ loops over the last two beats are shown in~\Cref{fig:_baseline}(B).
Parameter values of the baseline model are given in tables~\Cref{tab:ca_input_params,tab:ca_constants}
and~\Cref{tab:pde_input_params} for the 0D \ca and 3D PDE model, respectively.

\subsection{Effect of spatial resolution}
\begin{figure}[hpb]
	\centering
	\begin{subfigure}{0.39\textwidth}
		\includegraphics[width=\textwidth]{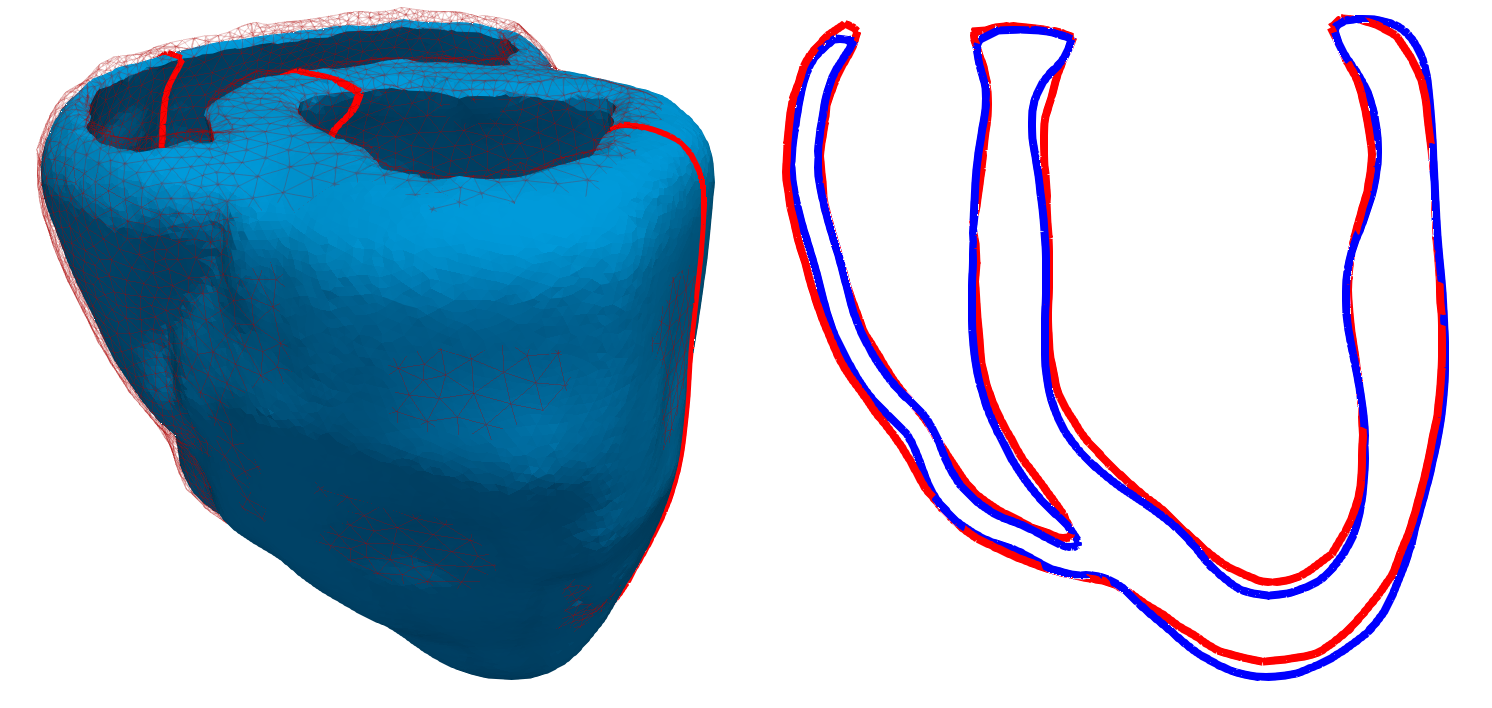}
		\subcaption{}
	\end{subfigure}
	\begin{subfigure}{0.39\textwidth}
		\includegraphics[width=\textwidth]{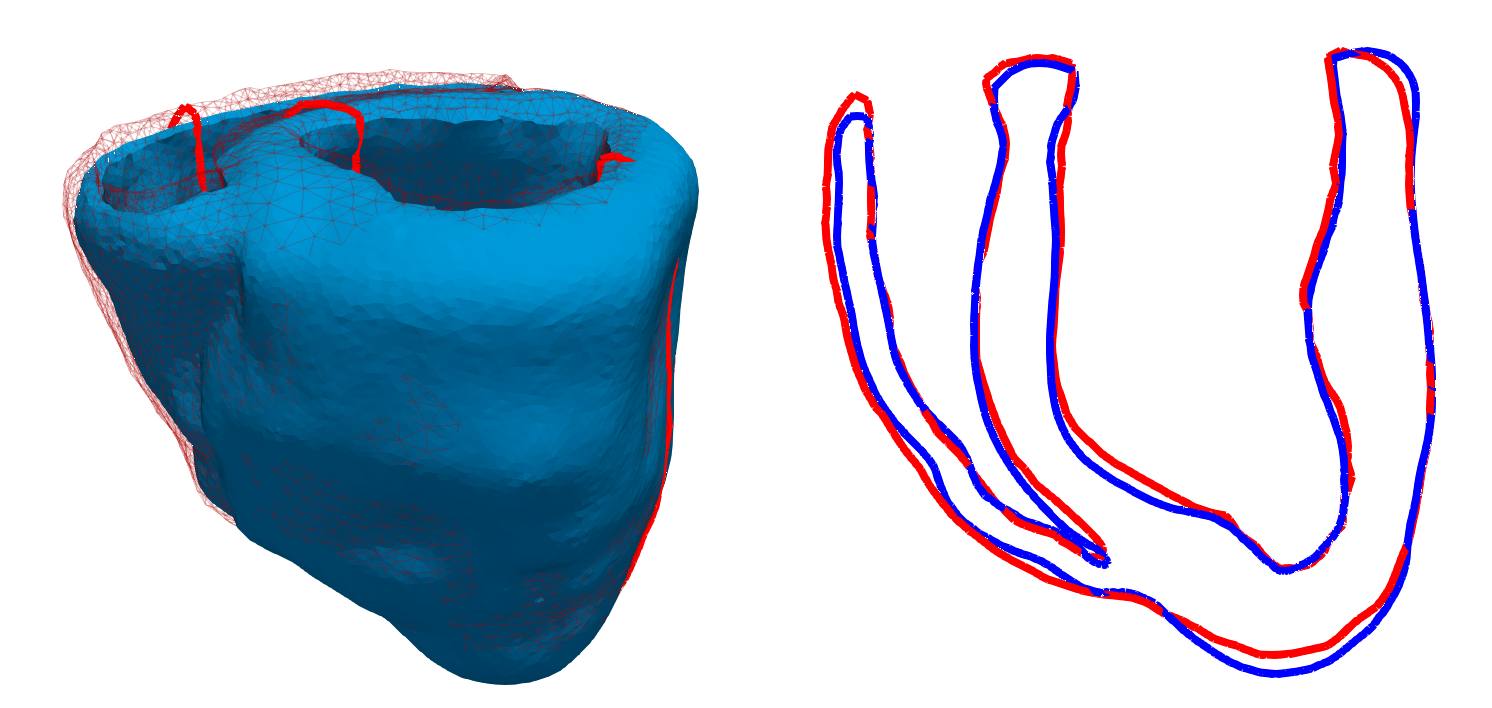}
		\subcaption{}
	\end{subfigure}
	\begin{subfigure}{0.2\textwidth}
  \includegraphics[width=\textwidth]{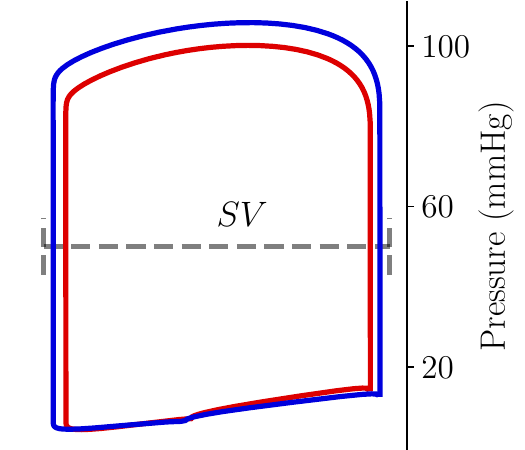}
  \subcaption{}
  \end{subfigure}
	\caption{Differences between the coarse (wireframe, red solid line)
		and higher resolution (solid, blue solid line) model
	are shown for (A) end-diasolic and (B) end-systolic configuration.
	(C) Dynamic behavior over the limit cycle protocol was comparable
        with minor difference in the stroke volume (SV) and peak pressure.}
	\label{fig:rescmp}
\end{figure}

The impact of the relatively coarse spatial resolution
($\sim\!\SI{3.4}{\milli \meter}$ and $\sim\!\SI{2.4}{\milli \meter}$ for LV and RV, respectively) used
was evaluated first by repeating the baseline limit cycle protocol
using a higher resolution mesh
($\sim\!\SI{1.3}{\milli\meter}$ and
$\sim\!\SI{1.2}{\milli\meter}$ for LV and RV, respectively).
Both simulations used the exact same set of parameters and initial state vectors.
With regard to $pV$ behavior that is governed by global deformation of the ventricles,
end-diastolic and end-systolic configurations were compared, see~\Cref{fig:rescmp}.
Observed discrepancies between coarse and higher resolution model were marginal
and well below the limits of experimental data uncertainty.
Differences in $\ppeak_{\mathrm{\LV}}$ and SV were less than $5.4$\% and $6.9$\%, respectively,
suggesting that the computationally efficient coarse model is suitable
for performing a physiological validation study.

\subsection{Physiological testing}

\begin{figure}[ht!]
	\centering
	\begin{subfigure}{0.32\textwidth}
	\includegraphics[width=\linewidth]{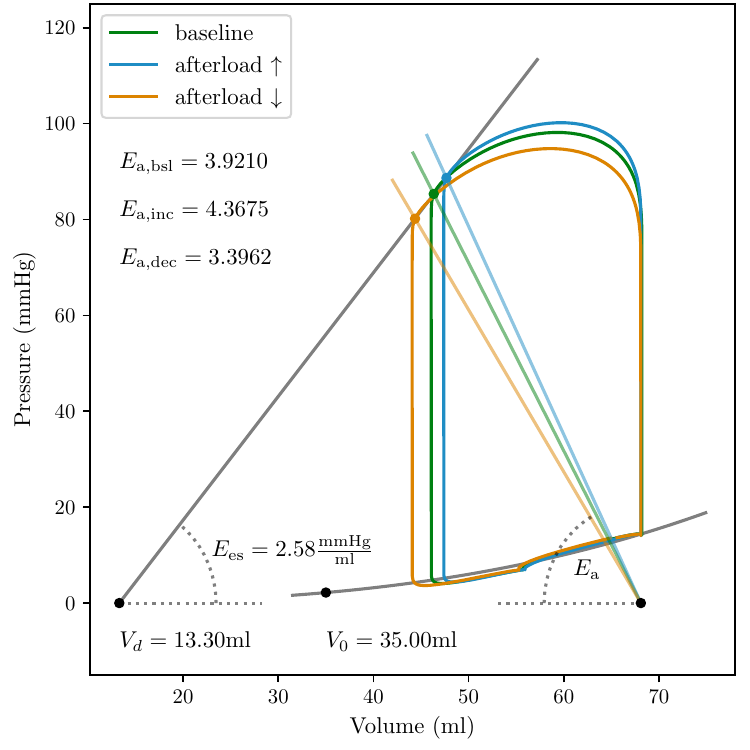}
	\subcaption{}
	\end{subfigure}
	\begin{subfigure}{0.32\textwidth}
	\includegraphics[width=\linewidth]{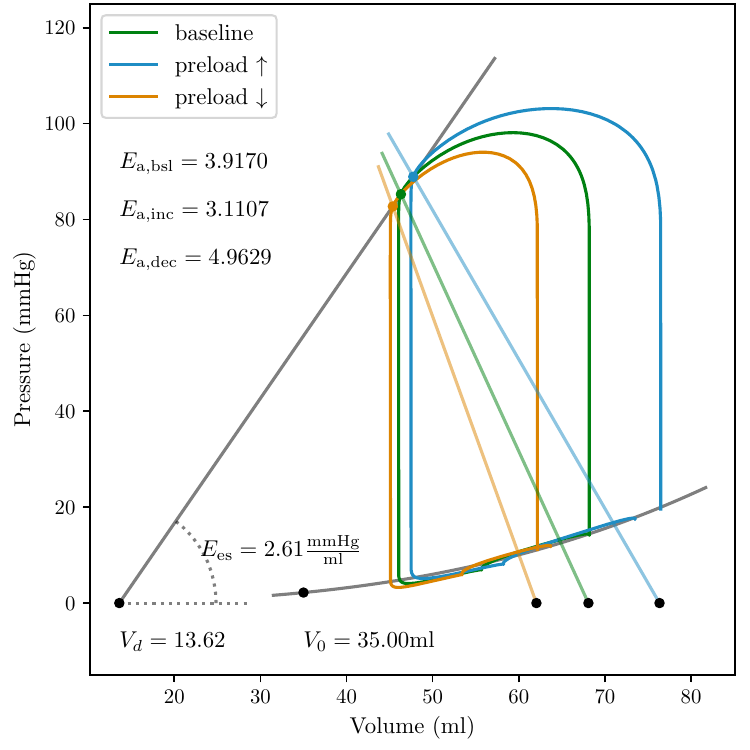}
	\subcaption{}
	\end{subfigure}
	\begin{subfigure}{0.32\textwidth}
	\includegraphics[width=\linewidth]{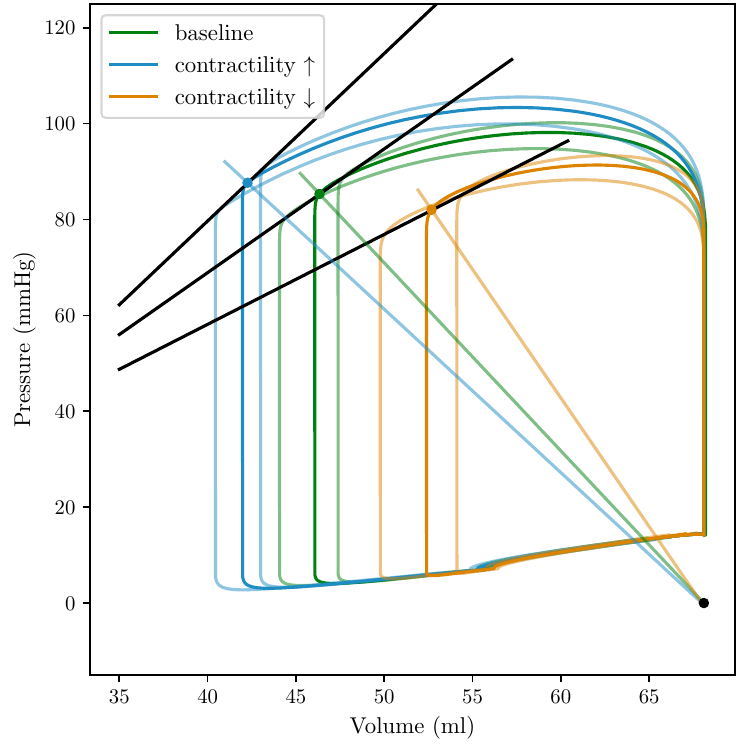}
	\subcaption{}
  \end{subfigure}\\
	\begin{subfigure}{0.32\textwidth}
  \includegraphics[width=\linewidth]{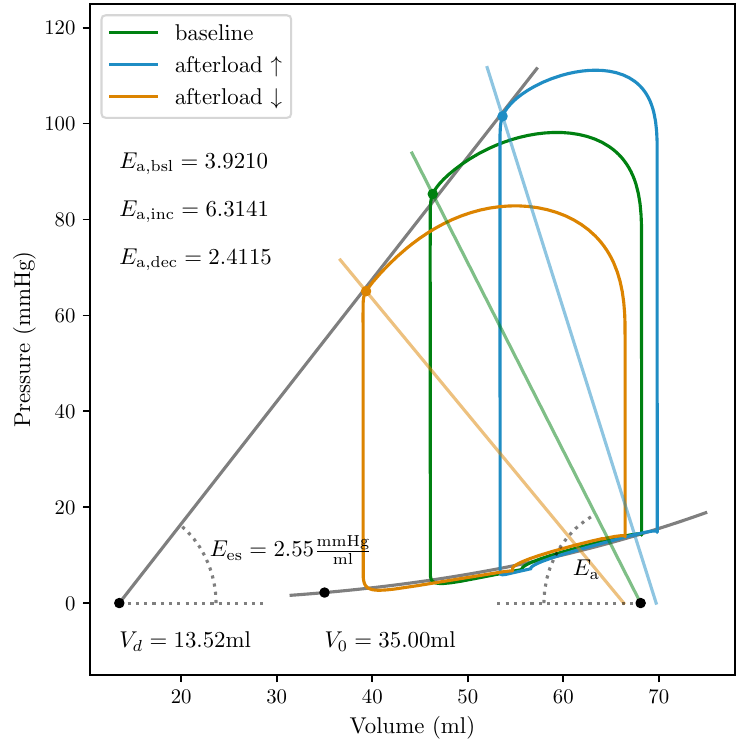}
  \subcaption{}
  \end{subfigure}
  \begin{subfigure}{0.32\textwidth}
  \includegraphics[width=\linewidth]{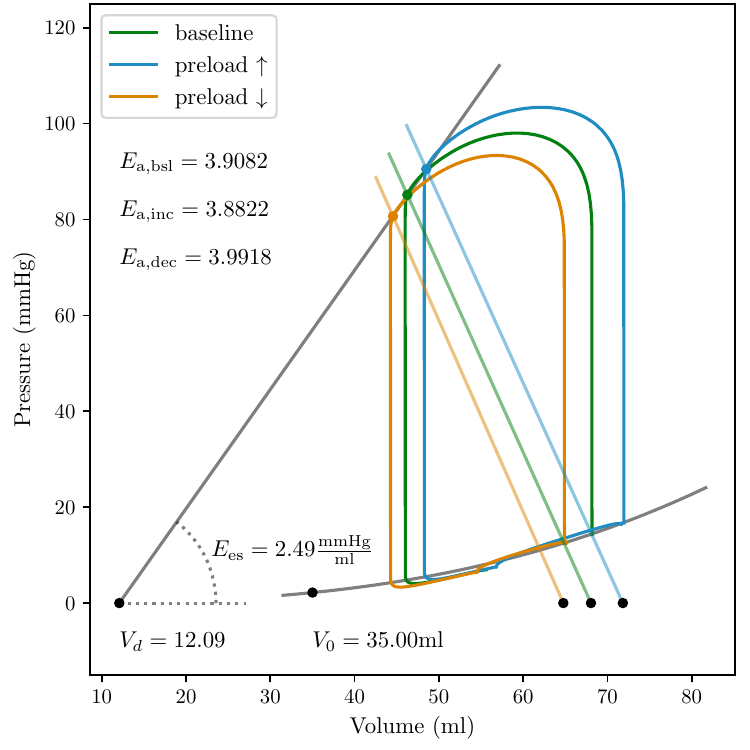}
  \subcaption{}
  \end{subfigure}
  \begin{subfigure}{0.32\textwidth}
  \includegraphics[width=\linewidth]{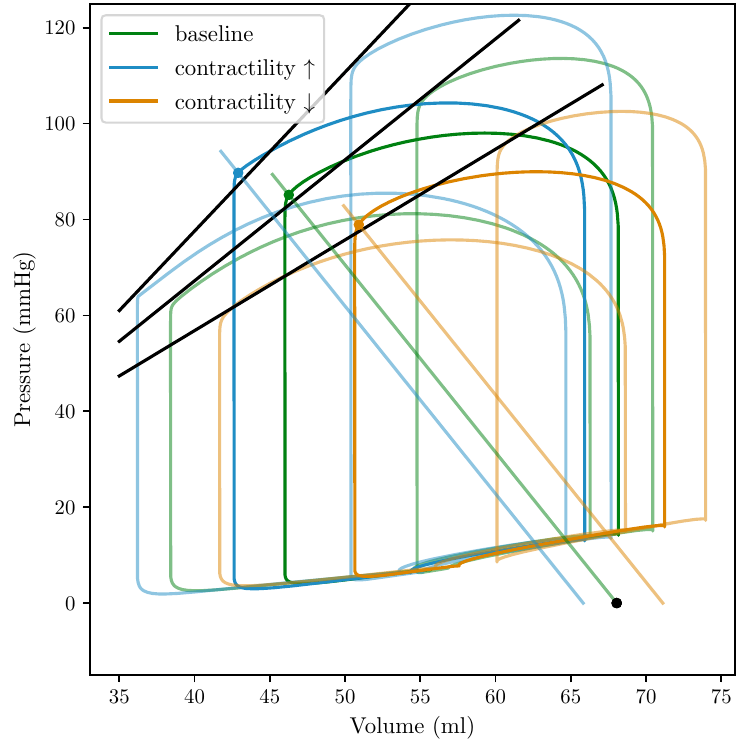}
  \subcaption{}
  \end{subfigure}\\
    \caption{Left ventricular $pV$ loops showing
      the initial response (A-C) and 4 cycles (D-F)
      after applying a step change in loading conditions and contractility.
      \textbf{(A)} Altering afterload by increasing/decreasing
      the systemic vascular resistance, \Rsys\ pivots
      arterial elastance $\Ea$ curve.
      Endsystolic elastance, $\Ees$ and intercept $\Vd$
      characterizing the ESPVR was determined by linear regression
      of end-systolic data points \Ves\ and \pes,
      marked by solid circles.
      \textbf{(B)} Increasing/decreasing preload shifts $\Ea$ curve
      and increases/decreases stroke volume via the Starling mechanism,
      mediated by the length-dependence of the active stress model.
      Determination of ESPVR was consistent with afterload protocol.
      \textbf{(C)} Increasing/decreasing contractility increases/decreases stroke volume,
      LV peak pressure and $p_{\rm {es}}$.
      For each contractile state afterload was also perturbed
      to determine end-systolic elastance $E_{\rm {es}}$ and $V_{\rm d}$.
	}%
	\label{fig:_delta}
\end{figure}

The response of the coupled 3D-0D system to changes in afterload was probed
by altering \Rsys\ in the range of $\pm 65\%$ around its nominal value of
\SI{6350}{\mmHg \per \milli \liter \per \milli \s}
These maneuvers alter the slope of the arterial elastance curve, $\Ea$,
pivoting \Ea\ around the point $V_{\rm {ed}}$ and $p=0$ in the $pV$ diagramme.
The initial response immediately after step changes in afterload
and the new limit cycle are shown in Fig.~\ref{fig:_delta}(A) and (D), respectively.

The response to altering LV preload was then probed
by stepwise reducing blood flow from the lungs into the LA
by varying the cross sectional area of the pulmonary veins.
Under such a \emph{walk down} protocol the \Ea\ curve is shifted to the left
towards smaller end-diastolic volumes $V_{\rm {ed}}$,
without altering its slope, i.e.\ $\Ea = p_{\rm {es}}/\text{SV} \approx \text{const}$,
leading to lower $p_{\rm {es}}$.
Stroke volumes under this protocol are assumed to gradually reduce
due to the Frank--Starling mechanism,
mediated by the length-dependence of active stress generation, $S_{\rm a}(\lambda)$.
Initial and limit cycle response under this preload perturbation protocol
are shown in Fig.~\ref{fig:_delta}(B) and (E), respectively.
As contractile properties remained unchanged,
the same slope $E_{\rm{es}}$ of the ESPVR was obtained
as before under step changes in afterload;
compare estimated $E_{\rm {es}}$ between~\Cref{fig:_delta}(D) and (E).

The effect of step changes in contractility was probed
by altering peak active stresses $S_{\rm {peak}}$ in the  LV by \SI{\pm 20}{\percent}
around the LV nominal value of \SI{100}{\kilo \pascal}
This maneuver steepened/flattened the ESPVR, see~\Cref{fig:_delta}(C) and (F).
In the initial response \Ves\ and \pes\ were affected,
with \Ved\ remaining constant, this led to a change in stroke volume
and an apparent change in arterial elastance estimated by $\Ea \approx p_{\rm {es}}/\text{SV}$,
with $\approx -13.09/+36.81 \%$ relative to baseline.
However, in the limit cycle response after readjustment of preload in all chambers,
all \Ea\ curves had the same slope and were only shifted
according to the new working \Ved\ for the given contractile state.
Transient $pV$ loops under this protocol are shown in~\Cref{fig:_delta}(C) and (F).

\begin{figure}[ht!]
  \centering
  \begin{subfigure}{0.32\textwidth}
  \includegraphics[width=\linewidth]{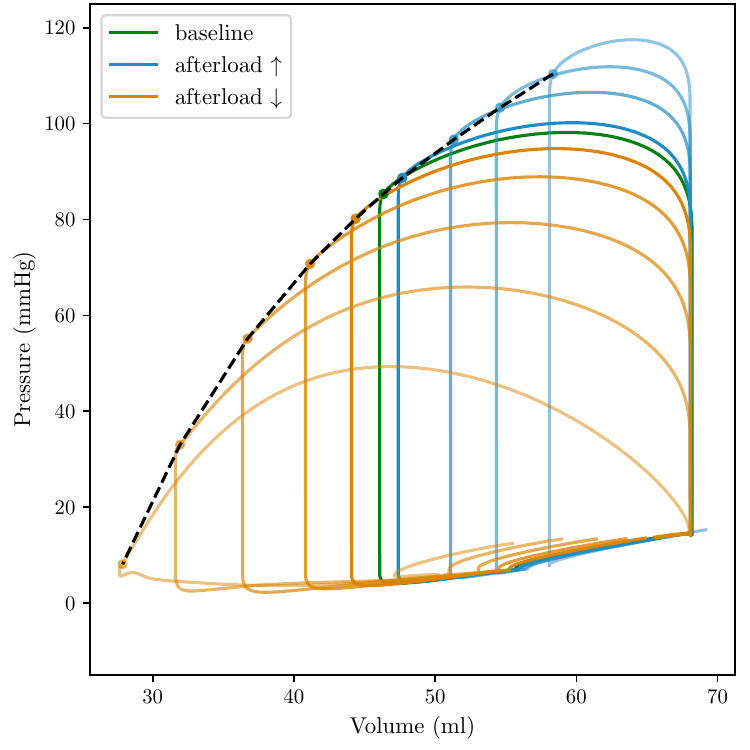}
  \subcaption{}
  \end{subfigure}
  \begin{subfigure}{0.32\textwidth}
  \includegraphics[width=\linewidth]{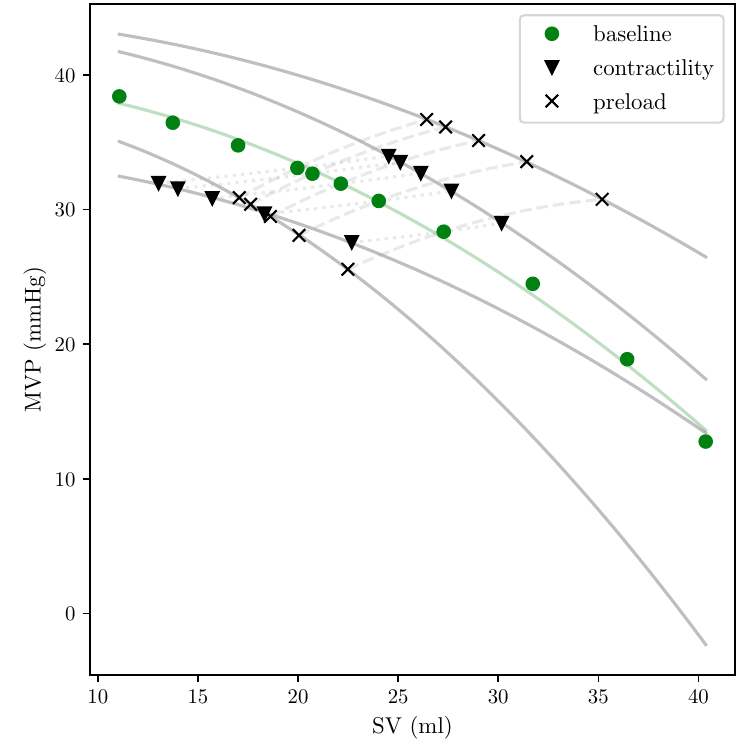}
  \subcaption{}
  \end{subfigure}
  \begin{subfigure}{0.32\textwidth}
      \includegraphics[width=\linewidth]{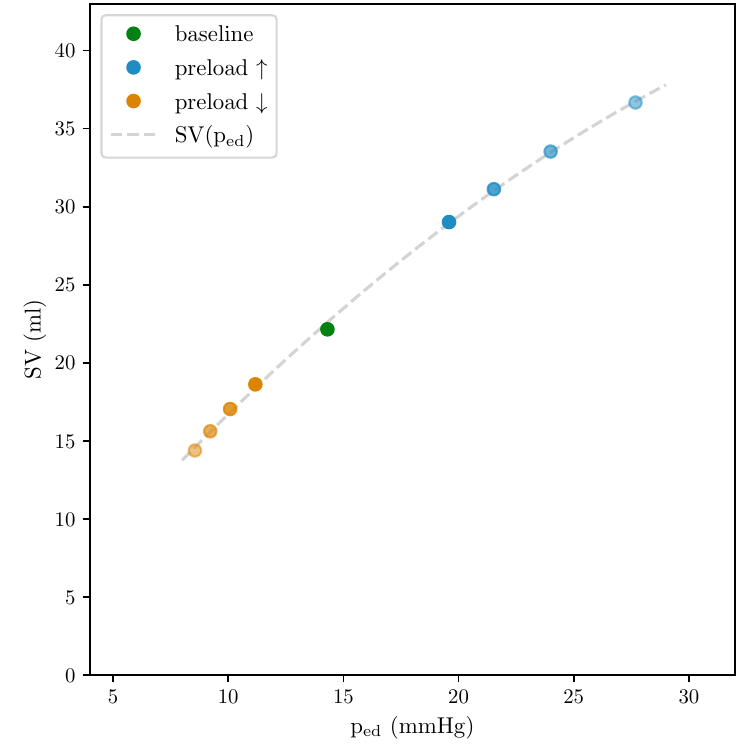}
  \subcaption{}
  \end{subfigure}
  \caption{
  	Pump function graph (PFG) and Frank-Starling curve of the LV.
    \textbf{(A)}
      $pV$ loops in LV under baseline conditions (green)
      and varying afterload conditions, ranging between unloaded, $\Ea \approx 0$,
      and isometric, $\Ea \approx \infty$, conditions.
      Note that $pV$ loops are plotted for the initial beat after altering afterload
      such that the end-diastolic volume is the same for all conditions.
      Thus, the system is not in a steady state and $pV$ loops are therefore not closed.
      \textbf{(B)} PFG, plotting mean ventricular pressure (MVP),~\Cref{eq:mvp},
      against stroke volume (SV),
      constructed from afterload variations
      with constant preload and contractility (solid circles),
      with increased preload (solid squares) shifting the PFG up and left
      towards higher flow and pressure.
      For both cases contractility was also perturbed
      which pivots the PFG (empty circles and squares, respectively),
      leading to a steeper/flatter slope $\text{MVP}/\Delta SV$
      for increased/decreased contractility.
      \textbf{(C)} Frank--Starling curve
      showing the relation between stroke volume and end-diastolic pressure, $SV(p_\mathrm{ed})$.
}%
    \label{fig:_pfg}
\end{figure}

A PFG was constructed by combining data from all tested protocols
with additional sampling of data within more extreme ranges of ventriculo-arterial coupling.
The models' PFG is in agreement with known cardiovascular physiology,
see~\Cref{fig:_pfg}.
Keeping contactility and preload constant,
the PFG with mean ventricular pressure (MVP),
as a function of flow or stroke volume
can be approximated by a quadratic function,
$\text{MVP}(q) = \hat{P}_{\rm {iso}}\left(1 - (q/q_{\rm{mx}})^2\right)$,
with $\hat{P}_{\rm {iso}}$ and $q_{\rm{mx}}$ being the maximum MVP
and maximum flow under isometric and unloaded conditions, respectively.
The MVP as a function of time is given by the integral
\begin{equation}\label{eq:mvp}
  \text{MVP}(t) = \frac{1}{t_\mathrm{cycle}} \;
  \int \limits_{0}^{t_{\mathrm{cycle}}}
  p_{\mathrm{LV}} (t + s) \; \dd s
\end{equation}
with a constant cycle length $t_{\mathrm{cycle}}$.
Increasing preload shifts the PFG towards higher flows and pressures.
Increasing/decreasing contractility pivots the PFG,
leading to a steeper/flatter slope $\Delta \text{MVP}/\Delta q$ of the PFG.

\begin{table}[htbp]
\centering
\footnotesize
\begin{tabularx}{0.8\textwidth}{lrlL}
\toprule
Parameter & Value & Unit & Description \\
\midrule
\multicolumn{4}{l}{\emph{Passive biomechanics}}\\
$\rho_0$        & \num{1060.0}  & \si{\kg/\m\cubed} & tissue density\\
$\kappa$        & 650  & \si{\kPa} & bulk modulus\\
$a$             & 0.7  & \si{\kPa} & stiffness scaling\\
$b_\mathrm{ff}$ & 5.0  & [-] & fiber strain scaling\\
$b_\mathrm{ss}$ & 6.0  & [-] & cross-fiber in-plain strain scaling\\
$b_\mathrm{nn}$ & 3.0  & [-] & radial strain scaling \\
$b_\mathrm{fs}$ & 10.0 & [-] & shear strain in fiber-sheet plane scaling \\
$b_\mathrm{fn}$ & 2.0  & [-] & shear strain in fiber-radial plane scaling \\
$b_\mathrm{ns}$ & 2.0  & [-] & shear strain in transverse plane scaling \\
\midrule
\multicolumn{4}{l}{\emph{Active biomechanics}}\\
$\lambda_0$           & 0.7   & \si{\ms}    &  minimum fiber stretch \\
$V_\mathrm{m,Thresh}$ & -60.0 & \si{\milli\volt} & membrane potential threshold \\
$t_\mathrm{emd}$      & 15.0  & \si{\ms}    &  EM delay\\
$S_\mathrm{peak}$     & 100 (\LV), 80 (\RV) & \si{\kPa} &  peak isometric tension\\
$t_\mathrm{dur}$      & 300.0 & \si{\ms}    &  duration of active contraction\\
$\tau_{c_0}$          & 100.0 & \si{\ms}    &  baseline time constant of contraction\\
$\mathrm{ld}$         &   5.0 & [-]         &  degree of length-dependence\\
$\mathrm{ld}_\mathrm{up}$ & 500.0 & \si{\ms}&  length-dependence of upstroke time\\
$\tau_\mathrm{r}$     & 100.0 & \si{\ms}    &  time constant of relaxation\\

\midrule
\multicolumn{4}{l}{\emph{Electrophysiology}}\\
$t_\mathrm{cycle}$ & \num{0.585}  & \si{\s}
    & cycle time ($=1/\mathrm{heartrate}$)\\
AA delay & 20.0   & \si{\ms} & inter-atrial conduction delay \\
AV delay & 100.0  & \si{\ms} & atrioventricular conduction delay \\
VV delay & 0.0    & \si{\ms} & inter-ventricular conduction delay \\
$(v_\mathrm{f},v_\mathrm{s},v_\mathrm{n})$ & (1.02, 0.68, 0.34)    & \si{\m/\s} & conduction velocities \\
$(g_\mathrm{f},g_\mathrm{s},g_\mathrm{n})$ & (0.44, 0.54, 0.54)  & \si{\m/\s} & conductivities in LV and RV \\
$\beta$ & 1/1400 & \si{\cm^{-1}} & membrane surface-to-volume ratio \\
$C_\mathrm{m}$ & 1 & \si{\micro\farad/cm^2} & membrane capacitance \\
\bottomrule
\end{tabularx}
\caption{Input parameters for the 3D PDE model of the left (\LV) and right (\RV) ventricle.
    Adjusted to match subject-specific data.}%
  \label{tab:pde_input_params}
\end{table}

\subsection{Numerical performance}
\label{sec:numerical_performances}
Computational times for a single heart beat of the lower and higher resolution
baseline models are given in~\Cref{tab:num_perf}.
Simulations were performed on the Vienna Scientific Cluster (VSC4) and we
distinguish between solver-time, $t_{\rm{s}}$,
which is the accummulated GMRES solver time over all loading/time steps;
and assembly-time, $t_{\rm{a}}$,
which is the time spent on the setup of
boundary conditions and on the assembly of matrices and vectors of the
linearized system~\eqref{eq:block_system}.
In total, for a full simulation with loading, 18 initialization beats,
and 2 final beats with a fully converging Newton method the computational costs were
\SI{4534.38}{\s} for the lower resolution model on 24 cores
and \SI{9393.73}{\s} for the higher resolution model on 256 cores.
Here, in addition to GMRES solver and assembly times,
also the solution of the R-E model governing EP, postprocessing,
\ca ODE times,
and input-output times are taken into account.
It is worth noting that the \ca ODE solver alone is very efficient:
for a simulation of 103 beats (i.e. 1 minute with a cycle length of \SI{0.585}{s})
computational costs were approximately \SI{2}{\s} on one core of a desktop computer.
Hence, \ca ODE times carry almost no weight in the coupled 3D-0D model.
\begin{table}[htbp]
	\centering
	\footnotesize
	\begin{tabularx}{0.95\textwidth}{lcccccccc}
		\toprule
                Model & cores & $\bar{h}_{\mathrm{\LV}}/\bar{h}_{\mathrm{\RV}}$ & Elems & Nodes &
                $t_{\rm{s},1}/t_{\rm{a},1}$ & $t_{\rm{s,c}}/t_{\rm{a,c}}$ & $T_{\rm{b,1}}/T_{\rm{b,c}}$ & $t_{\rm{s,ld}}/t_{\rm{a,ld}}$ \\
		      &  [-]  & [\si{\milli \meter}]   &  [-]  &   [-] & [\si{\second}] & [\si{\second}] & [\si{\second}] & [\si{\second}] \\
		\midrule
                coarse  & 24  & \num{3.4}/\num{2.4} & \num{45686}  & \num{11850}  & \num{ 91.4}/\num{42.1} & \num{ 620.2}/\num{299.0}  & \num{135.1}/\num{920.8} & \num{5.5}/\num{14.6} \\
                fine    & 256 & \num{1.3}/\num{1.2} & \num{557316} & \num{111234} & \num{165.6}/\num{62.8} & \num{1335.8}/\num{542.9}  & \num{231.6}/\num{1881.9} & \num{12.7}/\num{20.6} \\
		\bottomrule
  \end{tabularx}
  \caption{Summary of numerical  metrics for coarse and fine model.
    Given are the number of compute cores used on VSC4;
    the average spatial resolution in LV and RV,
    $\bar{h}_{\mathrm{\LV}}$ and $\bar{h}_{\mathrm{\RV}}$;
    the number of elements and nodes spanning the mesh; as well as
    solver, assembly, and total times for a
    single Newton iteration ($t_{\rm{s},1}$, $t_{\rm{a},1}$)
    and a fully converged Newton solution ($t_{\rm{s},c}$, $t_{\rm{a},c}$),
    and the total simulation time per heart beat
    for single iteration and fully converged Newton scenarios, $T_{\rm{b,1}}$ and $T_{\rm{b,c}}$, respectively.
    Timings refer to a single heart beat lasting \SI{0.585}{\second}
    at a time step size of \SI{1}{\milli \second}.
    In addition the cumulated solver ($t_{\rm{s,ld}}$)
    and assembly ($t_{\rm{a,ld}}$) times for the loading phase using 32 load steps are presented.}%
    \label{tab:num_perf}
\end{table}

\section{Discussion}%
\label{sec:discussion}
In this study, we report on the development of a monolithic 3D solid - 0D fluid coupling method
that allows to flexibly combine 3D PDE and non-linear 0D EM representations of cardiac cavities.
The hybrid 3D-0D model of the heart was coupled to a 0D closed-loop model of the cardiovascular system,
where all 0D components were based on the \ca model~\cite{walmsley2015fast}.
The combined model can be set up to represent one, two, three,
or all four cavities as 3D PDE model and all other elements
of the cardiovascular system as 0D models based on \ca.
In this study feasibility of this approach is demonstrated
by coupling a 3D PDE model of bi-ventricular EM
to 0D model representations of atrial EM function and circulation based on the \ca model.
The combined model was parameterized under baseline conditions
and subjected to comprehensive physiological testing
to demonstrate the model's ability to correctly predict known physiological behaviors.
A broad range of experimental protocols for altering loading conditions and contractility were simulated
to interrogate the models' transient responses to these maneuvers~\cite{burkhoff2005assessment}.
Overall, $pV$ analyses of the hemodynamic model output
showed close agreement with established knowledge on cardiovascular physiology.
The underlying numerical scheme is also represented in detail,
including a comprehensive mathematical representation of the \ca model.
Robustness -- in terms of stability and convergence properties --
and computational efficiency -- in terms of execution times -- are demonstrated.
These features combined render advanced EM modeling applications feasible.
The model facilitates the efficient and robust exploration of parameter spaces
over prolonged observation periods
which is pivotal for personalizing models to closely match observations.
Moreover, the model can be trusted to provide predictions of the acute transient response
to interventions or therapies altering loading conditions and contractility
that are valid within a commonly accepted physiological reference frame.

\subsection{Physiological validation}
Predictive modeling applications critically rely on the ability of models
to encapsulate the most relevant mechanisms
governing the cardiovascular response to a given intervention
that alters loading conditions or contractility.
In a closed-loop cardiovascular system as represented by \ca,
isolated changes to a single parameter entail transient adaption processes in the system as a whole.

Validation aimed at replicating, overall, known well established behaviors
and not at a 1:1 validation against experimental data.
For this sake, experimental standard protocols for altering afterload, preload and, contractility
were applied to the stabilized baseline model to study its response.
Two scenarios were analyzed, the initial acute response to a step change
in a single parameter -- afterload, preload, or contractility --
where effects on other unaltered parameters were minimal, see~\Cref{fig:_delta}(A)--(C),
and, the limit cycle response observed after a number of beats
where transients have largely subsided,
but indirect effects led to alteration of other parameters
due to the systemic inter-dependencies between these, see~\Cref{fig:_delta}(D)--(F).

Afterload was altered by varying the systemic resistance \Rsys\ and,
to mimick more extreme conditions closer to isometric contraction,
the resistance of the aortic valve.
Increasing/decreasing afterload is reflected in pivoting the arterial elastance curve \Ea\
around the point of end-diastolic volume and zero pressure.
This behavior is illustrated in~\Cref{fig:_delta} and for the more extensive protocols
used in constructing the PFG in~\Cref{fig:_pfg}(A).
\Ea\ was only marginally affected in the initial response,
but more significant changes were witnessed after stabilization to a new limit cycle.
As expected, changes in slope of \Ea\ were proportional to changes in \Rsys\
since, in absence of any regulatory mechanisms, heart rate remained unaltered.
Thus, $\Ea \approx \text{MAP} /(\text{SV} \cdot\text{HR}) \approx \Rsys$ holds.
The phenomenological active stress model as given in Eqs.~\eqref{eq:tanh_stress}--\eqref{eq:tanh_stress2}
accounted for length-dependent tension, but not for velocity dependence.
Thus, afterload effects on the velocity of fiber shortening were ignored.

Altering \Ved\ by changing the cross section of the pulmonary vein orifices
to increase/reduce preload increased/reduced SV.
This was due to the Frank--Starling mechanism,
as represented by the length-tension relationship of the active stress model in~\Cref{eq:tanh_stress2}.
In the immediate response to a step change in preload
the LV emptied to almost the same \Ves\,
with only minor deviations due to changes in arterial pressure induced by the change in SV.
After transients subsided, ESPVR and arterial elastance were the same
as under baseline conditions, with \Ea\ being shifted according to the changed $\Ved$.
In the new limit cycle \pes\ and MAP were increased,
but changes in SV were rather small.
This could be attributed to a rather flat slope of the Frank--Starling curve SV$(\ped)$ (not shown)
that is related to the significant heterogeneity in fiber stretch in end-diastolic state.
Such heterogeneity inevitably arises in biventricular EM models
that do not account for residual strains in the unpressurized configuration.
As fiber stretch in the reference configuration with $p=0$ is assumed $\lambda=1$,
increasing the filling pressure to \ped\ leads to a significant spread
in $\lambda$ around its mean.
Thus, while mean $\lambda$ in our simulations increased with \ped\,
the increasing spread of $\lambda$ around its mean led to low fiber stretch $\lambda < \lambda_0$ in various regions,
particularly in antero-septal and postero-septal segments of the LV.
These regions contributed increasingly less to contraction with increasing \ped\
due to the length-dependence of $\Sa(\lambda) \approx 0$,
thus, leading to an overall attenuation of the Frank--Starling effect.

Altering contractility by increasing/reducing the peak active stress \Sa\
led to an increase/decrease in SV and systolic pressures in terms of \ppeak\ and \pes.
Correspondingly, the ESPVR, as sampled by perturbing LV afterload,
was steepened/flattened as expected, see~\Cref{fig:_delta}(C).
In the initial response the slope of \Ea\ was affected,
but after stabilization \Ea\ was the same for all contractile states.
For the given parameterization ventriculo-arterial coupling
$\Ees/\Ea$ fell outside the optimal range of 1--2,
where external work is maximized around $\Ees/\Ea \approx 1$
and optimal efficiency is achieved at $\Ees/\Ea \approx 2$.
Since \ppeak\ and SV and as such $\Ea \approx \pes/\text{SV}$ were used to parameterize the model
the resulting slope of the ESPVR was to flat for the LV
to operate within this optimal range.
Reasons are multifactorial
and include the absence of velocity-dependence and, potentially,
also the role of mechanical boundary conditions.
The main culprit to blame for the limited slope of $\Ees$ is
the impairment of the Frank--Starling mechanisms
due to fiber stretch heterogeneity.
Our model deviates here from the general physiology-based assumption of uniform fiber-stretch
to be in place in an end-diastolic state~\cite{carruth2016,fung2013biomechanics_circulation,omens1990residual,Rodriguez1993}.

The constructed PFG was qualitatively in keeping with physiological expectations, see~\Cref{fig:_pfg}.
Data points obtained from varying afterload between close to unloaded and isometric conditions
agreed well with an assumed quadratic relation
between MVP and flow or SV.
Increasing preload shifted the PFG up and left
towards higher flows and pressures,
with the opposite trend being observed for decreasing preload.
Altering contractility rotated the PFG.

%
\subsection{Numerical aspects}
The computational cost imposed by higher resolution EM models
requires efficient numerical solvers.
Strong scaling characteristics of our numerical framework was reported in detail previously~\cite{Augustin2016anatomically,Karabelas2018towards}.
%
The compute times reported in~\Cref{sec:numerical_performances} indicate
that setup and assembly time were the dominating factors
during the initial passive inflation (loading) phase
while for the subsequent coupled 3D-0D EM simulations of a heart beat
solver time was the predominant part of total CPU time.
This is due to the Schur complement, see~\ref{sec:SchurComplement},
that needs to be solved during the 3D-0D active EM phase.
This involved -- in our case of a bi-ventricular model comprising two PDE cavities --
three applications of the GMRES solver while matrices were only assembled once per Newton step.

Further, significant savings in compute time could be achieved
using a semi-implicit approach, \reviewerOne{see~\Cref{sec:parameterization}},
until a limit cycle was reached.
A heartbeat using a fully converging Newton was about five times more expensive
compared to a heartbeat in the initialization phase.
As the deviations of the semi-implicit method from the implicit scheme
were negligible and the final two beats of the limit cycle protocol
were then computed using a fully converging Newton method
this gain in performance had no quantitative impact on any of the primary simulation outcomes.

\reviewerTwo{Reducing spatial resolution down to
    \SI{2.4}{milli meter} and \SI{3.4}{milli meter} in RV and LV, respectively,
    allowed for multibeat simulations within tractable time frames
    using a desktop computer.
    The impact of this reduction on the hemodynamic output variables
    was very minor, particularly when viewed in  context of the significant observational uncertainties
    the type of measured data used for model calibration are afflicted with.
    While probing only two resolutions by far cannot be considered a rigorous convergence study,
    our results suggest that the use of relatively coarse meshes,
    with resolutions in the range between 2 to \SI{4}{\milli\meter} are adequate.
    Indeed, this appears to fall within the range of resolutions
    used in other recent cardiac mechanics simulation studies.
    For instance, meshes comprising \num{208561}~\cite{baillargeon2014:_lhp} and \num{167323}~\cite{Pfaller2019}
    tetrahedral elements were used to represent human-sized hearts with all four chambers
    whereas our coarse model used \num{45686} of the same elements only for the ventricles
    of a much smaller sized canine heart.
    While our study cannot offer any conclusive recommendations
    on choosing an appropriate spatial resolution
    our results suggest that for solving cardiac mechanics problems significantly less
    spatial resolution is necessary to achieve acceptable accuracy \cite{land2015:_nversion_mech}
    relative to solving EP problems where spatial resolution is critical
    to resolve the steep propagating depolarization wavefronts \cite{niederer11:_nversion_ep}.
    The reaction-Eikonal model we used to represent electrical activation and repolarization
    is not sensitive to spatial resolution, as shown previously \cite{neic17:_efficient},
    and yields accurate activation patterns on coarse meshes,
    allowing to use the same grid for both EP and mechanics,
    without the need for projection of data between the physical grids.
}

Overall, the total simulation time incurring for 20 heart beats of the coarse model
was well below \SI{90}{\minute} on 24 cores.
Similar performance was achieved on a standard desktop computer
(AMD Ryzen Threadripper 2990X),
demonstrating that realistic multi-beat simulations of the presented 3D-0D cardiac models
deliver sufficient performance for advanced physiological simulation scenarios,
even on a small number of cores.
This is of paramount importance for future parameterization studies
where numerous simulations have to be carried out to personalize models to patient data.
With around 2 and a half hours on 256 cores for 20 beats also the higher resolution model
could be executed within a tractable time frame.

\reviewerOne{
\subsection{Relation to previous work}
The holistic framework described in this study constitutes a major step
towards a universal cardiac electro-mechanics simulation engine
that can be applied, in principle and after appropriate parameterization,
to a very broad range of applications.
Our study builds on and further advances various concepts
that have been reported previously in a number of excellent studies~\cite{fritz2014simulation,kerckhoffs2007coupling,Gurev2015high,Hirschvogel2017monolithic}.
While coupling 3D PDE models to a closed loop circulatory system is important
to ensure consistency, by allowing blood to redistribute between compliances in the system,
only a few have been reported~\cite{Gurev2015high,kerckhoffs2007coupling,Hirschvogel2017monolithic,sack2018construction}.
However, these were limited in some of the following regards
which restricted their universal applicability.
For instance, models were discretized with a small number of cubic Hermite elements
which led to anatomically stylized representations of the ventricles~\cite{Kerckhoffs2009bivscar,kerckhoffs2007coupling,niederer2011:_length,Xi2011},
with an artificially chopped base and a hole or collapsed elements~\cite{Lamata2014} in the apex,
owing to their limited ability to accurately accommodate more complex anatomical shapes without greatly increasing computational times~\cite{Pathmanathan2009}.
Often, artificial boundary conditions were used
that fixed the motion of the base~\cite{Gurev2015high,Wang2021} and, thus,
enforced a zero atrio-ventricular plane displacement.
These studies, with only a few exceptions~\cite{fritz2014simulation,Pfaller2019,Kariya2020},
were unable to replicate a physiological kinematics
characterized by the reciprocal filling properties of the heart
by maintaining a constant pericardial shape.
In other studies EP was not modeled at all,
assuming that contraction in the ventricles is initiated simultaneously~\cite{Hirschvogel2017monolithic,sack2018construction,Pfaller2019},
or non-physiological activation sequences were used to trigger contraction,
both of which impair length-dependent tension mechanisms~\cite{Gurev2015high}.
Computational cost of numerical methods is not addressed in most previous works;
notable exceptions include~\cite{Gurev2015high,Kariya2020,Hirschvogel2017monolithic,Pfaller2019,Wang2021}
where compute times for one heart beat range between 1.8 and 24 hours.
This is considerably less efficient compared to compute times presented in this study in~\Cref{sec:numerical_performances}.
Mostly simplified circuit models were
used~\cite{Gurev2015high,kerckhoffs2007coupling,Hirschvogel2017monolithic,sack2018construction}
for simulating a single heart beat or the simulated PV loops featured unphysiological edgy morphologies
due to the usage of (i) simple diode valve models not accounting for inertia, Bernoulli effects
and flow resistance between compartments and (ii)
time-varying elastance models for compliances such as the atria that yield fixed
pressure-volume relations and cannot contract in a load-dependent manner
as the sarcomere-based contractile 0D chambers used in the \ca model.
%
\\
None of these methodological limitations apply to the modeling framework presented in here.
}

\subsection{Limitations}
\reviewerTwo{Computational modeling relies on assumptions and approximations, especially
multiphysics simulations on organ scale level as presented in this study.
While the accuracy of most individual model components was already assessed
in previous studies, all these components have limitations and were chosen
(i) to achieve great computational performance while preserving almost the
full biophysical details of possibly more accurate, computationally costly models;
and
(ii) to ease parameterization, i.e., to achieve physiological results even with a
smaller amount of parameters compared to more accurate but also more complex models.
Dependent on the clinical application it might be required to consider other individual components
than those chosen in this study.
}

Efforts to parameterize the combined model were limited,
only a small subset of available data were used for model fitting.
\reviewerTwo{A generic five-fascicular representation of the cardiac conduction system was used
to drive ventricular activation, without attempting to match the recorded ECG \cite{gillette2021:_ep_twin}.}
More comprehensive and efficient procedures are required
to further enhance compatibility of the high dimensional combined 3D-0D model
with all available observations.
\reviewerTwo{Building on strategies for fitting the standalone \ca model to hemodynamic data~\cite{VanOsta2020a,VanOsta2021},
a hybrid parameterization approach appears a pragmatic solution
where the 0D model is fitted first to observed data
and in a subsequent parameterization step
the 3D cavities are fit to the $pV$ characteristics of the corresponding 0D cavities.}
However, for the sake of demonstrating overall compatibility with known cardiovascular physiology
a high fidelity match with experimental data, while desirable, is not crucial.
Future more advanced applications
that attempt to predict therapeutic responses in a patient-specific manner
will critically depend,
beyond a comprehensive representation of the most relevant therapy mechanisms,
on the fidelity of personalization.
\reviewerTwo{Given the large number of model parameters this is not a trivial task
that will require the development of dedicated parameter identification strategies.
In this regard the outstanding computational performance of the model is essential
to facilitate a detailed model personalization
which requires a large amount of forward simulations.}

While the framework used in this study \reviewerOne{can be considered universal and feature complete,
two important aspects remained unaccounted for.
First, active stresses generated by the phenomenological model featured length- but no velocity
dependence.
However, a velocity-dependent term could be incorporated,
as methods that avoid numerical instabilities due to velocity dependence have been reported~\cite{Regazzoni2021},
or a biophysically highly detailed model of excitation-contraction coupling could be used
instead as in our previous work~\cite{Augustin2016anatomically}.
Secondly, residual stresses in the unpressurized ventricles were ignored
which impairs, to some extent, the length-dependent Starling effect.}

Inflation of the unpressurized configuration to \ped\, \reviewerOne{without considering residual stresses},
inevitably introduces a significant spatial heterogeneity in fiber stretch.
This is in contrast to the common assumption of homogeneous fiber stretch
in the end-diastolic state ~\cite{carruth2016,fung2013biomechanics_circulation,omens1990residual,Rodriguez1993}.
This was reflected in a rather flat slope of the Starling curve SV$(\ped)$.
As shown in \Cref{fig:_pfg}C, for the given inotropic state and afterload,
the Starling curve was close to linear, with a slope of $\approx \SI{1.3}{ \milli \liter / \milli \meter Hg}$.
In humans, slopes in the range between $\approx 2.7-5.5 \si{\milli \liter / \milli \meter Hg}$
were measured for non-athletes and athletes, respectively \cite{Levine1991:_fsr}.


For the sake of saving computational costs,
a coarse spatial resolution was used for discretizing the bi-ventricular model.
Thus, models were lightweight enough to carry out the larger number of simulations
that were needed for parameterization, the determination of a limit cycle
as well as the fine grained testing of physiological maneuvers.
The use of such coarse spatial resolutions introduces inaccuracies
with regard to fibers and sheet arrangements
which are defined on a per element basis.
As parts of the model such as the RV wall were composed only of two element layers,
transmural fiber rotation was essentially reduced to two fiber families.
These were mostly aligned with the endocardial and epicardial fiber orientations
as prescribed on a per rule basis, see~\Cref{fig:model}.
Thus, the model's predictions of motion, strains and stresses may deviate quantitatively
from models discretized at higher resolution.
However, comparing between two models of different resolution revealed
that quantitative differences were minor and, qualitatively,
models showed essentially the same behavior.
Most differences stemmed from differences in stiffness
of the simple P1--P0 element types used
which tended to be more compliant for higher resolutions.
Nonetheless, errors in simulated hemodynamic outcomes were small enough
to be considered negligible
when put in context to observational uncertainties
clinical or experimental data are afflicted with.

\reviewerTwo{Finally, the presented model is not validated rigorously against
  experimental data. Typically, an independent validation is performed by the comparison
  of displacement~\cite{Ponnaluri2019, Pfaller2019, Sermesant2012} and/or
  strain~\cite{Genet2014,Finsberg2018,sack2018construction}
  to observations from cine MRI or 3D tagged MRI data.
  However, an accurate cardiac motion and deformation field can be obtained by
  tuning boundary conditions and \emph{in vivo} MRI strain measurements
  have major caveats~\cite{Amzulescu2019, Reichek2017}.
  Further, clinical validation can be conducted by comparing simulated
  ECG, pressure, and volume traces and derived quantities, e.g.,
  SV or ejection fraction, against clinically measured
  data~\cite{marx2020personalization,Finsberg2018,Kariya2020}.
  However, commonly, these measurements are all used for model calibration
  to optimize goodness of fit of simulated outputs to the data and, thus,
  these cannot be used for the purpose of model validation.
  Overall, difficulty of proper model validation remains a point of concern
  and in many cardiac modeling studies validation against actual patient data is
  limited~\cite{Niederer2020creation}.
  In this work, we focused on replicating known, well-established behaviors
  to show and prove a physiological predictive power under a broad range of
  experimental protocols.
  Together with a rigorous, independent validation against image data in
  future studies, this will set a new standard in cardiac modeling.
}

\section{Conclusions}%
\label{sec:conclusion}
This study reports on a flexible monolithic 3D solid - 0D fluid coupling method
for integrated models of cardiac EM and cardiovascular hemodynamics.
Feasibility of the approach is demonstrated by coupling a 3D PDE model of bi-ventricular EM
to 0D model representations of atrial EM and circulation based on the \ca model.
The coupled 3D-0D model is shown to be robust, computational efficient and
able to correctly replicate known physiological behaviors
in response to experimental protocols
for assessing systolic and diastolic ventricular properties based on $pV$ analysis.
These features combined render advanced EM modeling applications feasible.
The model facilitates the exploration of parameter spaces
over prolonged observation periods
which is pivotal for personalizing models to closely match observations.

\section*{Declaration of competing interest}
The authors declare that they have no known competing financial interests or
personal relationships that could have appeared to influence the work reported in this paper.

\section*{Acknowledgements}
This project has received funding from the European Union's Horizon 2020 research and
innovation programme under the Marie Sk{\l}odowska–Curie Action H2020-MSCA-IF-2016 InsiliCardio, GA No. 750835
and under the ERA-NET co-fund action No. 680969 (ERA-CVD SICVALVES, JTC2019) funded by the Austrian Science Fund (FWF), Grant I 4652-B to CMA.
Additionally, the research was supported by the Grants F3210-N18 and I2760-B30 from the Austrian Science Fund (FWF)
and a BioTechMed Graz flagship award ``ILearnHeart'' to GP.
Further, the project has received funding from the European Union's Horizon 2020 research and innovation programme under the ERA-LEARN co-fund
action No. 811171 (PUSHCART, JTC1\_27) funded by ERA-NET ERACoSysMed to
JL, FWP, EJV, and GP.

\clearpage
\begin{appendix}
\section{\ca equations summary}%
\label{sec:CircAdaptequations}
\subsection{\ca Tube Module}\label{sec:ca:tube}
The tube module represents the entrance of a compliant blood vessel capable of
propagating a pressure-flow wave component added to a constant flow,
see~\cite{arts2005adaptation}. Vessels directly attached to the heart,
aorta (\AO), arteria pulmonalis (\AP), venae cavae (\VC), and venae pulmonales (\VP)
are modeled in a similar fashion in \ca and for the whole section we
define the iterator $t\in\{\mathrm{\AO, \AP, \VC, \VP}\}$.

The current lumen cross sectional area is computed by
\begin{equation} \label{eq:ca:tube_cross_section}
  A_t = \frac{V_t}{l_t},
\end{equation}
with $V_t$ the cavity volume and $l_t$ the length of the vessel segment.

Using a model of a tube with a fibrous wall, see~\cite{arts1991relation},
gives the average extension, $\lambda_t$, of the fibers in the wall by
\[
    \lambda_t={\left(1+2V_t/V_t^\mathrm{wall}\right)}^{1/3}
    ={\left(1+2A_t/A_t^\mathrm{wall}\right)}^{1/3},
\]
with $A_t^\mathrm{wall}$ the wall cross-sectional area and
$V_t^\mathrm{wall}=A_t^\mathrm{wall}l_t$ the wall volume.
Cavity pressure depends on
$\lambda_t$
\[
  p_t=\sigma_t(\lambda_t)\,\lambda_t^{-3},
\]
with the mean Cauchy fiber stress $\sigma_t$ that is modeled
by the constitutive equation
\[ 
    \sigma_t(\lambda_t)=\sigma_t^\mathrm{ref}\cdot
    {\left(\lambda_t/\lambda_t^\mathrm{ref}\right)}^{k_t},
\]
see~\citet{arts1991relation}. Here, $k_t$ a stiffness exponent;
\(\lambda_t^\mathrm{ref}=
  {\left(1+2V_t^\mathrm{ref}/V_t^\mathrm{wall}\right)}^{1/3}\)
and
\(\sigma_t^\mathrm{ref}=
p_t^\mathrm{ref}\,{\left(\lambda_t^\mathrm{ref}\right)}^3\)
are the fiber extension and fiber state at normal physiological reference
state, respectively;
and $p_t^\mathrm{ref}$ is the reference tube pressure.
By combining above results the current tube pressure is computed as
\begin{align}
    p_t
    &= p_t^\mathrm{ref}\,
    {\left(\frac{\lambda}{\lambda_t^\mathrm{ref}}\right)}^k
    \lambda_t^{-3}{\left(\lambda_t^\mathrm{ref}\right)}^3
     = p_t^\mathrm{ref}\,
     {\left(\frac{\lambda_t}{\lambda_t^\mathrm{ref}}\right)}^{k-3}
       \nonumber\\
    &= p_t^\mathrm{ref}\,
       {\left(\frac{V_t^\mathrm{wall}+2V_t}
       {V_t^\mathrm{wall}+2V_t^\mathrm{ref}}\right)}^{\frac{k-3}{3}}
    = p_t^\mathrm{ref}\,{\left(\frac{A_t^\mathrm{wall}+2A_t}
    {A_t^\mathrm{wall}+2A_t^\mathrm{ref}}\right)}^{\frac{k-3}{3}},
    \label{eq:ca:ptube}
\end{align}
with $A_t^\mathrm{ref}$ the initial cross sectional area and
$V_t^\mathrm{ref}=A_t^\mathrm{ref}l_t$ the initial vessel volume.
The compliance is
\begin{align*}
 \frac{1}{C_t}
    &= \frac{\dd p_t}{\dd V_t}
    = \frac{\dd}{\dd V_t}\left[p_t^\mathrm{ref}{\left(\frac{V_t^\mathrm{wall}+2V_t}
                                                          {V_t^\mathrm{wall}
        +2V_t^\mathrm{ref}}\right)}^{\frac{k-3}{3}}\right]\\
    &= p_t^\mathrm{ref}\,\frac{k-3}{3}{\left(\frac{V_t^\mathrm{wall}+2V_t}
                                                 {V_t^\mathrm{wall}+2V_t^\mathrm{ref}}
                                       \right)}^{\left(\frac{k-3}{3}-1\right)}
        \frac{2}{V_t^\mathrm{wall}+2V_t^\mathrm{ref}}\\
        &= \frac{2 p_t\,(k-3)}{3\left(V_t^\mathrm{wall}+2V_t\right)}
    = \frac{2 p_t\,(k-3)}
      {3 l_t\left(A_t^\mathrm{wall}+2A_t\right)}.
\end{align*}
Finally, the basic relation between characteristic wave impedance $Z_t$
compliance $C_t$, and inertance $I_t$
\begin{equation}%
  \label{eq:impedance}
  Z_t^2=\frac{I_t}{C_t}
    =\frac{\rho_\mathrm{b}l_t}{A_t C_t}
\end{equation}
yields
\begin{equation}\label{eq:ca:impedance}
Z_t=\sqrt{\frac{2\rho_\mathrm{b}\, p_t\,l_t^2(k-3)}
    {3V_t\left(V_t^\mathrm{wall}+2V_t\right)}}
    =\sqrt{\frac{2\rho_\mathrm{b}\, p_t\,(k-3)}
    {3A_t\left(A_t^\mathrm{wall}+2A_t\right)}}
\end{equation}
with $\rho_\mathrm{b}$ the blood density.
\subsection{Sarcomere mechanics}%
\label{sec:ca:sarcomere}
In the following a sarcomere contraction model is described that is based on a
modified Hill model, see~\cite{lumens2009triseg,walmsley2015fast}, for all
tissue patches in the wall of the cavity: $c\in\{\mathrm{\LV,\RV,\SV,\LA,\RA}\}$, with the left (\LV) and right (\RV) ventricle, the septum (\SV), and the left (\LA) and right (\RA) atrium.
Natural strain $E_c^\mathrm{fib}$ of the myofiber is estimated as
\begin{equation} \label{eq:ca:sarc_strain}
  E_c^\mathrm{fib}
    = \ln\left(\frac{L_c^\mathrm{s}}{L^\mathrm{s,ref}}\right)
\end{equation}
and from this the total sarcomere length $L_c^\mathrm{s}$ can be computed as
\begin{equation} \label{eq:ca:sarc_length}
  L_c^\mathrm{s}
    = L^\mathrm{s,ref}\exp\left(E_c^\mathrm{fib}\right),
\end{equation}
with $L_c^\mathrm{s,ref}$ a constant describing the reference sarcomere length.
The sarcomere is supposed to be made up of a contractile element of length
$L_c^\mathrm{cont}$ in series with an elastic element of length
$L_c^\mathrm{elast}=L_c^\mathrm{s} - L_c^\mathrm{cont}$.

\subsection*{Sarcomere active stress}
Sarcomere contracting length $L_c^\mathrm{cont}$ varies over time according to
\begin{equation}\label{eq:ca:sarcomerelength}
  \dot{L}_c^\mathrm{cont}:=
  \frac{\dd L_c^\mathrm{cont}}{\dd t} = v^{\max}
    \left[\frac{L_c^\mathrm{s}
         - L_c^\mathrm{cont}}{L^\mathrm{elast,iso}}-1\right],
\end{equation}
where the constant $L^\mathrm{elast,iso}$ is the length of the series elastic
element during isovolumetric contraction and the constant $v^{\max}$
is the maximum velocity of contraction.

The governing equation for the contractility $C_c$, describing the density of
cross bridge formation within the fibers of the patch, is
\begin{equation} \label{eq:ca:contractility}
  \dot{C}_c :=
  \frac{\dd C_c}{\dd t} =
    f_c^\mathrm{rise}(t)
    C_c^\mathrm{s}\left(L_c^\mathrm{cont}\right)
    - f_c^\mathrm{decay}(t)C_c.
\end{equation}
In~\Cref{eq:ca:contractility} sarcomere contractility rises according to
the function $f_c^\mathrm{rise}$, a phenomenological representation of
the rate of cross bridge formation within the patch,
\begin{align*}
  f_c^\mathrm{rise}(t)
    &=\frac{1}{t_c^\mathrm{rise}}0.02 x_c^{3}{(8-x_c)}^{2} \exp(-x_c), \\
  x_c(t) &= \min \left(8, \max\left(0,\frac{t-t_c^\mathrm{act}}{t_c^\mathrm{rise}}\right)\right),
\end{align*}
depending on the time of onset of activation of the patch, $t_c^\mathrm{act}$,
and the rising time
\begin{align*}
  t_c^\mathrm{rise}=0.55 \tau^\mathrm{R}t_c^\mathrm{act,ref}.
\end{align*}
Here, $\tau^\mathrm{R}$ a constant and $t_c^\mathrm{act,ref}$ is
the reference duration of contraction for initial fiber length.

Sarcomere contractility in~\eqref{eq:ca:contractility} decays according to the
function $f_c^\mathrm{decay}$
\begin{align*}
  f_c^\mathrm{decay}(t) &= \frac{1}{2t_c^\mathrm{decay}}
  \left[1 + \sin \left(\operatorname{sign}(y_c)
    \min \left(\frac{\pi}{2},|y_c|\right)\right)\right], \\
    y_c(t) &=\frac{t - t_c^\mathrm{act} - t_c^\mathrm{act,dur}}{t_c^\mathrm{decay}},
\end{align*}
depending on the decay time
\begin{align*}
  t_c^\mathrm{decay}=0.33 \tau^\mathrm{D}t_c^\mathrm{act,ref},
\end{align*}
with $\tau^\mathrm{D}$ a constant and $t_c^\mathrm{act,dur}$
is the duration of contraction of the fiber that lengthens with sarcomere length
\begin{align*}
  t_c^\mathrm{act,dur}
  = \left(0.65+1.0570 L_c^\mathrm{norm}\right)t_c^\mathrm{act,ref}.
\end{align*}
Here, $L_c^\mathrm{norm}$ is the normalized sarcomere length for active
contraction
\begin{align*}
  L_c^\mathrm{norm} = \max\left(0.0001,
  L_c^\mathrm{cont}/L^\mathrm{act0,ref}-1\right),
\end{align*}
where $L^\mathrm{act0,ref}$ is the zero active stress sarcomere length.

$C_c^\mathrm{s}$ in~\eqref{eq:ca:contractility} describes the increase in
cross bridge formation with intrinsic sarcomere length due to an increase
in available binding sites,
\begin{align*}
  C_c^\mathrm{s}\left(L_c^\mathrm{cont}\right)
  &= \tanh \left(0.75*9.1204{\left(L_c^\mathrm{norm}\right)}^{2}\right).
\end{align*}
Contractility $C_c$~\eqref{eq:ca:contractility} and sarcomere contracting
length $L_c^\mathrm{cont}$~\eqref{eq:ca:sarcomerelength}
are used to compute the actively generated fiber stress
\begin{equation} \label{eq:ca:sarc_active}
  \sigma_c^\mathrm{fib,act} =
  L^\mathrm{act0,ref}\sigma^\mathrm{act,\max} C_c L_c^\mathrm{norm}
    \frac{L_c^\mathrm{s}- L_c^\mathrm{cont}}{L^\mathrm{elast,iso}},
\end{equation}
with constants $L^\mathrm{act0,ref}$, $\sigma^\mathrm{act,\max}$,
$L^\mathrm{elast,iso}$, see~\Cref{tab:ca_constants}.

\subsection*{Sarcomere passive stress}
Passive stress $\sigma_c^\mathrm{fib,pas}$ is considered to contain two components,
\begin{equation} \label{eq:ca:sarc_passive}
  \sigma_c^\mathrm{fib,pas} =\sigma_c^\mathrm{fib,tit}
  + \sigma_c^\mathrm{fib,ecm},
\end{equation}
first the stress arising from cellular structures such as titin,
a highly abundant structural protein of the sarcomere, anchoring to the Z-disc,
$\sigma_c^\mathrm{fib,tit}$,
and second the stress arising from the extracellular matrix (ECM),
$\sigma_c^\mathrm{fib,ecm}$. Both depend on the passive fiber stretch which is
computed as
\begin{align*}
  \lambda_c^\mathrm{pas}
  = \frac{L_c^\mathrm{s}}{L^\mathrm{pas0,ref}},
\end{align*}
where $L^\mathrm{pas0,ref}$ is sarcomere length with zero passive stress and
$L_c^\mathrm{s}$ the total sarcomere length, see above.
Using that we compute
\begin{align*}
    \sigma_c^\mathrm{fib,tit}
    = 0.01\sigma^\mathrm{act,\max}
    \left({\left[\lambda_c^\mathrm{pas}\right]}^{k^\mathrm{tit}}-1\right),
\end{align*}
with $\sigma^{\mathrm{act,\max}}$ the maximal
isometric stress and the constant exponent
\begin{align*}
  k^\mathrm{tit}=2\frac{L^\mathrm{s,ref}}{\dd L^\mathrm{s,pas}}.
\end{align*}
The ECM is modeled as being stiffer than the myocyte contribution using
\begin{align*}
  \sigma_c^\mathrm{fib,ecm} = 0.0349 \sigma^\mathrm{pas,\max}
  \left({\left(\lambda_c^\mathrm{pas}\right)}^{10}-1\right),
\end{align*}
where $\sigma^\mathrm{pas,\max}$ is an empirical parameter.

\subsection*{Sarcomere total stress}
Total myofiber stress $\sigma^\mathrm{fib}_c$ is the sum of an
active~\eqref{eq:ca:sarc_active} and a passive~\eqref{eq:ca:sarc_passive}
stress component
\begin{equation} \label{eq:ca:sarc_total}
  \sigma^\mathrm{fib}_c =\sigma_c^\mathrm{fib,act}
  + \sigma_c^\mathrm{fib,pas}.
\end{equation}
Sarcomere stiffness $\kappa_c^\mathrm{fib}$ is now computed as the derivative of total fiber
stress~\eqref{eq:ca:sarc_total} with respect to fiber
strain~\eqref{eq:ca:sarc_strain}
\begin{equation} \label{eq:ca:sarc_stiff}
  \kappa_c^\mathrm{fib} = \frac{\partial\sigma_c^\mathrm{fib}}{\partial E_c^\mathrm{fib}}
  = \frac{\partial\sigma_c^\mathrm{fib,act}}{\partial E_c^\mathrm{fib}}
  + \frac{\partial\sigma_c^\mathrm{fib,pas}}{\partial E_c^\mathrm{fib}},
\end{equation}
with
\begin{align*}
  \frac{\partial\sigma_c^\mathrm{fib,act}}{\partial E_c^\mathrm{fib}}
  &= L^\mathrm{act0,ref}\sigma^\mathrm{act,\max} C_c L_c^\mathrm{norm}
     \frac{L_c^\mathrm{s}}{L^\mathrm{elast,iso}},\\
  \frac{\partial\sigma_c^\mathrm{fib,pas}}{\partial E_c^\mathrm{fib}}
  &= 0.01 k^\mathrm{tit} \sigma_c^\mathrm{act,\max}
  {\left(\lambda_c^\mathrm{pas}\right)}^{k^\mathrm{tit}}
       + 0.0349*10 \sigma^\mathrm{pas,\max}
       {\left(\lambda_c^\mathrm{pas}\right)}^{10}.
\end{align*}

\subsection{\ca Chamber Module}%
\label{sec:ca:chamber}
An actively contracting chamber
$c\in\{\mathrm{\LV, \RV, \LA, \RA}\}$ is modeled using the
state variables volume $V_c$, length of the contractile element of the sarcomere $L_c^\mathrm{cont}$~\eqref{eq:ca:sarcomerelength}, and contractility
$C_c$~\eqref{eq:ca:contractility}.
Volume changes driven by inflow and outflow of blood induce changes in
midwall volume $V_c^\mathrm{mid}$ and area $A_c^\mathrm{mid}$.

\subsection*{Sphere mechanics}
Note that in \ca ventricles are usually modeled using the TriSeg
formulation, see~\ref{sec:ca:triseg}.
If TriSeg is turned on, the calculations in this chapter are only used for the
atria while ventricular values are computed as in~\ref{sec:ca:triseg}.

Midwall volume $V_c^\mathrm{mid}$ is estimated as
\begin{equation}
  V_c^\mathrm{mid} = V_c + \frac{1}{2}V_c^\mathrm{wall},
\end{equation}
where $V_c^\mathrm{wall}$ is constant wall volume.
If not set to a specific value the wall volume is estimated by extruding the
sphere enclosing the cavity volume $V_c$ by a constant wall thickness
$\operatorname{h}_c^\mathrm{wall}$, see~\Cref{tab:ca_input_params}.
Chambers are modeled as closed spheres, thus, the following equations
result from volume and surface formulas for spheres
\begin{align}
  \label{eq:ca:Ccmid}
  C_c^\mathrm{mid} &= {\left(\frac{4\pi}{3V_c^\mathrm{mid}}\right)}^{1/3}, \\
  A_c^\mathrm{mid,tot} &= \frac{4\pi}{{\left(C_c^\mathrm{mid}\right)}^2}, \\
  \label{eq:ca:Acmid}
  A_c^\mathrm{mid} &= A_c^\mathrm{mid,tot} - A_c^\mathrm{mid,dead},
\end{align}
where $C_c^\mathrm{mid}$ is midwall curvature, i.e., the inverse of radius;
and $A_c^\mathrm{mid,dead}$ is non-contractile area, i.e.,
valve openings and orifices.

\subsection*{Update fiber strain}
Natural fiber strain $E_c^\mathrm{fib}$ is calculated by
\begin{equation} \label{eq:cav_fiber_stress}
  E_c^\mathrm{fib}
  = \frac{1}{2}\ln\left(\frac{A_c^\mathrm{mid}}{A_c^\mathrm{mid,ref}}\right)
\end{equation}
with $A_c^\mathrm{mid,ref}$ the surface area in the reference state, see~\cite{walmsley2015fast}.
Note that this updated fiber strain is used in the
place of~\eqref{eq:ca:sarc_strain} to
update values in the sarcomere module~\ref{sec:ca:sarcomere}.

Cross-sectional area $A_c$ of chambers is estimated as
\begin{align}
  A_c &= \frac{V_c + 0.1V_c^\mathrm{wall}}{l_c}, \label{eq:ca:cav_cross_section}\\
  l_c &= 2{\left(V_c^\mathrm{mid}\right)}^{1/3}, \nonumber
\end{align}
with $l_c$ the long-axis length of the cavity.

The characteristic wave impedance $Z_c$ is approximated according
to~\eqref{eq:impedance}, see also~\cite{arts2005adaptation}, and by
applying the chain rule
\begin{equation} \label{eq:ca:wave_impedance}
  Z_c = \frac{1}{5A_c}\sqrt{\rho_\mathrm{b}l_c\left|\kappa_c^\mathrm{mid}\right|},
\end{equation}
with the sheet stiffness
\begin{equation}\label{eq:ca:sheet_stiffness}
\kappa_c^\mathrm{mid} = \frac{\partial T_c^\mathrm{mid}}{\partial A_c^\mathrm{mid}} =
  \frac{V_c^\mathrm{wall}}{4{\left(A_c^\mathrm{mid}\right)}^2}
  \left(\frac{\partial\sigma_c^\mathrm{fib}}{\partial E_c^\mathrm{fib}}
  -2\sigma_c^\mathrm{fib}\right)=
  \frac{V_c^\mathrm{wall}}{4{{\left(A_c^\mathrm{mid}\right)}^2}}
  \left(\kappa_c^\mathrm{fib}-2\sigma_c^\mathrm{fib}\right)
\end{equation}
and the updated fiber stiffness $\kappa_c^\mathrm{fib}$,
see~\Cref{eq:ca:sarc_stiff}.

\subsection*{Conservation of energy}

\ca connects midwall tension $T_c^\mathrm{mid}$ and
midwall area $A_c^\mathrm{mid}$
to fiber stress $\sigma_c^\mathrm{fib}$ and strain $E_c^\mathrm{fib}$
through the law of conservation of energy. With the law of Laplace we get
\begin{align}
  T_c^\mathrm{mid} \dd A_c^\mathrm{mid}
  = \sigma_c^\mathrm{fib}V_c^\mathrm{wall}\dd E_c^\mathrm{fib}
\end{align}
and with~\eqref{eq:cav_fiber_stress} we get for the midwall tension
\begin{equation} \label{eq:midwall_tension}
  T_c^\mathrm{mid}
    = \frac{\sigma_c^\mathrm{fib} V_c^\mathrm{wall}}
    {2 A_c^\mathrm{mid}}.
\end{equation}
Transmural pressure $p_c^\mathrm{trans}$ is finally computed as follows
\begin{equation}\label{eq:ca:pc_trans}
  p_c^\mathrm{trans} = 2 T_c^\mathrm{mid} C_c^\mathrm{mid}.
\end{equation}
Since at the moment external pressures are assumed to be zero,
the transmural pressure coincides with the internal pressure of the
contracting chamber
\begin{align*}
  p_c = p_c^\mathrm{trans}.
\end{align*}
\subsection{TriSeg model of ventricular interaction}%
\begin{figure}\tiny
  \begin{tikzpicture}
      \tikzstyle{filled}=[rectangle,draw,fill=black!10,rounded corners]

    \node[inner sep=0pt] (structure) at (0,0)
      {\includegraphics[width=\textwidth]{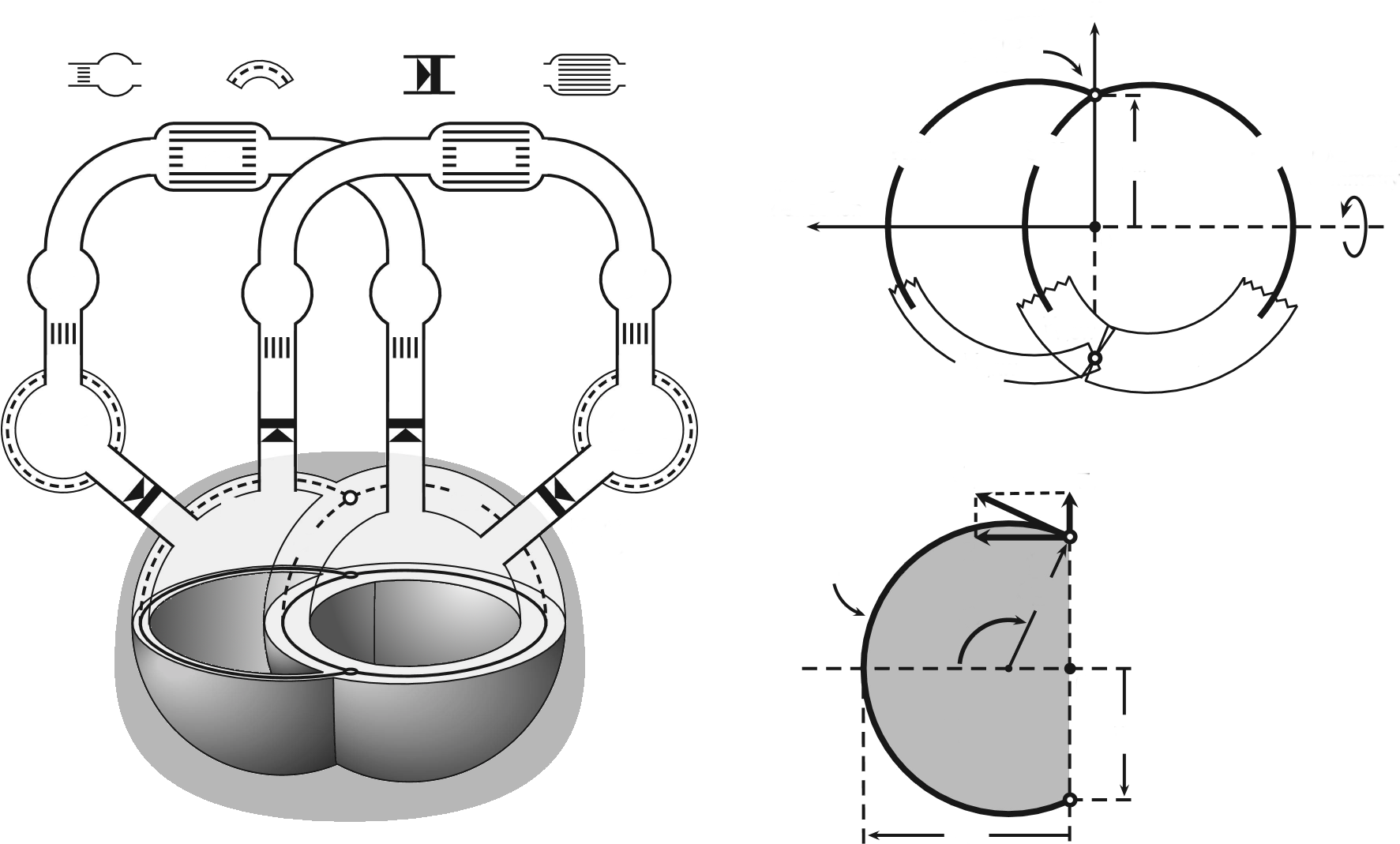}};

      \node[text height=1.5em] at (-7.5, 4.2) {\footnotesize\textbf{(a)}};
    \node[text height=1.5em] at (1,4.2) {\footnotesize\textbf{(b)}};
    \node[text height=1.5em] at (1,-0.5)  {\footnotesize\textbf{(c)}};

    \node[text height=1.5em] at (-1.25,4.4) {\tiny\textbf{Resistance}};
    \node[text height=1.5em] at (-2.95,4.4) {\tiny\textbf{Valve}};
    \node[text height=1.5em] at (-4.7,4.4) {\tiny\textbf{Wall Segment}};
    \node[text height=1.5em] at (-6.4,4.4)  {\tiny\textbf{Tube}};

    \node at (5.95, 2.9) {\tiny \textbf{\LV~Wall}};
    \node at (3.55, 2.9) {\tiny \textbf{Sep~Wall}};
    \node at (1.95, 2.9) {\tiny \textbf{\RV~Wall}};
    \node at (5.00, 4.4) {\tiny $y$-direction};
    \node at (4.80, 2.80) {\scriptsize $y^\mathrm{mid}$};
    \node at (4.45, 1.90) {\scriptsize $0$};
    \node at (5.55, 0.80) {\scriptsize $V_\mathrm{\LV}^\mathrm{wall}$};
    \node at (3.95, 1.05) {\scriptsize $V_\mathrm{\SV}^\mathrm{wall}$};
    \node at (3.10, 0.50) {\scriptsize $V_\mathrm{\RV}^\mathrm{wall}$};
    \node[align=center, text width=5em] at (7.10,1.4) {\tiny axis of rotational symmetry};
    \node[align=center, text width=7em] at (2.9,4.1) {\tiny intersection juction circle};
    \node[align=center, text width=5em] at (1.2,1.7) {\tiny positive $x$-direction};

    \node at (1.55, -1.55) {\scriptsize$A_c^\mathrm{mid}$};
    \node at (2.80, -2.15) {\scriptsize$\alpha$};
    \node at (2.95, -3.45) {\scriptsize$V_c^\mathrm{mid}$};
    \node at (2.90, -4.65) {\scriptsize$x_c^\mathrm{mid}$};
    \node at (4.65, -3.40) {\scriptsize$y^\mathrm{mid}$};
    \node at (4.20, -2.90) {\scriptsize$0$};
    \node at (3.55, -1.80) {\scriptsize$\frac{1}{C_c^\mathrm{mid}}$};
    \node at (2.95, -0.55) {\scriptsize$T_c^\mathrm{mid}$};
    \node at (4.25, -0.65) {\scriptsize$T_c^\mathrm{y}$};
    \node at (2.95, -1.50) {\scriptsize$T_c^\mathrm{x}$};

    \node[filled, align=center] at (-3.9,-4.4) {\footnotesize\textbf{TriSeg Model}};
    \node[filled, align=center, text width=2em] at (-3.1,-3.2) {\scriptsize\textbf{\LV}};
    \node[filled, align=center, text width=2em] at (-4.4,-2.2) {\scriptsize\textbf{\SV}};
    \node[filled, align=center, text width=2em] at (-5.1,-3.2) {\scriptsize\textbf{\RV}};
    \node at (-3.1, -2.2)  {\scriptsize $V_\mathrm{\LV}$};
    \node at (-2.5, -2.2)  {\scriptsize $p_\mathrm{\LV}$};
    \node at (-5.2, -2.05) {\scriptsize $V_\mathrm{\RV}$};
    \node at (-5.2, -2.4)  {\scriptsize $p_\mathrm{\RV}$};
    \node at (-4.80, -0.95) {\scriptsize $V_\mathrm{\RV}^\mathrm{wall}$};
    \node at (-4.05, -1.30) {\scriptsize $V_\mathrm{\SV}^\mathrm{wall}$};
    \node at (-2.37, -0.80) {\scriptsize $V_\mathrm{\LV}^\mathrm{wall}$};
    \node at (-5.25,2.82) {\tiny \textbf{Syst}};
    \node at (-2.29,2.82) {\tiny \textbf{Pulm}};
    \node[align=center, text width=4em] at (-6.89,-0.1) {\tiny \textbf{Right Atrium}};
    \node[align=center, text width=4em] at (-0.7,-0.1) {\tiny \textbf{Left Atrium}};
    \node[align=center, text width=4em] at (-6.9, 1.55) {\tiny \textbf{Syst Veins}};
    \node[align=center, text width=4em] at (-0.7, 1.55) {\tiny \textbf{Pulm Veins}};
    \node[align=center, text width=5em] at (-5.4, 1.0) {\tiny \textbf{Pulm Arteries}};
    \node[align=center, text width=5em] at (-2.4, 1.1) {\tiny \textbf{Syst Arteries}};
    \node[align=center, text width=4em] at (-2.5,-0.1) {\tiny \textbf{Aortic Valve}};
    \node[align=center, text width=4em] at (-5.3,-0.1) {\tiny \textbf{Pulm Valve}};
    \node[align=center, text width=4em] at (-6.9,-1.2) {\tiny \textbf{Tricuspid Valve}};
    \node[align=center, text width=4em] at (-0.9,-1.2) {\tiny \textbf{Mitral Valve}};
  \end{tikzpicture}
  \caption{TriSeg model of septal (\SV) and left (\LV) and right ventricular (\RV) mechanics.
    (a) The TriSeg model (gray shading) incorporated in the modular \ca
    model of the systemic (Syst) and pulmonary (Pulm) circulations.
    (b) Cross-section of the ventricular composite.
    (c) Cross-section of a single wall segment ($c\in\{\mathrm{\LV,\RV,\SV}\}$) through the axis of rotational symmetry.
  Adapted with permission from~\cite{lumens2009triseg}.}%
  \label{fig:triseg}
\end{figure}
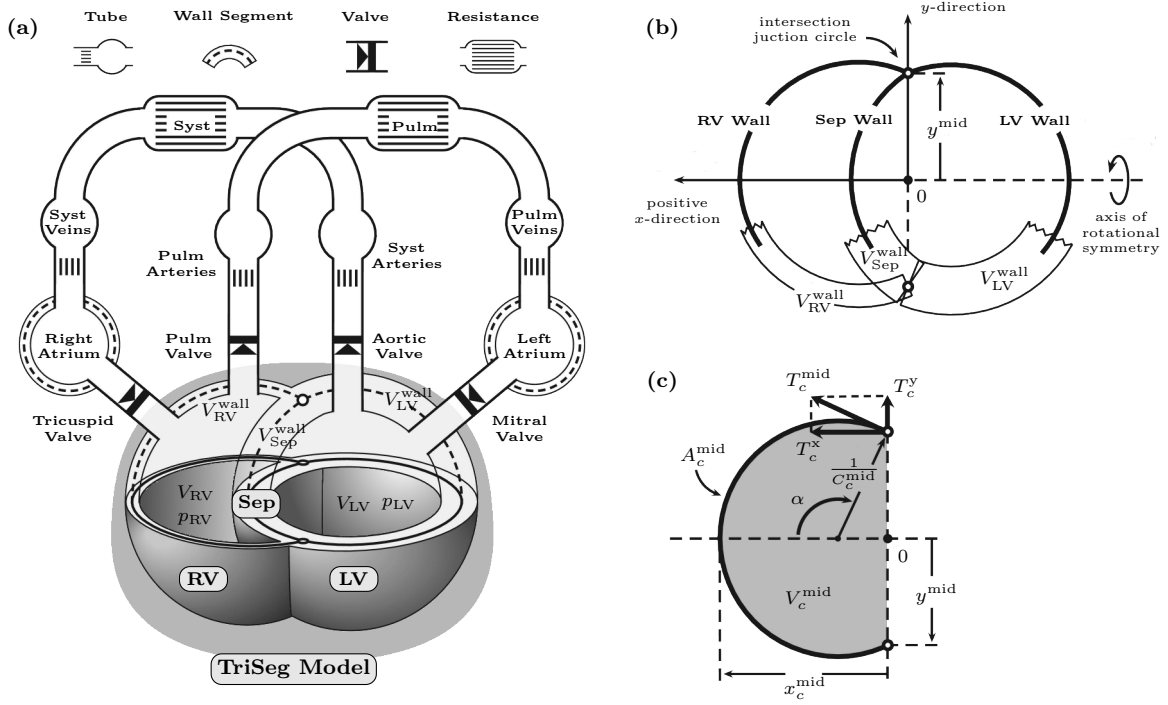

\label{sec:ca:triseg}
In case that one ODE and one PDE cavity is included in the model, ventricles
are modeled as atria above.
Otherwise, ventricular and septal midwall volumes are modeled as a
ventricular composite~\cite{lumens2009triseg} which
is defined by the common radius
$y^\mathrm{mid}$ of the wall junction and the enclosed midwall cap volumes,
see~\Cref{fig:triseg}c.
Midwall cap volumes of the right and the left ventricle are computed as
\begin{align*}
  V_\mathrm{\LV}^\mathrm{mid} &= - V_\mathrm{\LV} + V^\mathrm{mid}_\mathrm{\SV} - \frac{1}{2}
    \left(V_\mathrm{\LV}^\mathrm{wall} + V_\mathrm{\SV}^\mathrm{wall}\right),\\
  V_\mathrm{\RV}^\mathrm{mid} &= V_\mathrm{\RV} + V^\mathrm{mid}_\mathrm{\SV} + \frac{1}{2}
    \left(V_\mathrm{\RV}^\mathrm{wall} + V_\mathrm{\SV}^\mathrm{wall}\right).
\end{align*}
Here, the wall volumes of the left, $V_\mathrm{\LV}^\mathrm{wall}$, and right,
$V_\mathrm{\RV}^\mathrm{wall}$, ventricle are constants.
The blood pool volumes of the left, $V_\mathrm{\LV}$, and right,
$V_\mathrm{\RV}$, ventricle are ODE variables as well as the radius
$y^\mathrm{mid}$ and the septal midwall volume $V^\mathrm{mid}_\mathrm{\SV}$.
Note that the sign of midwall volume $V_c^\mathrm{mid}$ is positive if
wall curvature is convex to the positive x-direction and negative otherwise.

Distance $x^{\mathrm{mid}}_c$, see~\Cref{fig:triseg}c, is then computed
by the relation
\[
  V_c^\mathrm{mid}
= \frac{\pi}{6}x^{\mathrm{mid}}_c
\left( {(x^{\mathrm{mid}}_c)}^2 + 3{(y^\mathrm{mid})}^2\right),
  \quad\mbox{for } c\in\{\mathrm{\LV,\RV,\SV}\},
\]
hence
\[
    x^{\mathrm{mid}}_c = q_c - \frac{ {(y^\mathrm{mid})}^2}{q_c},
  \text{ with }
  q_c = \sqrt[3]{\sqrt{{\left(\frac{3}{\pi}V_c^\mathrm{mid}\right)}^2
      + {\left(y^\mathrm{mid}\right)}^6} + \frac{3}{\pi}V_c^\mathrm{mid}}.
\]
Midwall area and curvature are consequently computed
\begin{align*}
  A_c^\mathrm{mid} &= \pi\left({(x^{\mathrm{mid}}_{c})}^2
  + {({y^\mathrm{mid}})}^2\right),
  &&\mbox{for } c\in\{\mathrm{\LV,\RV,\SV}\}, \\
  C_c^\mathrm{mid}
  &= \frac{2x^{\mathrm{mid}}_c}{{(x^{\mathrm{mid}}_c)}^2 + {(y^\mathrm{mid})}^2},
  &&\mbox{for } c\in\{\mathrm{\LV,\RV,\SV}\},
\end{align*}
and used to calculate midwall tension
$T_c^\mathrm{mid}$~\eqref{eq:midwall_tension}.

Axial $T_c^\mathrm{x}$ and radial $T_c^\mathrm{y}$ tension
components are computed using laws of trigonometry
\begin{align*}
  T_c^\mathrm{x}
  &= T_c^\mathrm{mid}\sin\alpha, \quad \mbox{with }
  \sin\alpha = \frac{2x^{\mathrm{mid}}_c\,y^\mathrm{mid}}
               {{(x^{\mathrm{mid}}_c)}^2 + {(y^\mathrm{mid})}^2},
  &&\mbox{for } c\in\{\mathrm{\LV,\RV,\SV}\}, \\
  T_c^\mathrm{y}
  &= T_c^\mathrm{mid}\cos\alpha, \quad \mbox{with }
  \cos\alpha = \frac{-{(x^{\mathrm{mid}}_c)}^2 + {(y^\mathrm{mid})}^2}
                    {{(x^{\mathrm{mid}}_c)}^2 + {(y^\mathrm{mid})}^2},
                   &&\mbox{for } c\in\{\mathrm{\LV,\RV,\SV}\}.
\end{align*}
It is required that that the total midwall tension at junctions is zero, i.e.,
\begin{equation}\label{eq:tension_function}
  f(y^\mathrm{mid}, V_\mathrm{\SV}^\mathrm{mid}) :=
  \begin{pmatrix}
      T_\mathrm{\LV}^\mathrm{x} + T_\mathrm{\RV}^\mathrm{x}
     + T_\mathrm{\SV}^\mathrm{x}\\
     T_\mathrm{\LV}^\mathrm{y} + T_\mathrm{\RV}^\mathrm{y}
     + T_\mathrm{\SV}^\mathrm{y}
  \end{pmatrix}
  \stackrel{!}{=} 0.
\end{equation}
\Cref{eq:tension_function} is solved by an iterative Newton scheme
\begin{equation}\label{eq:tension_newton}
  f'(y_k^\mathrm{mid}, V_{k,\mathrm{\SV}}^\mathrm{mid})
  {\left(\Delta y_k^\mathrm{mid}, \Delta V_{k,\mathrm{\SV}}^\mathrm{mid}\right)}^\top
  = - f(y_k^\mathrm{mid}, V_{k,\mathrm{\SV}}^\mathrm{mid}),\quad
 k=1,2,\dots
\end{equation}
and increments $\Delta y_k^\mathrm{mid}$ and
$\Delta V_{k,\mathrm{\SV}}^\mathrm{mid}$
are added to $y_k^\mathrm{mid}$ and $V_{k,\mathrm{\SV}}^\mathrm{mid}$.
The solution of~\eqref{eq:tension_newton} in the first step, i.e., for
$k=0$ is used to define the ODE updates for the septum
\begin{equation}\label{eq:ca:ode_septum}
  \dot{V}^\mathrm{mid}_\mathrm{\SV} = \frac{1}{\tau_\mathrm{\SV}}\Delta V_{0,\mathrm{\SV}}^\mathrm{mid},\quad
  \dot{y}^\mathrm{mid} = \frac{1}{\tau_\mathrm{\SV}}\Delta y_0^\mathrm{mid},
\end{equation}
where $\tau_\mathrm{\SV}$ is a time constant.

Consequently, the values for the tensions discussed above are updated and the
scheme is iterated until convergence.
Midwall volumes are updated by
\begin{align*}
  V_c^\mathrm{mid} &= V_c + \frac{1}{2}
    \left(V_c^\mathrm{wall} + V_\mathrm{\SV}^\mathrm{wall}\right),
\end{align*}
long-axis length $l_c$ and cross-sectional area $A_c$ and of the cavity
are computed by
\begin{align}
  l_c &= 2{\left(V_c^\mathrm{mid}+\frac{1}{2}
  \left(V_c^\mathrm{wall} + V_\mathrm{\SV}^\mathrm{wall}\right)\right)}^{1/3},\\
  A_c &= \frac{V_c^\mathrm{mid} + \frac{1}{20}
  \left(V_c^\mathrm{wall} + V_\mathrm{\SV}^\mathrm{wall}\right)}{l_c}.
  \label{eq:ca:triseg_cross_section}
\end{align}
Finally, wave impedance $Z_c$ is computed according to~\Cref{eq:ca:wave_impedance}
and transmural pressure $p_c^\mathrm{trans}$ is computed as total axial force
\begin{align*}
  p_c^\mathrm{trans} = 2\, \frac{T_c^\mathrm{x}}{y^\mathrm{mid}},
  \quad\mbox{for } c\in\{\mathrm{\LV,\RV,\SV}\}.
\end{align*}
Assuming the pressure surrounding the ventricular composite to be zero,
internal chamber pressure for the ventricles is now found as
\begin{align*}
  p_\mathrm{\LV} &= -p_\mathrm{\LV}^\mathrm{trans}, \\
  p_\mathrm{\RV} &=  p_\mathrm{\RV}^\mathrm{trans}.
\end{align*}
\subsection{Pericardial mechanics}%
\label{sec:ca:pericardium}
The four cardiac chambers are supposed to have an additional pressure
component due to the pericardium.
Pressure $p_\mathrm{peri}$ exerted by the pericardial sack on atria and
ventricles was computed as a non-linear function of
pericardial volume $V_\mathrm{peri}$, computed as
the sum of blood pool and wall volumes of the four cardiac chambers:
\begin{align}
  V_\mathrm{peri} &= V_\mathrm{\LV} + V_\mathrm{\RV}
    + V_\mathrm{\LA} + V_\mathrm{\RA}
    + V_\mathrm{\LV}^\mathrm{wall} + V_\mathrm{\RV}^\mathrm{wall}
    + V_\mathrm{\LA}^\mathrm{wall} + V_\mathrm{\RA}^\mathrm{wall}\\
  \label{eq:ca:pperi}
  p_\mathrm{peri} &= p_\mathrm{peri}^\mathrm{ref}
  {\left(\frac{V_\mathrm{peri}}{V_\mathrm{peri}^\mathrm{ref}}\right)}^{k_\mathrm{peri}},
\end{align}
where $p_\mathrm{peri}^\mathrm{ref}$ and $V_\mathrm{peri}^\mathrm{ref}$
are constant reference pressure and volume, respectively, and $k_\mathrm{peri}$
defines the degree of non-linearity of the pressure-volume relation.

Cavity pressures are updated according to
\begin{equation}\label{eq:ca:periupdate}
    p_c = p_c + p_\mathrm{peri}, \quad\mbox{for } c\in\{\mathrm{\LV,\RV,\LA,\RA}\}.
\end{equation}

\subsection{Periphery}%
\label{sec:ca:periphery}
Pulmonary (pulm) and systemic (sys) periphery are modeled as resistances.
The current pressure drop $\Delta p_{py}$, for
$py\in\{\text{pulm}, \text{sys}\}$, is computed as the difference of the
pressures in the inflow artery $p_t^\mathrm{prox}$ and the outflow vein
$p_t^\mathrm{dist}$:
\[
  \Delta p_{py} = p_t^\mathrm{prox} - p_t^\mathrm{dist}.
\]
Using this, the current flow over the periphery is
\begin{equation}\label{eq:ca:qperiphery}
  q_{py} = q_{py}^\mathrm{ref}
           {\left(r_{py}\frac{\Delta p_{py}}
           {\Delta p_{py}^\mathrm{ref}}\right)}^{k_{py}},
\end{equation}
where $\Delta p_{py}^\mathrm{ref}$ is the reference arteriovenous pressure drop;
$q_{py}^\mathrm{ref}$ is the reference flow over the periphery;
$r_{py}$ is a scaling factor of the ateriovenous resistances; and
$k_{py}$ is a factor that accounts for the nonlinearity of the
arteriovenous resistances, see~\Cref{tab:ca_input_params,tab:ca_constants}.

\subsection{Connect modules}%
\label{sec:ca:connect}
Volume change of inflow arteries $\dot{V}_t^\mathrm{prox}$
and outflow veins $\dot{V}_t^\mathrm{dist}$ is now updated by
\begin{equation}\label{eq:ca:py_vol_dot}
\begin{aligned}
  \dot{V}_{t}^\mathrm{dist} &\mathrel{+}= q_{py}\\
  \dot{V}_{t}^\mathrm{prox} &\mathrel{+}= q_{py}
\end{aligned}
\end{equation}
Computation of time derivative of flow across
valves and venous-atrial inlet requires in input the cross-sectional area of
proximal and distal elements to the channel.

\begin{equation}\label{eq:ca:vlv_vol_dot}
\begin{aligned}
  \dot{V}_{c,t}^\mathrm{dist} &\mathrel{+}= q_v\\
  \dot{V}_{c,t}^\mathrm{prox} &\mathrel{+}= q_v
\end{aligned}
\end{equation}

\begin{equation}\label{eq:ca:vlv_p_up}
\begin{aligned}
  p_{c,t}^\mathrm{prox} & \mathrel{+}= \dot{V}_{c,t}^\mathrm{prox}
                                       Z_{c,t}^\mathrm{prox}\\
  p_{c,t}^\mathrm{dist} & \mathrel{+}= \dot{V}_{c,t}^\mathrm{dist}
                                       Z_{c,t}^\mathrm{dist}
\end{aligned}
\end{equation}

\subsection{Valve dynamics}%
\begin{figure}
    \begin{tikzpicture}[remember picture]
      \node[anchor=north west, inner sep=0pt] (structure) at (0.02*\textwidth,0)
  {\includegraphics[width=0.47\textwidth]{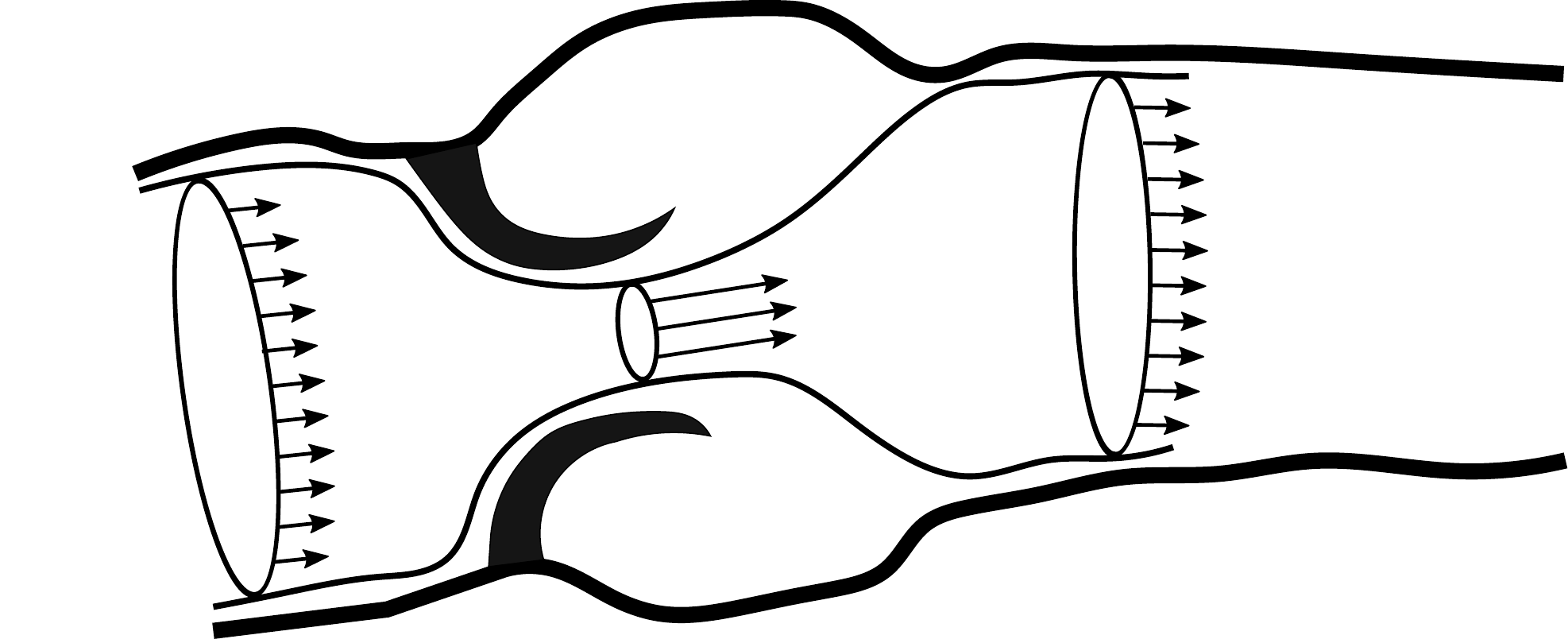}};
  \node[anchor=north west, inner sep=0pt] (structure) at (0.52*\textwidth,0)
  {\includegraphics[width=0.47\textwidth]{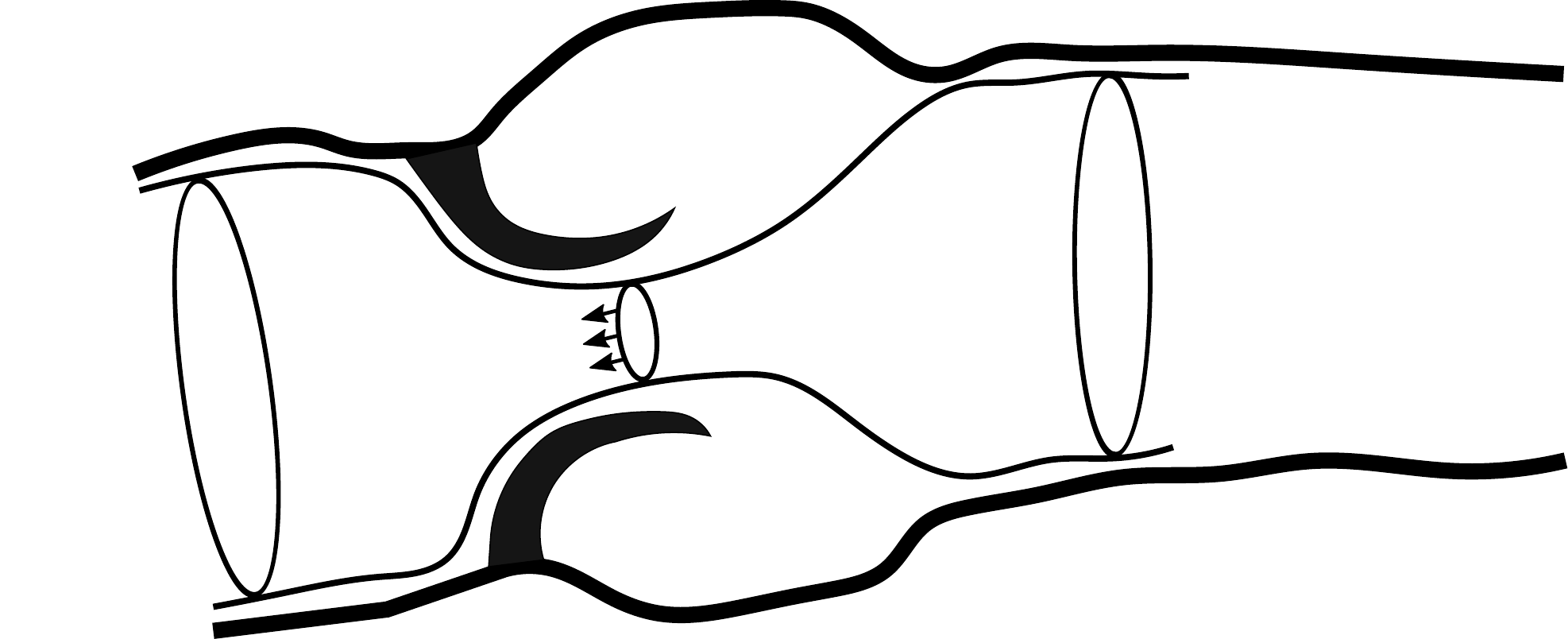}};
  \node[anchor=north west] at (0, 0) {\footnotesize\textbf{(a)}};
  \node[anchor=north west] at (0.51*\textwidth, 0) {\footnotesize\textbf{(b)}};
  \node at (0.7, -1.80) {\scriptsize $A_v^\mathrm{prox}$};
  \node at (2.85, -1.6) {\scriptsize $A_v$};
  \node at (4.77, -1.32) {\scriptsize $A_v^\mathrm{dist}$};
  \node at (8.25, -1.85) {\scriptsize $A_v^\mathrm{prox}$};
  \node at (10.3, -1.55) {\scriptsize $A_v$};
  \node at (12.4,-1.25) {\scriptsize $A_v^\mathrm{dist}$};
  \end{tikzpicture}
  \caption{Schematic of the (a) open and (b) regurging valve,
           based on~\cite{otto2006valvular}.}%
  \label{fig:ca_valve}
\end{figure}

\label{sec:ca:valve}
The pressure drop ($\Delta p_v$) across a valve is the sum of the effects of
inertia due to acceleration in time and the
Bernoulli effect, see~\cite{firstenberg2001noninvasive}
\begin{equation}
    \Delta p_v = \rho_\mathrm{b} \frac{l_v}{A_v}\dot{q}_v
    + \frac{\rho_\mathrm{b}}{2}\left({(v_v^\mathrm{out})}^2-{(v_v^\mathrm{in})}^2\right),
\end{equation}
where $\rho_\mathrm{b}$ is the density of blood, $A_v$ is the current cross-sectional
area of the valve, and $l_v$ is the length of the channel with inertia.
If not mentioned otherwise this values is estimated as
\[
    l_v = \sqrt{A_v^\mathrm{open}},
\]
with $A_v^\mathrm{open}$ the given cross-sectional area of the open valve,
see~\Cref{tab:ca_input_params}.
For $q_v\ge 0$, $v_v^\mathrm{in}$ is the velocity proximal to the valve
$v_v^\mathrm{prox}$. $v_v^\mathrm{out}$ is the maximum of the blood velocities in
the valve region
$v_v^{\max}=\max(v_v^\mathrm{dist},v_v,v_v^\mathrm{prox})$.
For $q_v< 0$ which indicates that the valve is leaking $v_v^\mathrm{in}$ is the
velocity distal to the valve $v_v^\mathrm{dist}$ and the outflow velocity is the
maximum of the blood velocities in the valve region
$v_v^\mathrm{out}=v_v^{\max}$.
Using $v_v=q_v/A_v$ we can write
\begin{equation} \label{eq:_dp_valve_ca}
  \Delta p_v = p_v^\mathrm{prox} - p_v^\mathrm{dist} = \alpha_v \dot{q}_v + \beta_v q_v^2
\end{equation}
with
\begin{equation}
  \alpha_v = \rho_\mathrm{b} \frac{l_v}{A_v}
\end{equation}
the inertia of the channel.
The open/closed status of the valve is a function of pressure drop and flow.
Valves are clearly open/closed if both pressure drop and flow point in the same
direction.
With forward pressure drop, the valve opens immediately.
With backward pressure and forward flow, the valve is
closing softly by a continuous function
\begin{align}
A_{v}^\mathrm{closing} &= \sqrt{\frac{x_v}{x_v^2+\Delta p_v^2}}
  \left(A_{v}^\mathrm{open}- A_{v}^\mathrm{leak}\right) + A_{v}^\mathrm{leak}
  \label{eq:valve_transition} \\
  x_v &= \frac{40\rho_\mathrm{b}\,q_v\left|q_v\right|}{{\left(A_{v}^\mathrm{open}\right)}^2}, \nonumber
\end{align}
where $A_v^\mathrm{leak}$ is the given valve cross-sectional
areas of the closed (regurging) valve.
Using this the current cross sectional area of the valve is
\begin{equation}
  A_v =
  \begin{cases}
    A_v^\mathrm{open}    &\ \text{for}\ \Delta p_v > 0, \\
    A_v^\mathrm{leak}    &\ \text{for}\ \Delta p_v < 0\ \text{and}\ q_v < 0,  \\
    A_v^\mathrm{closing} &\ \text{for}\ \Delta p_v < 0\ \text{and}\ q_v > 0.  \\
  \end{cases}
\end{equation}
We define
\begin{equation}
  A_{v}^{\min} = \min \left(A_{v}^{\mathrm{prox}}, A_v,
  A_{v}^{\mathrm{dist}} \right),
\end{equation}
with $A_{v}^{\mathrm{prox}}$ and $A_{v}^{\mathrm{dist}}$ the cross-sectional
area of the proximal and distal cavities or tubes respectively,
see (\ref{eq:ca:tube_cross_section},\ref{eq:ca:cav_cross_section},\ref{eq:ca:triseg_cross_section}) and~\Cref{fig:ca_valve}.
Using this $\beta_v$ is given as
\begin{equation}
  \beta_v  =
  \begin{cases}
      \frac{1}{2} \rho_\mathrm{b} \left[{\left(\frac{1}{A_{v}^{\min}}\right)}^2
    - {\left(\frac{1}{A_{v}^\mathrm{prox}}\right)}^2 \right] &\ \text{for}\ q_v \ge 0, \\[1em]
    \frac{1}{2} \rho_\mathrm{b} \left[ {\left(\frac{1}{A_{v}^\mathrm{dist}}\right)}^2
    - {\left(\frac{1}{A_{v}^{\min}}\right)}^2 \right] &\ \text{for}\ q_v < 0.
  \end{cases}
\end{equation}
Flow over the valve is finally updated using~\eqref{eq:_dp_valve_ca} by
\begin{equation}\label{eq:ca:valve_flow}
    \dot{q}_v = \frac{\Delta p_v - \beta_v q_v^2}{\alpha_v}.
\end{equation}
\subsection{Solve ODE system}%
\label{sec:ca:solve}
A Runge--Kutta--Fehlberg method (RKF45), \reviewerOne{see, e.g., \cite{hairer1993solving}},
is used to solve the system of 26 ordinary differential equations (ODEs):
\begin{itemize}
  \item[8] ODEs: for each of the four tubes and the four cavities we get an
    ODE to update the volume using~\Cref{eq:ca:py_vol_dot,eq:ca:vlv_vol_dot}.
  \item[2] ODEs: for the septum we update midwall volume and the radius according
      to~\eqref{eq:ca:ode_septum}.
  \item[10] ODEs: for the sarcomeres of each cavity and the septum we update
    sarcomere contracting length and contractility using
    (\ref{eq:ca:sarcomerelength}--\ref{eq:ca:contractility}).
  \item[6] ODEs: for each of the four valves and the two outlets we
    update flow by~\eqref{eq:ca:valve_flow}.
\end{itemize}

\begin{table}[htbp]
\centering
\footnotesize
\begin{tabularx}{\textwidth}{l>{\hsize=.99\hsize}RlL}
\toprule
Parameter\hspace*{-3em} & Value & Unit & Description \\
\midrule
\multicolumn{4}{l}{\emph{General}}\\
$\rho_\mathrm{b}$  & \num{1050.0}  & \si{\kg/\m\cubed}
    & blood density \\
$t_\mathrm{cycle}$ & \num{0.585}  & \si{\s}
    & cycle time ($=1/\mathrm{heart rate}$)\\
\midrule
\multicolumn{4}{l}{\emph{Tubes:} aorta (\AO), arteria pulmonalis (\AP), venae cavae (\VC), and venae pulmonales (\VP)}\\
$A_\mathrm{t}^\mathrm{wall}$  & 274 (\AO), 141 (\AP), 58 (\VC), 85 (\VP)
                              & \si{\mm^2} & cross-sectional wall area\\
$l_\mathrm{t}$  & 500 (\AO), 400 (\VC),  200 (\AP, \VP) & \si{\mm} & length of vessel\\
$A_\mathrm{t}^\mathrm{ref}$  & adjacent valve area & \si{\mm^2} & initial cross sectional area\\
$k_t$  & 5 (\AO), 8 (\AP), 10 (\VC, \VP) & [-] & stiffness exponent\\
\midrule
\multicolumn{4}{l}{\emph{Chambers:} left (\LV) and right (\RV) ventricle; left (\LA) and right atrium (\RA)}\\
$V_\mathrm{c}$  & \mbox{57.0 (\LV), 75.3 (\RV)}, \mbox{44.2 (\LA), 54.4 (\RA)}
                              & \si{\mL} & cavity volume\\
$h_\mathrm{c}^\mathrm{wall}$  & 15.0 (\LV), 4.0 (\RV), 2.0 (\LA), 2.0 (\RA)
                              & \si{\mm} & constant wall thickness\\
$\Delta t_\mathrm{c}^\mathrm{act}$  & 0.1 (\LV), 0.1 (\RV), 0.02 (\LA), 0.0 (\RA)
                              & \si{\s} & delays of onset of activation in each
                                beat starting at $t^\mathrm{bt}$; $t_c^\mathrm{act}=t^\mathrm{bt}+\Delta t_\mathrm{c}^\mathrm{act}$\\
\midrule
\multicolumn{4}{l}{\emph{Valves:} aortic (\AV), pulmonary (\PV), mitral (\MV), and tricuspid (\TV) valve;}\\
\multicolumn{4}{l}{\phantom{\emph{Valves:} }pulmonary (\PO) and systemic (\SO) outlet}\\
$A_v^\mathrm{open}$  & 500 (\MV, \TV) 400 (\AV, \PV, \SO, \PO)
& \si{\mm^2} & valve cross-sectional area\\
$A_v^\mathrm{leak}$  & 0 (\AV, \PV, \MV, \TV) $A_v^\mathrm{open}$ (\SO, \PO)
& \si{\mm^2} & cross-sectional area of closed/regurging valve\\
\midrule
\multicolumn{4}{l}{\emph{Periphery:} systemic (sys) and pulmonary (pulm) circulation}\\
$\Delta p_{py}^\mathrm{ref}$  & 1.5 (pulm) 10.0 (sys)  & \si{\kPa}
    & blood pressure drop in pulmonary/systemic circulation\\
    $q_{py}^\mathrm{ref}$  & \num{85} (pulm, sys)& \si{mL/s}
    & reference pulmonary/systemic flow\\
$r_{py}$ & 1 (pulm) 2 (sys) & [-] & resistance scaling factor \\
\bottomrule
\end{tabularx}
  \caption{Input parameters for the \ca model. Adjusted to match
    patient-specific data.}%
  \label{tab:ca_input_params}
\end{table}

\begin{table}[htbp]
\centering
\footnotesize
\begin{tabularx}{0.9\textwidth}{lRlL}
\toprule
Parameter\hspace*{-3em} & Value & Unit & Description \\
\midrule
\multicolumn{4}{l}{\emph{Tubes:} aorta (\AO), arteria pulmonalis (\AP), venae cavae (\VC), and venae pulmonales (\VP)}\\
$p_t^\mathrm{ref}$  & 12.0 (\AO) 0.5 (\VP) \mbox{0.12 (\VC) 1.8 (\AP)} & \si{\kPa} &
    reference tube pressure\\
\midrule
\multicolumn{4}{l}{\emph{Sarcomeres} in left (\LV) and right (\RV) ventricle; left (\LA) and right atrium (\RA)}\\
$L^\mathrm{s,ref}$  & 2.00 & \si{\um} & reference sarcomere length\\
$L^\mathrm{elast,iso}$ & 0.04 & \si{\um} & length of isometrically stressed series
                                          elastic element\\
$v^{\max}$ & 7 (\LV, \RV) 14 (\LA, \RA) & \si{\um/\s} & reference shortening velocity\\
$t_c^\mathrm{act,ref}$ & $0.5\,t_\mathrm{cycle}$ (\LV, \RV) $0.15\, t_\mathrm{cycle}$ (\LA, \RA) & \si{s} & reference duration of contraction\\
$\tau^\mathrm{R}$ & 0.25 (\LV, \RV) 0.4 (\LA, \RA) & [-] & ratio rise time to $t_c^\mathrm{act,ref}$ \\
$\tau^\mathrm{D}$ & 0.25 (\LV, \RV) 0.4 (\LA, \RA) & [-] & ratio decay time to $t_c^\mathrm{act,ref}$\\
$L^\mathrm{act0,ref}$ & 1.51 & \si{\um} & contractile element length with zero
                                            active stress\\
$L^\mathrm{pas0,ref}$ & 1.80 & \si{\um} & sarcomere length with zero passive
                                            stress\\
$\sigma^\mathrm{act,\max}$ & 120 (\LV, \RV) 84 (\LA, \RA) & \si{\kPa} & maximal isometric stress \\
$\sigma^\mathrm{pas,\max}$ & 22 (\LV, \RV) 50 (\LA, \RA) & \si{\kPa} & maximal passive stress  \\
$\dd L^\mathrm{s,pas}$  & 0.6 & \si{\um} & \\
\midrule
\multicolumn{4}{l}{\emph{TriSeg Module}}\\
$\tau_\mathrm{\SV}$ & 0.005 & [-] & time constant \\
\midrule
\multicolumn{4}{l}{\emph{Pericardium}}\\
$p_\mathrm{peri}^\mathrm{ref}$ & 0.005 & [-] & constant reference pressure \\
$V_\mathrm{peri}^\mathrm{ref}$ & 0.005 & [-] & constant reference volume \\
\midrule
\multicolumn{4}{l}{\emph{Periphery:} systemic (sys) and pulmonary (pulm) circulation}\\
$k_{py}$ & 2 (pulm) 1 (sys) & [-] & nonlinearity exponent \\
\bottomrule
\end{tabularx}
\caption{Default parameters for the \ca model, fitted to general
  experimental data in~\cite{walmsley2015fast}.}%
\label{tab:ca_constants}
\end{table}

\section{Finite Element Formulation}
\subsection{Variational Formulation}
We first ignore the acceleration term in $\Cref{eq:current}$ and look at the
stationary version of the boundary value problem (\ref{eq:current}--\ref{eq:neumann_bc}) and~\eqref{eq:VolumeEqu}. For the full nonlinear elastodynamics problem see~\ref{sec:generalized_alpha}.
The stationary boundary value problem is formally equivalent to the equations
\begin{align}
   \langle \mathcal{A}_0(\vec{u}),\vec{v}\rangle_{\Omega_0}
  - \langle \mathcal{F}_0(\vec{u},p_c),\vec{v}\rangle_{\Omega_0} &=\vec{0}
  ,\label{eq:saddlePCVSystem1}\\
  \langle V_c^\mathrm{PDE}(\vec{u}),q\rangle_{\Omega_0}
  - \langle V_c^\mathrm{ODE}(p_c), q\rangle_{\Omega_0}  &= 0,
  \label{eq:saddlePCVSystem2}
\end{align}
which is valid for all smooth enough vector fields $\vec{v}$ vanishing on
the Dirichlet boundary ${\Gamma}_{0,\mathrm{D}}$,
testfunctions $q$ that are $1$ for the cavity $c$ and $0$ otherwise,
the duality pairing $\langle \cdot ,\cdot \rangle_{\Omega_0}$,
and cavities $c\in\{\mathrm{\LV,\RV,\LA,\RA}\}$.
The second term on the left hand side of the variational
equation~\eqref{eq:saddlePCVSystem1} has the physical interpretation of the
rate of internal mechanical work and is given by
\begin{equation}\label{eq:lhs_stiffness}
    \langle \mathcal{A}_0(\vec{u}) , \vec{v} \rangle_{\Omega_0}
    :=\int_{\Omega_0} \tensor{S}(\vec{u}) :
  \tensor{\Sigma}(\vec{u},\vec{v}) \dd \vec{X},
\end{equation}
with the second Piola--Kirchhoff stress tensor $\tensor{S}$,  
see~\eqref{eq:additiveSplit}, and the directional derivative of the
Green--Lagrange strain tensor $\tensor{\Sigma}(\vec{u},\vec{v})$,  
see~\cite{Augustin2016anatomically, holzapfel2000nonlinear}.
The weak form of the contribution of pressure
loads~\eqref{eq:saddlePCVSystem1}, right term,
is computed using~\eqref{eq:neumann_bc}
\begin{equation} \label{eq:pressureLoads1}
  \langle \mathcal{F}_0(\vec{u},p_c),\vec{v} \rangle_{\Omega_0} =
  -p_c \int\limits_{\Gamma_{0,\mathrm{N}}}J\,\tensor{F}^{-\top}(\vec{u})\,\normalout
  \cdot\vec{v} \, \dsX.
\end{equation}
The first term of the coupling equation~\eqref{eq:saddlePCVSystem2} is
computed from~\eqref{eq:volumeOmega} using Nanson's formula and
$\vec x = \Xvec X + \vec u$ by
\begin{equation}\label{eq:VCAV}
  \langle V_c^\mathrm{PDE}(\vec{u}), q\rangle_{\Omega_0} =
  \frac{1}{3} \int\limits_{\Gamma_{0, \mathrm{N}}} (\Xvec X + \vec u)
  \cdot J \tensor F^{-\top} \normalout\, q \dsX,
\end{equation}
The second term of~\eqref{eq:saddlePCVSystem2} is computed using
the lumped \ca model, see~\ref{sec:CircAdapt},
for $c\in\{\mathrm{\LV,\RV,\LA,\RA}\}$.
\subsection{Consistent Linearization}%
\label{sec:linearization}
To solve the nonlinear variational
equations~\eqref{eq:saddlePCVSystem1}--\eqref{eq:saddlePCVSystem2},
with a FE approach we first apply a Newton--Raphson scheme,  
see~\cite{deuflhard2011newton}.
Given a nonlinear and continuously differentiable operator \(F\colon X\to Y\)
a solution to \(F(x)=0\) can be approximated by
\begin{align*}
  x^{k+1} &= x^{k} + \Delta x, \\
  \left.\frac{\partial F}{\partial x}\right|_{x = x^k} \Delta x &= -F(x^k),
\end{align*}
which is looped until convergence.
In our case, we have \(X = \left[H^1(\Omega_0,\Gamma_{0,\mathrm{D}})\right]^3 \times \mathbb R\), \(Y = \mathbb R^2\),
\(\Delta x = {(\Delta \vec u, \Delta p_c)}^\top\), \(x^k = {(\vec u^k, p_c^k)}^\top\),
and \(F = {(R_\vec{u},R_\mathrm{p})}^\top\).
We obtain the following linearized saddle-point problem for each
\((\vec u^k, p_c^k) \in \left[H^1(\Omega_0,\Gamma_{0,\mathrm{D}})\right]^3
\times \mathbb R\),
find \((\Delta \vec u, \Delta p_c) \in \left[H^1_0(\Omega_0)\right]^3
\times \mathbb R\) such that
\begin{align}
  \label{eq:saddle_point_nl:1}
  \langle \Delta\vec{u},A_0'(\vec{u}^k)\, \vec{v}\rangle_{\Omega_0}
  + \langle \Delta\vec{u},\mathcal{F}_0'(\vec{u}^k, p_c^k)\,\vec{v}\rangle_{\Omega_0}&\nonumber
  \\+ \langle \Delta p_c,\mathcal{F}_0'(\vec{u}^k, p_c^k)\,\vec{v}\rangle_{\Omega_0}
  &= -\langle R_\vec{u}(\vec u^k, p_c^k), \vec v\rangle_{\Omega_0},\\
  \label{eq:saddle_point_nl:2}
  \langle \Delta\vec{u},V_c^\mathrm{PDE}(\vec{u}^k)\, q\rangle_{\Omega_0}
  - \langle \Delta p_c, V_c^\mathrm{ODE}(p_c^k)\, q \rangle_{\Omega_0} &=
  -\langle R_\mathrm{p}(\vec u^k, p_c^k), q\rangle_{\Omega_0},
\end{align}
with the updates
\begin{align}
  \vec{u}^{k+1}&=\vec{u}^k+\Delta\vec{u},\\
  p_c^{k+1}&=p_c^k+\Delta p_c,
\end{align}
and the particular terms are introduced below.
The G\^ateaux derivative of~\eqref{eq:saddlePCVSystem1} with respect to the
displacement change update $\Delta \vec{u}$ yields the first
\begin{align}
  \langle \Delta\vec{u},A_0'(\vec{u}^k)\, \vec{v}\rangle_{\Omega_0}
  :&= \left.D_{\Delta\vec{u}} \langle \mathcal{A}_0(\vec{u}),
        \vec{v}\rangle_{\Omega_0}\right|_{\vec u =\vec u^k} \nonumber\\
  &= \int\limits_{\Omega_0} \tensor{S}_k: \tensor \Sigma(\Delta \vec u, \vec v)\dX
    + \int\limits_{\Omega_0}\tensor \Sigma(\vec u^k, \Delta \vec u)
    : \mathbb C_k : \tensor \Sigma(\vec u^k, \vec v)\dX,
    \label{eq:linearizedSaddlePCV1}
\end{align}
and second term of~\eqref{eq:saddle_point_nl:1}
\begin{align}
  \langle \Delta\vec{u},\mathcal{F}_0'(\vec{u}^k, p_c^k)\,\vec{v}\rangle_{\Omega_0}
  :&= \left.D_{\Delta\vec{u}} \langle \mathcal{F}_0(\vec{u},p_c),
    \vec{v}\rangle_{\Omega_0}\right|_{\vec u =\vec u^k, p_c = p_c^k}\nonumber\\
  &= p^k_c \int_{\Gamma_{0,\mathrm{N}}}J_k\tensor{F}_k^{-\top}\Grad^\top\!\Delta\vec{u}\,
  \tensor{F}_k^{-\top}\normalout\cdot\vec{v}\, \dsX \nonumber \\
  &-p^k_c \int_{\Gamma_{0,\mathrm{N}}}J_k(\tensor{F}_k^{-\top}:\Grad\Delta\vec{u})\,
  \tensor{F}_k^{-\top}\normalout\cdot\vec{v}\, \dsX,
    \label{eq:linearizedSaddlePCV2}
\end{align}
with abbreviations
\begin{align*}
  \tensor F_k &:= \tensor F(\vec u^k),\
  J_k := \det(\tensor F^k),\
  \tensor{S}_k:= \left.\tensor S\right|_{\vec u =\vec u^k},\
  \mathbb{C}_k:= \left.\mathbb{C}\right|_{\vec u =\vec u^k}.
\end{align*}
The G\^ateaux derivative of~\eqref{eq:saddlePCVSystem1} with respect to the
pressure change update $\Delta p_c$ yields the third term of~\eqref{eq:saddle_point_nl:1}
\begin{align}
  \langle \Delta p_c,\mathcal{F}_0'(\vec{u}^k, p_c^k)\,\vec{v}\rangle_{\Omega_0}
:&=\left.D_{\Delta p_c} \langle \mathcal{F}_0(\vec{u},p_c),
        \vec{v}\rangle_{\Omega_0}\right|_{\vec u =\vec u^k, p_c = p_c^k} \nonumber\\
        &=-\Delta p_c  \int_{\Gamma_{0,\mathrm{N}}}J_k\,\tensor{F}_k^{-\top}\,\normalout\cdot\vec{v} \, \dd s_{\vec{X}}.
    \label{eq:linearizedSaddlePCV3}
\end{align}
The residual $R_\vec{u}$, i.e.,
the right hand side of~\eqref{eq:saddle_point_nl:1}, is computed as
\begin{equation}\label{eq:upper_rhs_variational}
\langle R_\vec{u}(\vec u^k, p_c^k), \vec v\rangle_{\Omega_0}
  :=\langle A_0(\vec{u}^k),\vec{v}\rangle_{\Omega_0}
  - \langle \vec{\mathcal{F}}_0(\vec{u}^k, p_c^k),\vec{v} \rangle_{\Omega_0}.
\end{equation}
From~\eqref{eq:VCAV},
using the known relations, see, e.g.,~\cite{holzapfel2000nonlinear},
\begin{align*}
  \frac{\partial J}{\partial \tensor F} : \Grad \Delta \vec u
    &= J \tensor F^{-\top} : \Grad \Delta \vec u \\
  \frac{\partial \tensor F^{-\top}}{\partial \tensor F} : \Grad \Delta \vec u
    &= -\tensor F^{-\top}{(\Grad \Delta \vec u)}^\top \tensor F^{-\top}
\end{align*}
we can calculate the first term of~\eqref{eq:saddle_point_nl:2} as the
G\^ateaux derivative with respect to the update $\Delta\vec{u}$
\begin{align}
  \langle \Delta\vec{u},V_c^\mathrm{PDE}(\vec{u}^k)\, q\rangle_{\Omega_0}
  :&=\left.D_{\Delta \vec{u}}
  \langle V_c^\mathrm{PDE}(\vec{u}),q\rangle_{\Omega_0}
    \right|_{\vec u =\vec u^k} \nonumber \\
    &= D_{\Delta \vec u} \frac{1}{3}\int_{\Gamma_{0,\mathrm{N}}}
    \left(\vec X + \vec{u}^k\right) \cdot J_k \tensor F_k^{-\top}
    \normalout\,q \dsX \nonumber \\
    &= \frac{1}{3} \int_{\Gamma_{0,\mathrm{N}}} J_k (\tensor F_k^{-\top}
  : \Grad \Delta \vec u) \vec x \cdot \tensor F_k^{-\top}
    \normalout\,q \dsX \nonumber \\
    &\quad - \frac{1}{3} \int_{\Gamma_{0,\mathrm{N}}} J_k \vec x \cdot
    \tensor F_k^{-\top}{(\Grad \Delta \vec u)}^\top
    \tensor F_k^{-\top} \normalout\,q \dsX \nonumber \\
    &\quad + \frac{1}{3} \int_{\Gamma_{0,\mathrm{N}}} J_k \Delta \vec u \cdot
    \tensor F_k^{-\top}\normalout\,q \dsX,  \label{eq:var_volume_cav}
\end{align}
with $q$  a testfunction that is $1$ for the surface of cavity $c$, $\Gamma_{0,c}$,
and $0$ otherwise.

The second term of~\eqref{eq:saddle_point_nl:2} is computed as a
numerical derivative
\begin{align}
  \langle \Delta p_c, V_c^\mathrm{ODE}(p_c^k)\, q \rangle_{\Omega_0}
  :&= \left.D_{\Delta p_c} \langle V_c^\mathrm{ODE}(p_c)\, q \rangle_{\Omega_0}
    \right|_{p_c = p_c^k} \nonumber \\
    &= \frac{1}{\epsilon}\left(V_c^\mathrm{ODE}(p_c^k + \epsilon) -
    V_c^\mathrm{ODE}(p_c^k)\right)q, \label{eq:var_compliance}
\end{align}
where $\epsilon=p_c^k\sqrt{\epsilon_\mathrm{m}}$ is chosen according
to~\cite[Chapter 5.7]{Press2007} with
$\epsilon_\mathrm{m}=2^{-52}\approx2.2*10^{-16}$ the machine accuracy.

Finally, the residual $R_\mathrm{p}$, i.e.,
the right hand side of~\eqref{eq:saddle_point_nl:2}, is computed as
\begin{equation} \label{eq:lower_rhs_variational}
\langle R_\mathrm{p}(\vec u^k, p_c^k), q\rangle_{\Omega_0}
  := \langle V_c^\mathrm{PDE}(\vec{u}),q\rangle_{\Omega_0}
  - \langle V_c^\mathrm{ODE}(p_c), q\rangle_{\Omega_0}.
\end{equation}

\subsection{Assembling of the block matrices}%
\label{sec:assembling}
To apply the finite element method (FEM) we consider an admissible decomposition
of the computational domain $\Omega \subset \mathbb{R}^3$
into $M$ tetrahedral elements $\tau_j$ and introduce a conformal finite
element space
\begin{align*}
X_h \subset H^1(\Omega_0), \ N = \text{dim}\, X_h
\end{align*}
of piecewise polynomial continuous basis functions $\varphi_i$.
The linearized variational problem~\eqref{eq:saddle_point_nl:1}--\eqref{eq:saddle_point_nl:2}
and a Galerkin FE discretization result in solving the
block system to find $\delta\Rvec{u}\in\mathbb{R}^{3N}$ and
$\delta \pcfem\in\mathbb{R}^{N_\mathrm{cav}}$ such that
\[
  \tensor{K}'(\Rvec{u}^k, \pcfem^k)
  \begin{pmatrix} \delta\Rvec{u}\\ \delta \pcfem \end{pmatrix}
  = -\Rvec{K}(\Rvec{u}^k, \pcfem^k), \quad
  \Rvec{K}(\Rvec{u}^k, \pcfem^k):= -\begin{pmatrix}\Rvec{R}_{\vec{u}}(\Rvec{u}^k, \pcfem^k) \\ \Rvec{R}_p(\Rvec{u}^k, \pcfem^k)
      \end{pmatrix},
\]
i.e.,
\begin{align}\label{eq:blockCVSystem}
  \begin{pmatrix}
    (\tensor{A}'-\tensor{M}')(\Rvec{u}^k, \pcfem^k)  & \tensor{B}'_\mathrm{p}(\Rvec{u}^k) \\
        \tensor{B}'_\vec{u}(\Rvec{u}^k) & \tensor{C}'(\pcfem^k) \\
 \end{pmatrix}
\begin{pmatrix}
  \delta\Rvec{u}\\ \delta \pcfem
\end{pmatrix}
&=-
\begin{pmatrix}
  \Rvec{A}(\Rvec{u}^k)-\Rvec{B}_\mathrm{p}(\Rvec{u}^k,\pcfem^k) \\
  \Rvec{V}_c^\mathrm{PDE}(\Rvec{u}^k) -\Rvec{V}_c^\mathrm{ODE}(\pcfem^k)
\end{pmatrix},\\
    \Rvec{u}^{k+1}   &= \Rvec{u}^{k} + \delta\Rvec{u},\\
    \pcfem^{k+1} &= \pcfem^k + \delta\pcfem
\end{align}
with the solution vectors $\Rvec{u}^k\in\mathbb{R}^{3N}$ and
$\pcfem^k\in\mathbb{R}^{N_\mathrm{cav}}$ at the $k$-th Newton
step.
The tangent stiffness matrix $\tensor{A}'\in\mathbb{R}^{3N\times 3N}$
is calculated from~\eqref{eq:linearizedSaddlePCV1} according to
\begin{equation}\label{eq:lhs_assemblingv}
  \tensor{A}'(\Rvec{u}^k)[j,i] :=
    \langle \boldsymbol{\varphi}_i, \mathcal{A}_0'(\vec{u}^k)\,
            \boldsymbol{\varphi}_j \rangle_{\Omega_0}
  \end{equation}
and the mass matrix $\tensor{M}'\in\mathbb{R}^{3N\times 3N}$
is calculated from~\eqref{eq:linearizedSaddlePCV2} according to
\begin{equation}\label{eq:lhs_mass}
  \tensor{M}'(\Rvec{u}^k, \pcfem^k)[j,i] :=
    \langle \boldsymbol{\varphi}_i, \mathcal{F}_0'(\vec{u}^k, p_c^k)\,
            \boldsymbol{\varphi}_j \rangle_{\Omega_0},
\end{equation}
see also~\cite{Augustin2016anatomically, holzapfel2000nonlinear}.

The off-diagonal matrices
$\tensor{B}'_{\vec{u}}\in\mathbb{R}^{3N\times N_\mathrm{cav}}$ and
$\tensor{B}'_{\mathrm{p}}\in\mathbb{R}^{N_\mathrm{cav}\times 3N}$
in~\eqref{eq:blockCVSystem} are assembled using~\eqref{eq:var_volume_cav}
\begin{equation}\label{eq:Bu_assembling}
  \tensor{B}'_{\vec{u}}(\Rvec{u}^k,\pcfem^k)[i,j]
  = \langle \boldsymbol{\varphi}_j,
  V_c^\mathrm{PDE}(\vec{u}^k)\hat{\varphi_i}\rangle_{\Omega_{0}},\quad
            i=1,\dots,N_\mathrm{cav}
\end{equation}
and using~\eqref{eq:linearizedSaddlePCV3}
\begin{equation}\label{eq:Bp_assembling}
  \tensor{B}'_{\mathrm{p}}(\Rvec{u}^k,\pcfem^k)[i,j]
  = \langle \hat{\varphi_j},\mathcal{F}_0'(\vec{u}^k, p_c^k)
  \boldsymbol{\varphi}_i\rangle_{\Omega_0}, \quad
  j=1,\dots,N_\mathrm{cav},
\end{equation}
with the constant shape function $\hat{\varphi_j}=1$ if $\tau_j\in\Gamma_{0,c}$
and $\hat{\varphi_j}=0$ if $\tau_j\notin\Gamma_{0,c}$
for $c\in\{\mathrm{\LV,\RV,\LA,\RA}\}$.

Using a technique as described in~\cite[Sect.~4.2]{rumpel2003volume} this
assembling procedure can be simplified for closed cavities such that
\[
  \tensor{B}'_{\mathrm{p}}(\Rvec{u}^k,\pcfem^k) =
  {\left[\tensor{B}'_{\vec{u}}(\Rvec{u}^k,\pcfem^k)\right]}^\top.
\]
The circulatory compliance matrix
$\tensor{C}'(\pcfem^k)\in\mathbb{R}^{N_\mathrm{cav}\times N_\mathrm{cav}}$
is computed from~\eqref{eq:var_compliance} as
\begin{equation}\label{eq:Cp_assembling}
  \tensor{C}'(\pcfem^k)[i,j]=
  \langle \hat{\varphi_j}, V_c^\mathrm{ODE}(p_c^k)\,
    \hat{\varphi_i}\rangle_{\Omega_0}, \quad
    i,j=1,\dots,N_\mathrm{cav},
\end{equation}
with the constant shape function $\hat{\varphi_i}, \hat{\varphi_j}=1$ for
cavity c and 0 otherwise, leading to a diagonal matrix.

The terms on the upper right hand side $\Rvec{A}\in\mathbb{R}^{3N}$,
$\Rvec{B}_\mathrm{p}\in\mathbb{R}^{3N}$ are constructed using~\eqref{eq:upper_rhs_variational}
resulting in
$\Rvec{R}_{\vec{u}}(\Rvec{u}^k, \pcfem^k) =
\Rvec{A}(\Rvec{u}^k)-\Rvec{B}_\mathrm{p}(\Rvec{u}^k, \pcfem^k)$ with
\begin{equation}\label{eq:residual_K}
  \Rvec{A}(\Rvec{u}^k)[i] := \langle
  \mathcal{A}_0(\vec{u}^k),\boldsymbol{\varphi}_i \rangle_{\Omega_0}
\end{equation}
and
\begin{equation}\label{eq:residual_B}
  \Rvec{B}_\mathrm{p}(\Rvec{u}^k, \pcfem^k)[i] := \langle
  \mathcal{F}_0(\vec{u}^k, p_c^k),\boldsymbol{\varphi}_i \rangle_{\Omega_0}.
\end{equation}
Finally, the lower right hand side in~\eqref{eq:blockCVSystem},
$\Rvec{R}_p(\Rvec{u}^k,\Rvec{p}_\mathrm{c}^k)=\Rvec{V}^\mathrm{PDE}(\Rvec{u}^k)-\Rvec{V}^\mathrm{ODE}(\Rvec{p}_\mathrm{c}^k)\in\mathbb{R}^{N_\mathrm{cav}}$,
is assembled from~\eqref{eq:lower_rhs_variational} with
\begin{equation}\label{eq:VcPDE_assembling}
  \Rvec{V}^\mathrm{PDE}(\Rvec{u}^k)[i]
  =\langle V_c^\mathrm{PDE}(\vec{u}),\hat{\varphi_i}\rangle_{\Omega_0},\quad
    i=1,\dots,N_\mathrm{cav},
\end{equation}
and
\begin{equation}\label{eq:VcODE_assembling}
  \Rvec{V}^\mathrm{ODE}(\Rvec{p}_\mathrm{c}^k)[i]
  =\langle V_c^\mathrm{ODE}(p_c), \hat{\varphi_i}\rangle_{\Omega_0},\quad
    i=1,\dots,N_\mathrm{cav}.
\end{equation}

\section{Generalized-$\alpha$ Scheme}\label{sec:generalized_alpha}
After standard discretization we rewrite~\Cref{eq:current}
using~\Cref{eq:residual_K,eq:residual_B} as a nonlinear ODE reading
\begin{equation} \label{eq:ode_elast}
  \rho_0\tensor{M}_\alpha \ddot{\Rvec u}(t) + \Rvec R_{\vec{u}}(\Rvec{u}, t) = \Rvec{0},
\end{equation}
with the mass matrix
\begin{align*}
  \tensor M_\alpha[i,j] &:=
    \int_{\Omega_0} \boldsymbol{\varphi}_i(\vec X) \cdot \boldsymbol{\varphi}_j(\vec X)\dd\vec X.
\end{align*}
Following \cite{kadapa2017advantages} we reformulate \Cref{eq:ode_elast} as a
first order ODE system by introducing the velocity $\Rvec v$
\begin{alignat}{2}\label{eq:ode_elast_first_order}
  \rho_0\tensor{M}_\alpha \dot{\Rvec v}(t) &+ \Rvec R_{\vec{u}}(\Rvec{u}, t) &&= \Rvec{0}, \\
  \tensor{M}_\alpha \dot{\Rvec u}(t) &- \tensor{M}_\alpha \Rvec v(t) &&= \Rvec 0
\end{alignat}
and apply a generalized-$\alpha$ approach~\cite{chung1993time}.
To this end we define three parameters
\begin{align*}
  \alpha_\mathrm{f} := \frac{1}{1+\rho_\infty},\quad
  \alpha_\mathrm{m} := \frac{3-\rho_\infty}{2(1+\rho_\infty)},\quad
  \gamma := \frac{1}{2}+\alpha_\mathrm{m} - \alpha_\mathrm{f},
\end{align*}
where the \emph{spectral radius} $\rho_\infty$ is a parameter between $0$ and $1$.
With this we introduce
\begin{alignat*}{4}
  \dot{\Rvec v}_{n+\alpha_{\mathrm{m}}} &:=
    \alpha_{\mathrm{m}} &&\dot{\Rvec v}_{n+1} &&+ (1-\alpha_{\mathrm{m}}) &&\dot{\Rvec v}_{n},\\
  \dot{\Rvec u}_{n+\alpha_{\mathrm{m}}} &:=
    \alpha_{\mathrm{m}} &&\dot{\Rvec u}_{n+1} &&+ (1-\alpha_{\mathrm{m}}) &&\dot{\Rvec u}_{n},\\
  \Rvec v_{n+\alpha_{\mathrm{f}}} &:=
    \alpha_{\mathrm{f}} &&\Rvec v_{n+1} &&+ (1-\alpha_{\mathrm{f}}) &&\Rvec v_{n},\\
  \Rvec u_{n+\alpha_{\mathrm{f}}} &:=
    \alpha_{\mathrm{f}} &&\Rvec u_{n+1} &&+ (1-\alpha_{\mathrm{f}}) &&\Rvec u_{n},
\end{alignat*}
and reformulate~\Cref{eq:ode_elast_first_order} as
\begin{alignat}{2} \label{eq:GA_elast_system:1}
  \rho_0\tensor{M}_\alpha\dot{\Rvec v}_{n+\alpha_\mathrm{m}} &+ \Rvec R_{\vec{u}}(\Rvec{u}_{n+\alpha_\mathrm{f}}) &&= \Rvec{0},\\ \label{eq:GA_elast_system:2}
  \tensor{M}_\alpha \dot{\Rvec u}_{n+\alpha_\mathrm{m}} &- \tensor{M}_\alpha \Rvec v_{n+\alpha_\mathrm{f}} &&=
  \Rvec 0.
\end{alignat}
Here, the second equation gives us
\begin{align*}
  \dot{\Rvec u}_{n+\alpha_\mathrm{m}} = \Rvec v_{n+\alpha_\mathrm{f}}
\end{align*}
and we get for the velocity update
\begin{align}\label{eq:vel_update}
  \Rvec v_{n+1} = \frac{\alpha_\mathrm{m}}{\alpha_\mathrm{f} \gamma \Delta t}\left(\Rvec u_{n+1}
  - \Rvec u_n\right) + \frac{\gamma - \alpha_\mathrm{m}}{\gamma \alpha_\mathrm{f}} \dot{\Rvec u}_n
  + \frac{\alpha_\mathrm{f} - 1}{\alpha_\mathrm{f}} \Rvec{v}_n.
\end{align}
From this and the relationship by \citet{newmark1959method}
\begin{alignat*}{3}
  \Rvec{u}_{n+1} &= \Rvec{u}_n &&+ \Delta t \left(\gamma \dot{\Rvec u}_{n+1} +
  (1-\gamma) \dot{\Rvec u}_n \right),\\
  \Rvec{v}_{n+1} &= \Rvec v_n &&+ \Delta t \left(\gamma \dot{\Rvec v}_{n+1} +
(1-\gamma) \dot{\Rvec v}_n \right),
\end{alignat*}
we obtain
\begin{align}\label{eq:dvel_update}
  \dot{\Rvec v}_{n+1} = \frac{\alpha_{\mathrm{m}}}{\alpha_{\mathrm{f}} \gamma^2 \Delta t^2}
  \left(\Rvec u_{n+1}-\Rvec u_n\right) -\frac{1}{\alpha_{\mathrm{f}} \gamma \Delta t}
  \Rvec v_n + \frac{\gamma-1}{\gamma}\dot{\Rvec v}_n
  + \frac{\gamma-\alpha_{\mathrm{m}}}{\alpha_{\mathrm{f}} \gamma^2 \Delta t}.
\end{align}
Hence, we can rewrite the whole first order system only dependent on the
unknowns $\Rvec u_{n+1}$.
\paragraph{Newton's method for the generalized-$\alpha$ scheme}
For the implementation of Newton's method we compute
\begin{equation}\label{eq:ga:derivatives}
  \frac{\partial \dot{\Rvec v}_{n+\alpha_{\mathrm{m}}}}{\partial \Rvec u_{n+1}} =
    \frac{\alpha_{\mathrm{m}}^2}{\alpha_{\mathrm{f}} \gamma^2 \Delta t^2},\quad
  \frac{\partial \Rvec v_{n+\alpha_{\mathrm{f}}}}{\partial \Rvec u_{n+1}} =
    \frac{\alpha_{\mathrm{f}}\alpha_{\mathrm{m}}}{\alpha_{\mathrm{f}} \gamma \Delta t} =
    \frac{\alpha_{\mathrm{m}}}{\gamma \Delta t},\quad
  \frac{\partial \Rvec u_{n+\alpha_{\mathrm{f}}}}{\partial \Rvec u_{n+1}} = \alpha_{\mathrm{f}}.
\end{equation}
To calculate the solution at the current timestep we assume that we know
$\Rvec{u}_n$, $\dot{\Rvec{u}}_n$, $\Rvec{v}_n$
and $\dot{\Rvec v}_n$ from the previous time step $n$ and get from~\Cref{eq:GA_elast_system:1}
for the residual
\begin{equation} \label{eq:ga:residual}
  \Rvec R_\alpha(\Rvec u_{n+1}^k) :=
  - \rho_0\tensor{M}_\alpha \dot{\Rvec v}^k_{n+\alpha_{\mathrm{m}}}
  - \Rvec{R}_\vec{u}(\Rvec u^k_{n+\alpha_{\mathrm{f}}}),
\end{equation}
with $\dot{\Rvec v}^k_{n+\alpha_{\mathrm{m}}}:=\dot{\Rvec v}_{n+\alpha_{\mathrm{m}}}(\Rvec u^k_{n+1})$ and
$\Rvec u^k_{n+\alpha_{\mathrm{f}}}:= \Rvec u_{n+\alpha_{\mathrm{f}}}(\Rvec u^k_{n+1})$.
To increase stability we consider Rayleigh damping by adding the two matrices
\begin{alignat}{4}
  \tensor{D}_\mathrm{mass}(\Rvec u_{n+1}^k)
    &= \rho_0 &&\beta_\mathrm{mass} &&\tensor{M}_\alpha &&\Rvec{v}^k_{n+\alpha_\mathrm{f}},\label{eq:ga:mass_damping}\\
  \tensor{D}_\mathrm{stiff}(\Rvec u_{n+1}^k)
  &=&&\beta_\mathrm{stiff} &&\tensor{A}'(u_n) && \Rvec{v}^k_{n+\alpha_\mathrm{f}}, \label{eq:ga:stiffness_damping}
\end{alignat}
to the residual~\eqref{eq:ga:residual}
with $\Rvec{v}^k_{n+\alpha_{\mathrm{f}}}:=\Rvec{v}_{n+\alpha_{\mathrm{f}}}(\Rvec u^k_{n+1})$ and Rayleigh
damping parameters $\beta_\mathrm{mass}\ge\SI{0}{\ms^{-1}}$, $\beta_\mathrm{stiff}\ge\SI{0}{\ms}$.

The tangent stiffness matrix is now calculated using \eqref{eq:ga:derivatives} as
\begin{align}
  \tensor{A}'_\alpha(\Rvec u^k_{n+1}, \pcfem[n+1]^k) &:=
  \rho_0 \frac{\partial \dot{\Rvec v}_{n+\alpha_{\mathrm{m}}}}{\partial \Rvec u_{n+1}} \tensor{M}_\alpha
    + \frac{\partial \Rvec u_{n+\alpha_{\mathrm{f}}}}{\partial \Rvec u_{n+1}}
    \left(\tensor{A}'(\Rvec{u}^k_{n+\alpha_{\mathrm{f}}})-\tensor{M}'(\Rvec{u}^k_{n+\alpha_{\mathrm{f}}},\pcfem[n+1]^k)\right)\nonumber\\
    &= \rho_0 \frac{\alpha_{\mathrm{m}}^2}{\alpha_{\mathrm{f}} \gamma^2 \Delta t^2} \tensor{M}_\alpha
    + \alpha_{\mathrm{f}}\left(\tensor{A}'(\Rvec{u}^k_{n+\alpha_{\mathrm{f}}})-\tensor{M}'(\Rvec{u}^k_{n+\alpha_{\mathrm{f}}},\pcfem[n+1]^k)\right), \label{eq:ga:stiffness}
\end{align}
with $\tensor{A}'$, and $\tensor{M}'$ being the known tangent
stiffness matrices from the quasi-stationary elasticity case,
see~\Cref{eq:lhs_assemblingv,eq:lhs_mass}.
When using a coupling with the circulatory system we compute the off diagonal
matrices and lower right hand side,
see~\Cref{eq:Bu_assembling,eq:Bp_assembling,eq:VcPDE_assembling},
in terms of $\Rvec{u}^k_{n+\alpha_{\mathrm{f}}}$.

\section{Direct Schur Complement Solver for a Small Number of Constraints}%
\label{sec:SchurComplement}
Given the block system $\tensor{A}\in\mathbb{R}^{n\times n}$, $\tensor{D}\in\mathbb{R}^{m\times m}$
\begin{align*}
  \begin{pmatrix}
    \tensor{A}&\tensor{B}\\
    \tensor{C}&\tensor{D}
  \end{pmatrix}
  \begin{pmatrix}
    \Rvec{x}\\ \Rvec{y}
  \end{pmatrix}=-
  \begin{pmatrix}
    \Rvec{f}\\ \Rvec{g}
  \end{pmatrix}
\end{align*}
with
\begin{align*}
   \tensor{B}=\begin{pmatrix} \Rvec{b}_1 & \aug & \cdots & \aug &\Rvec{b}_m\end{pmatrix}\in\mathbb{R}^{n\times m}, \quad
  \tensor{C}=\begin{pmatrix} \Rvec{c}_1 & \aug & \cdots & \aug & \Rvec{c}_m\end{pmatrix}^\top\in\mathbb{R}^{m\times n},
\end{align*}
we can write the Schur complement system as
\begin{align*}
  (\tensor{C}\tensor{A}^{-1}\tensor{B}-\tensor{D})\Rvec{y}&=\Rvec{g}-\tensor{C}\tensor{A}^{-1}\Rvec{f}\\
\Rvec{x}&=\tensor{A}^{-1}\Rvec{f}-\tensor{A}^{-1}\tensor{B}\Rvec{y}.
\end{align*}
With
\begin{equation}
  \Rvec{r}=\tensor{A}^{-1}\Rvec{f},\quad
  \tensor{S}=\tensor{A}^{-1}\tensor{B}=
    \begin{pmatrix}\Rvec{s}_1 & \aug & \cdots & \aug & \Rvec{s}_m\end{pmatrix}\in\mathbb{R}^{n\times m},\
  \Rvec{s}_i=\tensor{A}^{-1}\Rvec{b}_i,\ i=1,\dots,m
\end{equation}
we get
\begin{align}
  (\tensor{C}\tensor{S}-\tensor{D})\Rvec{y}&=\Rvec{g}-\tensor{C}\Rvec{r}\nonumber\\
\Rvec{x}&=\Rvec{r}-\tensor{S}\Rvec{y}.\label{eq:SchurSolve}
\end{align}
The realization of~\eqref{eq:SchurSolve} involves $m+1$ solves and the inversion of an $m\times m$ matrix. Since $m$ is generally small this can be done symbolically.
\begin{align*}
  {[\tensor{C}\tensor{S}]}_{ij}=\Rvec{c}_i\cdot\Rvec{s}_j,\ \mbox{for}\ i,j=1,\dots,m.
\end{align*}
\end{appendix}

\clearpage
\bibliographystyle{elsarticle-num-names}
\bibliography{bibliography.bib}

\begin{thebibliography}{109}
\expandafter\ifx\csname natexlab\endcsname\relax\def\natexlab#1{#1}\fi
\providecommand{\url}[1]{\texttt{#1}}
\providecommand{\href}[2]{#2}
\providecommand{\path}[1]{#1}
\providecommand{\DOIprefix}{doi:}
\providecommand{\ArXivprefix}{arXiv:}
\providecommand{\URLprefix}{URL: }
\providecommand{\Pubmedprefix}{pmid:}
\providecommand{\doi}[1]{\href{http://dx.doi.org/#1}{\path{#1}}}
\providecommand{\Pubmed}[1]{\href{pmid:#1}{\path{#1}}}
\providecommand{\bibinfo}[2]{#2}
\ifx\xfnm\relax \def\xfnm[#1]{\unskip,\space#1}\fi
\bibitem[{Laslett et~al.(2012)Laslett, Alagona, Clark, Drozda, Saldivar,
  Wilson, Poe, and Hart}]{laslett2012worldwide}
\bibinfo{author}{L.~J. Laslett}, \bibinfo{author}{P.~Alagona},
  \bibinfo{author}{B.~A. Clark}, \bibinfo{author}{J.~P. Drozda},
  \bibinfo{author}{F.~Saldivar}, \bibinfo{author}{S.~R. Wilson},
  \bibinfo{author}{C.~Poe}, \bibinfo{author}{M.~Hart},
\newblock \bibinfo{title}{The worldwide environment of cardiovascular disease:
  prevalence, diagnosis, therapy, and policy issues: a report from the american
  college of cardiology},
\newblock \bibinfo{journal}{Journal of the American College of Cardiology}
  \bibinfo{volume}{60} (\bibinfo{year}{2012}) \bibinfo{pages}{S1--S49}.
\bibitem[{Timmis et~al.(2020)Timmis, Townsend, Gale, Torbica, Lettino,
  Petersen, Mossialos, Maggioni, Kazakiewicz, May et~al.}]{timmis2020european}
\bibinfo{author}{A.~Timmis}, \bibinfo{author}{N.~Townsend},
  \bibinfo{author}{C.~P. Gale}, \bibinfo{author}{A.~Torbica},
  \bibinfo{author}{M.~Lettino}, \bibinfo{author}{S.~E. Petersen},
  \bibinfo{author}{E.~A. Mossialos}, \bibinfo{author}{A.~P. Maggioni},
  \bibinfo{author}{D.~Kazakiewicz}, \bibinfo{author}{H.~T. May}, et~al.,
\newblock \bibinfo{title}{European society of cardiology: cardiovascular
  disease statistics 2019},
\newblock \bibinfo{journal}{European Heart Journal} \bibinfo{volume}{41}
  (\bibinfo{year}{2020}) \bibinfo{pages}{12--85}.
\bibitem[{Wilkins et~al.(2017)Wilkins, Wilson, Wickramasinghe, Bhatnagar, Leal,
  Luengo-Fernandez, Burns, Rayner, and Townsend}]{wilkins2017european}
\bibinfo{author}{E.~Wilkins}, \bibinfo{author}{L.~Wilson},
  \bibinfo{author}{K.~Wickramasinghe}, \bibinfo{author}{P.~Bhatnagar},
  \bibinfo{author}{J.~Leal}, \bibinfo{author}{R.~Luengo-Fernandez},
  \bibinfo{author}{R.~Burns}, \bibinfo{author}{M.~Rayner},
  \bibinfo{author}{N.~Townsend},
\newblock \bibinfo{title}{European cardiovascular disease statistics 2017},
\newblock \bibinfo{journal}{European Heart Network}  (\bibinfo{year}{2017}).
\bibitem[{Smith et~al.(2011)Smith, de~Vecchi, McCormick, Nordsletten, Camara,
  Frangi, Delingette, Sermesant, Relan, Ayache, Krueger, Schulze, Hose,
  Valverde, Beerbaum, Staicu, Siebes, Spaan, Hunter, Weese, Lehmann, Chapelle,
  and Rezavi}]{smith2011euHeart}
\bibinfo{author}{N.~P. Smith}, \bibinfo{author}{A.~de~Vecchi},
  \bibinfo{author}{M.~McCormick}, \bibinfo{author}{D.~A. Nordsletten},
  \bibinfo{author}{O.~Camara}, \bibinfo{author}{a.~F. Frangi},
  \bibinfo{author}{H.~Delingette}, \bibinfo{author}{M.~Sermesant},
  \bibinfo{author}{J.~Relan}, \bibinfo{author}{N.~Ayache},
  \bibinfo{author}{M.~W. Krueger}, \bibinfo{author}{W.~H.~W. Schulze},
  \bibinfo{author}{R.~Hose}, \bibinfo{author}{I.~Valverde},
  \bibinfo{author}{P.~Beerbaum}, \bibinfo{author}{C.~Staicu},
  \bibinfo{author}{M.~Siebes}, \bibinfo{author}{J.~Spaan},
  \bibinfo{author}{P.~J. Hunter}, \bibinfo{author}{J.~Weese},
  \bibinfo{author}{H.~Lehmann}, \bibinfo{author}{D.~Chapelle},
  \bibinfo{author}{R.~Rezavi},
\newblock \bibinfo{title}{{euHeart: personalized and integrated cardiac care
  using patient-specific cardiovascular modelling}},
\newblock \bibinfo{journal}{Interface Focus} \bibinfo{volume}{1}
  (\bibinfo{year}{2011}) \bibinfo{pages}{349--364}.
\bibitem[{Arevalo et~al.(2016)Arevalo, Vadakkumpadan, Guallar, Jebb, Malamas,
  Wu, and Trayanova}]{arevalo2016arrhythmia}
\bibinfo{author}{H.~J. Arevalo}, \bibinfo{author}{F.~Vadakkumpadan},
  \bibinfo{author}{E.~Guallar}, \bibinfo{author}{A.~Jebb},
  \bibinfo{author}{P.~Malamas}, \bibinfo{author}{K.~C. Wu},
  \bibinfo{author}{N.~A. Trayanova},
\newblock \bibinfo{title}{{Arrhythmia risk stratification of patients after
  myocardial infarction using personalized heart models}},
\newblock \bibinfo{journal}{Nature Communications}  (\bibinfo{year}{2016}).
\bibitem[{Prakosa et~al.(2018)Prakosa, Arevalo, Deng, Boyle, Nikolov, Ashikaga,
  Blauer, Ghafoori, Park, Blake, Han, MacLeod, Halperin, Callans, Ranjan,
  Chrispin, Nazarian, and Trayanova}]{prakosa2018personalized}
\bibinfo{author}{A.~Prakosa}, \bibinfo{author}{H.~J. Arevalo},
  \bibinfo{author}{D.~Deng}, \bibinfo{author}{P.~M. Boyle},
  \bibinfo{author}{P.~P. Nikolov}, \bibinfo{author}{H.~Ashikaga},
  \bibinfo{author}{J.~J. Blauer}, \bibinfo{author}{E.~Ghafoori},
  \bibinfo{author}{C.~J. Park}, \bibinfo{author}{R.~C. Blake},
  \bibinfo{author}{F.~T. Han}, \bibinfo{author}{R.~S. MacLeod},
  \bibinfo{author}{H.~R. Halperin}, \bibinfo{author}{D.~J. Callans},
  \bibinfo{author}{R.~Ranjan}, \bibinfo{author}{J.~Chrispin},
  \bibinfo{author}{S.~Nazarian}, \bibinfo{author}{N.~A. Trayanova},
\newblock \bibinfo{title}{{Personalized virtual-heart technology for guiding
  the ablation of infarct-related ventricular tachycardia}},
\newblock \bibinfo{journal}{Nature Biomedical Engineering}
  (\bibinfo{year}{2018}).
\bibitem[{Strocchi et~al.(2020)Strocchi, Gsell, Augustin, Razeghi, Roney,
  Prassl, Vigmond, Behar, Gould, Rinaldi, Bishop, Plank, and
  Niederer}]{strocchi2020simulating}
\bibinfo{author}{M.~Strocchi}, \bibinfo{author}{M.~A. Gsell},
  \bibinfo{author}{C.~M. Augustin}, \bibinfo{author}{O.~Razeghi},
  \bibinfo{author}{C.~H. Roney}, \bibinfo{author}{A.~J. Prassl},
  \bibinfo{author}{E.~J. Vigmond}, \bibinfo{author}{J.~M. Behar},
  \bibinfo{author}{J.~S. Gould}, \bibinfo{author}{C.~A. Rinaldi},
  \bibinfo{author}{M.~J. Bishop}, \bibinfo{author}{G.~Plank},
  \bibinfo{author}{S.~A. Niederer},
\newblock \bibinfo{title}{{Simulating ventricular systolic motion in a
  four-chamber heart model with spatially varying robin boundary conditions to
  model the effect of the pericardium}},
\newblock \bibinfo{journal}{Journal of Biomechanics} \bibinfo{volume}{101}
  (\bibinfo{year}{2020}) \bibinfo{pages}{109645}.
\bibitem[{Toorop et~al.(1988)Toorop, van~den Horn, Elzinga, and
  Westerhof}]{toorop1988matching}
\bibinfo{author}{G.~P. Toorop}, \bibinfo{author}{G.~J. van~den Horn},
  \bibinfo{author}{G.~Elzinga}, \bibinfo{author}{N.~Westerhof},
\newblock \bibinfo{title}{{Matching between feline left ventricle and arterial
  load: Optimal external power or efficiency}},
\newblock \bibinfo{journal}{American Journal of Physiology - Heart and
  Circulatory Physiology}  (\bibinfo{year}{1988}).
\bibitem[{Nordsletten et~al.(2011)Nordsletten, Mccormick, Kilner, Hunter, Kay,
  and Smith}]{nordsletten2011fluid}
\bibinfo{author}{D.~Nordsletten}, \bibinfo{author}{M.~Mccormick},
  \bibinfo{author}{P.~J. Kilner}, \bibinfo{author}{P.~Hunter},
  \bibinfo{author}{D.~Kay}, \bibinfo{author}{N.~P. Smith},
\newblock \bibinfo{title}{{Fluid-solid coupling for the investigation of
  diastolic and systolic human left ventricular function}},
\newblock \bibinfo{journal}{International Journal for Numerical Methods in
  Biomedical Engineering} \bibinfo{volume}{27} (\bibinfo{year}{2011})
  \bibinfo{pages}{1017--1039}.
\bibitem[{Karabelas et~al.(2020)Karabelas, Haase, Plank, and
  Augustin}]{karabelas2019towards}
\bibinfo{author}{E.~Karabelas}, \bibinfo{author}{G.~Haase},
  \bibinfo{author}{G.~Plank}, \bibinfo{author}{C.~M. Augustin},
\newblock \bibinfo{title}{{Versatile stabilized finite element formulations for
  nearly and fully incompressible solid mechanics}},
\newblock \bibinfo{journal}{Computational Mechanics} \bibinfo{volume}{65}
  (\bibinfo{year}{2020}) \bibinfo{pages}{193--215}.
\bibitem[{Heusinkveld et~al.(2019)Heusinkveld, Huberts, Lumens, Arts, Delhaas,
  and Reesink}]{Heusinkveld2019}
\bibinfo{author}{M.~H.~G. Heusinkveld}, \bibinfo{author}{W.~Huberts},
  \bibinfo{author}{J.~Lumens}, \bibinfo{author}{T.~Arts},
  \bibinfo{author}{T.~Delhaas}, \bibinfo{author}{K.~D. Reesink},
\newblock \bibinfo{title}{{Large vessels as a tree of transmission lines
  incorporated in the CircAdapt whole-heart model: A computational tool to
  examine heart-vessel interaction}},
\newblock \bibinfo{journal}{PLOS Computational Biology} \bibinfo{volume}{15}
  (\bibinfo{year}{2019}) \bibinfo{pages}{e1007173}.
\bibitem[{Shi et~al.(2011)Shi, Lawford, and Hose}]{Shi2011}
\bibinfo{author}{Y.~Shi}, \bibinfo{author}{P.~Lawford},
  \bibinfo{author}{R.~Hose},
\newblock \bibinfo{title}{{Review of Zero-D and 1-D Models of Blood Flow in the
  Cardiovascular System}},
\newblock \bibinfo{journal}{BioMedical Engineering OnLine} \bibinfo{volume}{10}
  (\bibinfo{year}{2011}) \bibinfo{pages}{33}.
\bibitem[{Elzinga and Westerhof(1973)}]{elzinga1973pressure}
\bibinfo{author}{G.~Elzinga}, \bibinfo{author}{N.~Westerhof},
\newblock \bibinfo{title}{Pressure and flow generated by the left ventricle
  against different impedances},
\newblock \bibinfo{journal}{Circulation Research} \bibinfo{volume}{32}
  (\bibinfo{year}{1973}) \bibinfo{pages}{178--186}.
\bibitem[{Liu et~al.(2015)Liu, Liang, Wong, Fujiwara, Ye, Tsubota, and
  Sugawara}]{liu2015multi}
\bibinfo{author}{H.~Liu}, \bibinfo{author}{F.~Liang},
  \bibinfo{author}{J.~Wong}, \bibinfo{author}{T.~Fujiwara},
  \bibinfo{author}{W.~Ye}, \bibinfo{author}{K.-i. Tsubota},
  \bibinfo{author}{M.~Sugawara},
\newblock \bibinfo{title}{Multi-scale modeling of hemodynamics in the
  cardiovascular system},
\newblock \bibinfo{journal}{Acta Mechanica Sinica} \bibinfo{volume}{31}
  (\bibinfo{year}{2015}) \bibinfo{pages}{446--464}.
\bibitem[{Segers et~al.(2008)Segers, Rietzschel, De~Buyzere, Stergiopulos,
  Westerhof, Van~Bortel, Gillebert, and Verdonck}]{segers2008three}
\bibinfo{author}{P.~Segers}, \bibinfo{author}{E.~Rietzschel},
  \bibinfo{author}{M.~De~Buyzere}, \bibinfo{author}{N.~Stergiopulos},
  \bibinfo{author}{N.~Westerhof}, \bibinfo{author}{L.~Van~Bortel},
  \bibinfo{author}{T.~Gillebert}, \bibinfo{author}{P.~Verdonck},
\newblock \bibinfo{title}{Three-and four-element windkessel models: assessment
  of their fitting performance in a large cohort of healthy middle-aged
  individuals},
\newblock \bibinfo{journal}{Proceedings of the Institution of Mechanical
  Engineers, Part H: Journal of Engineering in Medicine} \bibinfo{volume}{222}
  (\bibinfo{year}{2008}) \bibinfo{pages}{417--428}.
\bibitem[{Stergiopulos et~al.(1999)Stergiopulos, Westerhof, and
  Westerhof}]{stergiopulos1999total}
\bibinfo{author}{N.~Stergiopulos}, \bibinfo{author}{B.~E. Westerhof},
  \bibinfo{author}{N.~Westerhof},
\newblock \bibinfo{title}{Total arterial inertance as the fourth element of the
  windkessel model},
\newblock \bibinfo{journal}{American Journal of Physiology-Heart and
  Circulatory Physiology} \bibinfo{volume}{276} (\bibinfo{year}{1999})
  \bibinfo{pages}{H81--H88}.
\bibitem[{Wang et~al.(2003)Wang, O'Brien, Shrive, Parker, and
  Tyberg}]{wang2003time}
\bibinfo{author}{J.-J. Wang}, \bibinfo{author}{A.~B. O'Brien},
  \bibinfo{author}{N.~G. Shrive}, \bibinfo{author}{K.~H. Parker},
  \bibinfo{author}{J.~V. Tyberg},
\newblock \bibinfo{title}{Time-domain representation of ventricular-arterial
  coupling as a windkessel and wave system},
\newblock \bibinfo{journal}{American Journal of Physiology-Heart and
  Circulatory Physiology} \bibinfo{volume}{284} (\bibinfo{year}{2003})
  \bibinfo{pages}{H1358--H1368}.
\bibitem[{Westerhof and Elzinga(1991)}]{westerhof1991normalized}
\bibinfo{author}{N.~Westerhof}, \bibinfo{author}{G.~Elzinga},
\newblock \bibinfo{title}{Normalized input impedance and arterial decay time
  over heart period are independent of animal size},
\newblock \bibinfo{journal}{American Journal of Physiology-Regulatory,
  Integrative and Comparative Physiology} \bibinfo{volume}{261}
  (\bibinfo{year}{1991}) \bibinfo{pages}{R126--R133}.
\bibitem[{Westerhof and Stergiopulos(2000)}]{westerhof2000models}
\bibinfo{author}{N.~Westerhof}, \bibinfo{author}{N.~Stergiopulos},
\newblock \bibinfo{title}{Models of the arterial tree.},
\newblock \bibinfo{journal}{Studies in health technology and informatics}
  \bibinfo{volume}{71} (\bibinfo{year}{2000}) \bibinfo{pages}{65--77}.
\bibitem[{Alastruey et~al.(2012)Alastruey, Parker, and
  Sherwin}]{Alastruey2012a}
\bibinfo{author}{J.~Alastruey}, \bibinfo{author}{K.~H. Parker},
  \bibinfo{author}{S.~J. Sherwin},
\newblock \bibinfo{title}{{Arterial pulse wave haemodynamics}},
\newblock \bibinfo{journal}{BHR Group - 11th International Conferences on
  Pressure Surges}  (\bibinfo{year}{2012}) \bibinfo{pages}{401--442}.
\bibitem[{Blanco et~al.(2014)Blanco, Watanabe, Dari, Passos, and
  Feij{\'{o}}o}]{Blanco2014}
\bibinfo{author}{P.~J. Blanco}, \bibinfo{author}{S.~M. Watanabe},
  \bibinfo{author}{E.~A. Dari}, \bibinfo{author}{M.~A.~R. Passos},
  \bibinfo{author}{R.~A. Feij{\'{o}}o},
\newblock \bibinfo{title}{{Blood flow distribution in an anatomically detailed
  arterial network model: criteria and algorithms}},
\newblock \bibinfo{journal}{Biomechanics and Modeling in Mechanobiology}
  \bibinfo{volume}{13} (\bibinfo{year}{2014}) \bibinfo{pages}{1303--1330}.
\bibitem[{Formaggia et~al.(2003)Formaggia, Lamponi, and
  Quarteroni}]{Formaggia2003b}
\bibinfo{author}{L.~Formaggia}, \bibinfo{author}{D.~Lamponi},
  \bibinfo{author}{A.~Quarteroni},
\newblock \bibinfo{title}{{One-dimensional models for blood flow in arteries}},
\newblock \bibinfo{journal}{Journal of Engineering Mathematics}
  \bibinfo{volume}{47} (\bibinfo{year}{2003}) \bibinfo{pages}{251--276}.
\bibitem[{Mynard and Nithiarasu(2008)}]{Mynard2008}
\bibinfo{author}{J.~P. Mynard}, \bibinfo{author}{P.~Nithiarasu},
\newblock \bibinfo{title}{{A 1D arterial blood flow model incorporating
  ventricular pressure, aortic vaive ana regional coronary flow using the
  locally conservative Galerkin (LCG) method}},
\newblock \bibinfo{journal}{Communications in Numerical Methods in Engineering}
   (\bibinfo{year}{2008}).
\bibitem[{Müller and Toro(2014)}]{Muller2013}
\bibinfo{author}{L.~O. Müller}, \bibinfo{author}{E.~F. Toro},
\newblock \bibinfo{title}{A global multiscale mathematical model for the human
  circulation with emphasis on the venous system},
\newblock \bibinfo{journal}{International Journal for Numerical Methods in
  Biomedical Engineering} \bibinfo{volume}{30} (\bibinfo{year}{2014})
  \bibinfo{pages}{681--725}.
\bibitem[{Marx et~al.(2020)Marx, Gsell, Rund, Caforio, Prassl, Toth-Gayor,
  Kuehne, Augustin, and Plank}]{marx2020personalization}
\bibinfo{author}{L.~Marx}, \bibinfo{author}{M.~A.~F. Gsell},
  \bibinfo{author}{A.~Rund}, \bibinfo{author}{F.~Caforio},
  \bibinfo{author}{A.~J. Prassl}, \bibinfo{author}{G.~Toth-Gayor},
  \bibinfo{author}{T.~Kuehne}, \bibinfo{author}{C.~M. Augustin},
  \bibinfo{author}{G.~Plank},
\newblock \bibinfo{title}{{Personalization of electro-mechanical models of the
  pressure-overloaded left ventricle: fitting of Windkessel-type afterload
  models}},
\newblock \bibinfo{journal}{Philosophical Transactions of the Royal Society A:
  Mathematical, Physical and Engineering Sciences} \bibinfo{volume}{378}
  (\bibinfo{year}{2020}) \bibinfo{pages}{20190342}.
\bibitem[{Niederer et~al.(2011)Niederer, Plank, Chinchapatnam, Ginks, Lamata,
  Rhode, Rinaldi, Razavi, and Smith}]{niederer2011:_length}
\bibinfo{author}{S.~A. Niederer}, \bibinfo{author}{G.~Plank},
  \bibinfo{author}{P.~Chinchapatnam}, \bibinfo{author}{M.~Ginks},
  \bibinfo{author}{P.~Lamata}, \bibinfo{author}{K.~S. Rhode},
  \bibinfo{author}{C.~A. Rinaldi}, \bibinfo{author}{R.~Razavi},
  \bibinfo{author}{N.~P. Smith},
\newblock \bibinfo{title}{Length-dependent tension in the failing heart and the
  efficacy of cardiac resynchronization therapy},
\newblock \bibinfo{journal}{Cardiovascular Research} \bibinfo{volume}{89}
  (\bibinfo{year}{2011}) \bibinfo{pages}{336}.
\bibitem[{Arts et~al.(2005)Arts, Delhaas, Bovendeerd, Verbeek, and
  Prinzen}]{arts2005adaptation}
\bibinfo{author}{T.~Arts}, \bibinfo{author}{T.~Delhaas},
  \bibinfo{author}{P.~Bovendeerd}, \bibinfo{author}{X.~Verbeek},
  \bibinfo{author}{F.~Prinzen},
\newblock \bibinfo{title}{{Adaptation to mechanical load determines shape and
  properties of heart and circulation: the CircAdapt model}},
\newblock \bibinfo{journal}{Am. J. Physiol. Heart Circ. Physiol.}
  \bibinfo{volume}{288} (\bibinfo{year}{2005}) \bibinfo{pages}{1943--1954}.
\bibitem[{Blanco et~al.(2010)Blanco, Feij{\'o}o
  et~al.}]{Blanco2010computational}
\bibinfo{author}{P.~J. Blanco}, \bibinfo{author}{R.~A. Feij{\'o}o}, et~al.,
\newblock \bibinfo{title}{{A 3D-1D-0D Computational model for the entire
  cardiovascular system}},
\newblock \bibinfo{journal}{Computational Mechanics, eds. E. Dvorking, M.
  Goldschmit, M. Storti} \bibinfo{volume}{29} (\bibinfo{year}{2010})
  \bibinfo{pages}{5887--5911}.
\bibitem[{Guidoboni et~al.(2019)Guidoboni, Sala, Enayati, Sacco, Szopos,
  Keller, Popescu, Despins, Huxley, and Skubic}]{guidoboni2019cardiovascular}
\bibinfo{author}{G.~Guidoboni}, \bibinfo{author}{L.~Sala},
  \bibinfo{author}{M.~Enayati}, \bibinfo{author}{R.~Sacco},
  \bibinfo{author}{M.~Szopos}, \bibinfo{author}{J.~M. Keller},
  \bibinfo{author}{M.~Popescu}, \bibinfo{author}{L.~Despins},
  \bibinfo{author}{V.~H. Huxley}, \bibinfo{author}{M.~Skubic},
\newblock \bibinfo{title}{Cardiovascular function and ballistocardiogram: a
  relationship interpreted via mathematical modeling},
\newblock \bibinfo{journal}{IEEE Transactions on Biomedical Engineering}
  \bibinfo{volume}{66} (\bibinfo{year}{2019}) \bibinfo{pages}{2906--2917}.
\bibitem[{Neal and Bassingthwaighte(2007)}]{neal2007subject}
\bibinfo{author}{M.~L. Neal}, \bibinfo{author}{J.~B. Bassingthwaighte},
\newblock \bibinfo{title}{{Subject-specific model estimation of cardiac output
  and blood volume during hemorrhage}},
\newblock \bibinfo{journal}{Cardiovascular Engineering} \bibinfo{volume}{7}
  (\bibinfo{year}{2007}) \bibinfo{pages}{97--120}.
\bibitem[{Paeme et~al.(2011)Paeme, Moorhead, Chase, Lambermont, Kolh, D'orio,
  Pierard, Moonen, Lancellotti, Dauby et~al.}]{paeme2011mathematical}
\bibinfo{author}{S.~Paeme}, \bibinfo{author}{K.~T. Moorhead},
  \bibinfo{author}{J.~G. Chase}, \bibinfo{author}{B.~Lambermont},
  \bibinfo{author}{P.~Kolh}, \bibinfo{author}{V.~D'orio},
  \bibinfo{author}{L.~Pierard}, \bibinfo{author}{M.~Moonen},
  \bibinfo{author}{P.~Lancellotti}, \bibinfo{author}{P.~C. Dauby}, et~al.,
\newblock \bibinfo{title}{Mathematical multi-scale model of the cardiovascular
  system including mitral valve dynamics. application to ischemic mitral
  insufficiency},
\newblock \bibinfo{journal}{Biomedical engineering online} \bibinfo{volume}{10}
  (\bibinfo{year}{2011}) \bibinfo{pages}{86}.
\bibitem[{Eriksson et~al.(2013)Eriksson, Prassl, Plank, and
  Holzapfel}]{eriksson2013influence}
\bibinfo{author}{T.~S.~E. Eriksson}, \bibinfo{author}{A.~J. Prassl},
  \bibinfo{author}{G.~Plank}, \bibinfo{author}{G.~A. Holzapfel},
\newblock \bibinfo{title}{{Influence of myocardial fiber/sheet orientations on
  left ventricular mechanical contraction}},
\newblock \bibinfo{journal}{Math Mech Solids} \bibinfo{volume}{18}
  (\bibinfo{year}{2013}) \bibinfo{pages}{592--606}.
\bibitem[{Kerckhoffs et~al.(2007)Kerckhoffs, Neal, Gu, Bassingthwaighte, Omens,
  and McCulloch}]{kerckhoffs2007coupling}
\bibinfo{author}{R.~C. Kerckhoffs}, \bibinfo{author}{M.~L. Neal},
  \bibinfo{author}{Q.~Gu}, \bibinfo{author}{J.~B. Bassingthwaighte},
  \bibinfo{author}{J.~H. Omens}, \bibinfo{author}{A.~D. McCulloch},
\newblock \bibinfo{title}{{Coupling of a 3D finite element model of cardiac
  ventricular mechanics} to lumped systems models of the systemic and pulmonic
  circulation},
\newblock \bibinfo{journal}{Annals of biomedical engineering}
  \bibinfo{volume}{35} (\bibinfo{year}{2007}) \bibinfo{pages}{1--18}.
\bibitem[{Usyk et~al.(2002)Usyk, LeGrice, and McCulloch}]{usyk2002compuational}
\bibinfo{author}{T.~P. Usyk}, \bibinfo{author}{I.~J. LeGrice},
  \bibinfo{author}{A.~D. McCulloch},
\newblock \bibinfo{title}{{Computational model of three-dimensional cardiac
  electromechanics}},
\newblock \bibinfo{journal}{Computing and Visualization in Science}
  \bibinfo{volume}{4} (\bibinfo{year}{2002}) \bibinfo{pages}{249--257}.
\bibitem[{Fritz et~al.(2014)Fritz, Wieners, Seemann, Steen, and
  D{\"{o}}ssel}]{fritz2014simulation}
\bibinfo{author}{T.~Fritz}, \bibinfo{author}{C.~Wieners},
  \bibinfo{author}{G.~Seemann}, \bibinfo{author}{H.~Steen},
  \bibinfo{author}{O.~D{\"{o}}ssel},
\newblock \bibinfo{title}{{Simulation of the contraction of the ventricles in a
  human heart model including atria and pericardium}},
\newblock \bibinfo{journal}{Biomechanics and Modeling in Mechanobiology}
  \bibinfo{volume}{13} (\bibinfo{year}{2014}) \bibinfo{pages}{627--641}.
\bibitem[{Gurev et~al.(2011)Gurev, Lee, Constantino, Arevalo, and
  Trayanova}]{Gurev2011a}
\bibinfo{author}{V.~Gurev}, \bibinfo{author}{T.~Lee},
  \bibinfo{author}{J.~Constantino}, \bibinfo{author}{H.~J. Arevalo},
  \bibinfo{author}{N.~A. Trayanova},
\newblock \bibinfo{title}{{Models of cardiac electromechanics based on
  individual hearts imaging data}},
\newblock \bibinfo{journal}{Biomechanics and Modeling in Mechanobiology}
  \bibinfo{volume}{10} (\bibinfo{year}{2011}) \bibinfo{pages}{295--306}.
\bibitem[{Gurev et~al.(2015)Gurev, Pathmanathan, Fattebert, Wen, Magerlein,
  Gray, Richards, and Rice}]{Gurev2015high}
\bibinfo{author}{V.~Gurev}, \bibinfo{author}{P.~Pathmanathan},
  \bibinfo{author}{J.-L. Fattebert}, \bibinfo{author}{H.-F. Wen},
  \bibinfo{author}{J.~Magerlein}, \bibinfo{author}{R.~a. Gray},
  \bibinfo{author}{D.~F. Richards}, \bibinfo{author}{J.~J. Rice},
\newblock \bibinfo{title}{{A high-resolution computational model of the
  deforming human heart}},
\newblock \bibinfo{journal}{Biomechanics and Modeling in Mechanobiology}
  \bibinfo{volume}{14} (\bibinfo{year}{2015}) \bibinfo{pages}{829--849}.
\bibitem[{Hirschvogel et~al.(2017)Hirschvogel, Bassilious, Jagschies, Wildhirt,
  and Gee}]{Hirschvogel2017monolithic}
\bibinfo{author}{M.~Hirschvogel}, \bibinfo{author}{M.~Bassilious},
  \bibinfo{author}{L.~Jagschies}, \bibinfo{author}{S.~M. Wildhirt},
  \bibinfo{author}{M.~W. Gee},
\newblock \bibinfo{title}{{A monolithic 3D-0D coupled closed-loop model of the
  heart and the vascular system: Experiment-based parameter estimation for
  patient-specific cardiac mechanics}},
\newblock \bibinfo{journal}{International Journal for Numerical Methods in
  Biomedical Engineering} \bibinfo{volume}{33} (\bibinfo{year}{2017}).
\bibitem[{Sainte-Marie et~al.(2006)Sainte-Marie, Chapelle, Cimrman, and
  Sorine}]{Sainte-Marie2006modeling}
\bibinfo{author}{J.~Sainte-Marie}, \bibinfo{author}{D.~Chapelle},
  \bibinfo{author}{R.~Cimrman}, \bibinfo{author}{M.~Sorine},
\newblock \bibinfo{title}{{Modeling and estimation of the cardiac
  electromechanical activity}},
\newblock \bibinfo{journal}{Computers and Structures} \bibinfo{volume}{84}
  (\bibinfo{year}{2006}) \bibinfo{pages}{1743--1759}.
\bibitem[{Sack et~al.(2018)Sack, Aliotta, Ennis, Choy, Kassab, Guccione, and
  Franz}]{sack2018construction}
\bibinfo{author}{K.~L. Sack}, \bibinfo{author}{E.~Aliotta},
  \bibinfo{author}{D.~B. Ennis}, \bibinfo{author}{J.~S. Choy},
  \bibinfo{author}{G.~S. Kassab}, \bibinfo{author}{J.~M. Guccione},
  \bibinfo{author}{T.~Franz},
\newblock \bibinfo{title}{{Construction and validation of subject-specific
  biventricular finite-element models of healthy and failing swine hearts from
  high-resolution DT-MRI}},
\newblock \bibinfo{journal}{Frontiers in Physiology} \bibinfo{volume}{9}
  (\bibinfo{year}{2018}) \bibinfo{pages}{1--19}.
\bibitem[{Augustin et~al.(2016{\natexlab{a}})Augustin, Neic, Liebmann, Prassl,
  Niederer, Haase, and Plank}]{Augustin2016anatomically}
\bibinfo{author}{C.~M. Augustin}, \bibinfo{author}{A.~Neic},
  \bibinfo{author}{M.~Liebmann}, \bibinfo{author}{A.~J. Prassl},
  \bibinfo{author}{S.~A. Niederer}, \bibinfo{author}{G.~Haase},
  \bibinfo{author}{G.~Plank},
\newblock \bibinfo{title}{{Anatomically accurate high resolution modeling of
  human whole heart electromechanics: A strongly scalable algebraic multigrid
  solver method for nonlinear deformation}},
\newblock \bibinfo{journal}{Journal of Computational Physics}
  \bibinfo{volume}{305} (\bibinfo{year}{2016}{\natexlab{a}})
  \bibinfo{pages}{622--646}.
\bibitem[{Augustin et~al.(2016{\natexlab{b}})Augustin, Crozier, Neic, Prassl,
  Karabelas, {Ferreira da Silva}, Fernandes, Campos, Kuehne, Plank, da~Silva,
  Fernandes, Campos, Kuehne, and Plank}]{Augustin2016patient}
\bibinfo{author}{C.~M. Augustin}, \bibinfo{author}{A.~Crozier},
  \bibinfo{author}{A.~Neic}, \bibinfo{author}{A.~J. Prassl},
  \bibinfo{author}{E.~Karabelas}, \bibinfo{author}{T.~{Ferreira da Silva}},
  \bibinfo{author}{J.~F. Fernandes}, \bibinfo{author}{F.~Campos},
  \bibinfo{author}{T.~Kuehne}, \bibinfo{author}{G.~Plank},
  \bibinfo{author}{T.~da~Silva}, \bibinfo{author}{J.~F. Fernandes},
  \bibinfo{author}{F.~Campos}, \bibinfo{author}{T.~Kuehne},
  \bibinfo{author}{G.~Plank},
\newblock \bibinfo{title}{{Patient-specific modeling of left ventricular
  electromechanics as a driver for haemodynamic analysis}},
\newblock \bibinfo{journal}{Europace} \bibinfo{volume}{18}
  (\bibinfo{year}{2016}{\natexlab{b}}) \bibinfo{pages}{iv121--iv129}.
\bibitem[{Crozier et~al.(2016)Crozier, Augustin, Neic, Prassl, Holler, Fastl,
  Hennemuth, Bredies, Kuehne, Bishop, Niederer, and Plank}]{crozier2016image}
\bibinfo{author}{A.~Crozier}, \bibinfo{author}{C.~M. Augustin},
  \bibinfo{author}{A.~Neic}, \bibinfo{author}{A.~J. Prassl},
  \bibinfo{author}{M.~Holler}, \bibinfo{author}{T.~E. Fastl},
  \bibinfo{author}{A.~Hennemuth}, \bibinfo{author}{K.~Bredies},
  \bibinfo{author}{T.~Kuehne}, \bibinfo{author}{M.~J. Bishop},
  \bibinfo{author}{S.~A. Niederer}, \bibinfo{author}{G.~Plank},
\newblock \bibinfo{title}{{Image-Based Personalization of Cardiac Anatomy for
  Coupled Electromechanical Modeling}},
\newblock \bibinfo{journal}{Annals of Biomedical Engineering}
  \bibinfo{volume}{44} (\bibinfo{year}{2016}) \bibinfo{pages}{58--70}.
\bibitem[{Walmsley et~al.(2015)Walmsley, Arts, Derval, Bordachar, Cochet,
  Ploux, Prinzen, Delhaas, and Lumens}]{walmsley2015fast}
\bibinfo{author}{J.~Walmsley}, \bibinfo{author}{T.~Arts},
  \bibinfo{author}{N.~Derval}, \bibinfo{author}{P.~Bordachar},
  \bibinfo{author}{H.~Cochet}, \bibinfo{author}{S.~Ploux},
  \bibinfo{author}{F.~W. Prinzen}, \bibinfo{author}{T.~Delhaas},
  \bibinfo{author}{J.~Lumens},
\newblock \bibinfo{title}{{Fast Simulation of Mechanical Heterogeneity in the
  Electrically Asynchronous Heart Using the MultiPatch Module}},
\newblock \bibinfo{journal}{PLOS Computational Biology} \bibinfo{volume}{11}
  (\bibinfo{year}{2015}) \bibinfo{pages}{e1004284}.
\bibitem[{Mills et~al.(2009)Mills, Cornelussen, Mulligan, Strik, Rademakers,
  Skadsberg, van Hunnik, Kuiper, Lampert, Delhaas, and Prinzen}]{prinzen2009}
\bibinfo{author}{R.~W. Mills}, \bibinfo{author}{R.~N. Cornelussen},
  \bibinfo{author}{L.~J. Mulligan}, \bibinfo{author}{M.~Strik},
  \bibinfo{author}{L.~M. Rademakers}, \bibinfo{author}{N.~D. Skadsberg},
  \bibinfo{author}{A.~van Hunnik}, \bibinfo{author}{M.~Kuiper},
  \bibinfo{author}{A.~Lampert}, \bibinfo{author}{T.~Delhaas},
  \bibinfo{author}{F.~W. Prinzen},
\newblock \bibinfo{title}{{Left Ventricular Septal and Left Ventricular Apical
  Pacing Chronically Maintain Cardiac Contractile Coordination, Pump Function
  and Efficiency}},
\newblock \bibinfo{journal}{Circulation: Arrhythmia and Electrophysiology}
  \bibinfo{volume}{2} (\bibinfo{year}{2009}) \bibinfo{pages}{571--579}.
\bibitem[{Verbeek et~al.(2003)Verbeek, Vernooy, Peschar, Cornelussen, and
  Prinzen}]{verbeek2003}
\bibinfo{author}{X.~A. Verbeek}, \bibinfo{author}{K.~Vernooy},
  \bibinfo{author}{M.~Peschar}, \bibinfo{author}{R.~N. Cornelussen},
  \bibinfo{author}{F.~W. Prinzen},
\newblock \bibinfo{title}{{Intra-ventricular resynchronization for optimal left
  ventricular function during pacing in experimental left bundle branch
  block}},
\newblock \bibinfo{journal}{Journal of the American College of Cardiology}
  \bibinfo{volume}{42} (\bibinfo{year}{2003}) \bibinfo{pages}{558--567}.
\bibitem[{CIBC(2016)}]{sci:seg3D}
\bibinfo{author}{CIBC}, \bibinfo{year}{2016}. \URLprefix
  \url{http://www.seg3d.org}, \bibinfo{note}{seg3D: Volumetric Image
  Segmentation and Visualization. Scientific Computing and Imaging}.
\bibitem[{Neic et~al.(2020)Neic, Gsell, Karabelas, Prassl, and
  Plank}]{neic2020:meshtool}
\bibinfo{author}{A.~Neic}, \bibinfo{author}{M.~A.~F. Gsell},
  \bibinfo{author}{E.~Karabelas}, \bibinfo{author}{A.~J. Prassl},
  \bibinfo{author}{G.~Plank},
\newblock \bibinfo{title}{{Automating image-based mesh generation and
  manipulation tasks in cardiac modeling workflows using Meshtool}},
\newblock \bibinfo{journal}{SoftwareX} \bibinfo{volume}{11}
  (\bibinfo{year}{2020}) \bibinfo{pages}{100454}.
\bibitem[{Bayer et~al.(2012)Bayer, Blake, Plank, and
  Trayanova}]{bayer2012a:fibers}
\bibinfo{author}{J.~D. Bayer}, \bibinfo{author}{R.~C. Blake},
  \bibinfo{author}{G.~Plank}, \bibinfo{author}{N.~A. Trayanova},
\newblock \bibinfo{title}{{A novel rule-based algorithm for assigning
  myocardial fiber orientation to computational heart models}},
\newblock \bibinfo{journal}{Annals of Biomedical Engineering}
  \bibinfo{volume}{40} (\bibinfo{year}{2012}) \bibinfo{pages}{2243--2254}.
\bibitem[{Streeter et~al.(1969)Streeter, Spotnitz, Patel, Ross, and
  Sonnenblick}]{streeter1969fiber}
\bibinfo{author}{D.~D. Streeter}, \bibinfo{author}{H.~M. Spotnitz},
  \bibinfo{author}{D.~P. Patel}, \bibinfo{author}{J.~Ross},
  \bibinfo{author}{E.~H. Sonnenblick},
\newblock \bibinfo{title}{{Fiber orientation in the canine left ventricle
  during diastole and systole.}},
\newblock \bibinfo{journal}{Circulation research}  (\bibinfo{year}{1969}).
\bibitem[{Bayer et~al.(2018)Bayer, Prassl, Pashaei, Gomez, Frontera, Neic,
  Plank, and Vigmond}]{bayer2018universal}
\bibinfo{author}{J.~Bayer}, \bibinfo{author}{A.~J. Prassl},
  \bibinfo{author}{A.~Pashaei}, \bibinfo{author}{J.~F. Gomez},
  \bibinfo{author}{A.~Frontera}, \bibinfo{author}{A.~Neic},
  \bibinfo{author}{G.~Plank}, \bibinfo{author}{E.~J. Vigmond},
\newblock \bibinfo{title}{Universal ventricular coordinates: A generic
  framework for describing position within the heart and transferring data},
\newblock \bibinfo{journal}{Medical Image Analysis} \bibinfo{volume}{45}
  (\bibinfo{year}{2018}) \bibinfo{pages}{83--93}.
\bibitem[{Flory(1961)}]{flory1961thermodynamic}
\bibinfo{author}{P.~J. Flory},
\newblock \bibinfo{title}{Thermodynamic relations for high elastic materials},
\newblock \bibinfo{journal}{Trans Faraday Soc} \bibinfo{volume}{57}
  (\bibinfo{year}{1961}) \bibinfo{pages}{829--838}.
\bibitem[{Land and Niederer(2018)}]{land2018influence}
\bibinfo{author}{S.~Land}, \bibinfo{author}{S.~A. Niederer},
\newblock \bibinfo{title}{Influence of atrial contraction dynamics on cardiac
  function},
\newblock \bibinfo{journal}{International journal for numerical methods in
  biomedical engineering} \bibinfo{volume}{34} (\bibinfo{year}{2018})
  \bibinfo{pages}{e2931}.
\bibitem[{Usyk et~al.(2000)Usyk, Mazhari, and McCulloch}]{usyk2000effect}
\bibinfo{author}{T.~P. Usyk}, \bibinfo{author}{R.~Mazhari},
  \bibinfo{author}{A.~D. McCulloch},
\newblock \bibinfo{title}{{Effect of laminar orthotropic myofiber architecture
  on regional stress and strain in the canine left ventricle}},
\newblock \bibinfo{journal}{J. Elast.} \bibinfo{volume}{61}
  (\bibinfo{year}{2000}) \bibinfo{pages}{143--164}.
\bibitem[{Genet et~al.(2014)Genet, Lee, Nguyen, Haraldsson, Acevedo-Bolton,
  Zhang, Ge, Ordovas, Kozerke, and Guccione}]{Genet2014}
\bibinfo{author}{M.~Genet}, \bibinfo{author}{L.~C. Lee},
  \bibinfo{author}{R.~Nguyen}, \bibinfo{author}{H.~Haraldsson},
  \bibinfo{author}{G.~Acevedo-Bolton}, \bibinfo{author}{Z.~Zhang},
  \bibinfo{author}{L.~Ge}, \bibinfo{author}{K.~Ordovas},
  \bibinfo{author}{S.~Kozerke}, \bibinfo{author}{J.~M. Guccione},
\newblock \bibinfo{title}{{Distribution of normal human left ventricular
  myofiber stress at end diastole and end systole: a target for in silico
  design of heart failure treatments}},
\newblock \bibinfo{journal}{Journal of Applied Physiology}
  \bibinfo{volume}{117} (\bibinfo{year}{2014}) \bibinfo{pages}{142--152}.
\bibitem[{Walker et~al.(2005)Walker, Ratcliffe, Zhang, Wallace, Fata, Hsu,
  Saloner, and Guccione}]{walker2005mri}
\bibinfo{author}{J.~C. Walker}, \bibinfo{author}{M.~B. Ratcliffe},
  \bibinfo{author}{P.~Zhang}, \bibinfo{author}{A.~W. Wallace},
  \bibinfo{author}{B.~Fata}, \bibinfo{author}{E.~W. Hsu},
  \bibinfo{author}{D.~Saloner}, \bibinfo{author}{J.~M. Guccione},
\newblock \bibinfo{title}{{MRI-based finite-element analysis of left
  ventricular aneurysm}},
\newblock \bibinfo{journal}{American Journal of Physiology-Heart and
  Circulatory Physiology} \bibinfo{volume}{289} (\bibinfo{year}{2005})
  \bibinfo{pages}{H692--H700}.
\bibitem[{Neic et~al.(2017)Neic, Campos, Prassl, Niederer, Bishop, Vigmond, and
  Plank}]{neic17:_efficient}
\bibinfo{author}{A.~Neic}, \bibinfo{author}{F.~O. Campos},
  \bibinfo{author}{A.~J. Prassl}, \bibinfo{author}{S.~A. Niederer},
  \bibinfo{author}{M.~J. Bishop}, \bibinfo{author}{E.~J. Vigmond},
  \bibinfo{author}{G.~Plank},
\newblock \bibinfo{title}{Efficient computation of electrograms and {ECGs} in
  human whole heart simulations using a reaction-eikonal model},
\newblock \bibinfo{journal}{Journal of Computational Physics}
  \bibinfo{volume}{346} (\bibinfo{year}{2017}) \bibinfo{pages}{191--211}.
\bibitem[{Costa et~al.(2013)Costa, Hoetzl, Rocha, Prassl, and
  Plank}]{costa13:_automatic_parameterization}
\bibinfo{author}{C.~M. Costa}, \bibinfo{author}{E.~Hoetzl},
  \bibinfo{author}{B.~M. Rocha}, \bibinfo{author}{A.~J. Prassl},
  \bibinfo{author}{G.~Plank},
\newblock \bibinfo{title}{{Automatic Parameterization Strategy for Cardiac
  Electrophysiology Simulations.}},
\newblock \bibinfo{journal}{Computing in cardiology} \bibinfo{volume}{40}
  (\bibinfo{year}{2013}) \bibinfo{pages}{373--376}.
\bibitem[{Niederer et~al.(2011)Niederer, Kerfoot, Benson, Bernabeu, Bernus,
  Bradley, Cherry, Clayton, Fenton, Garny, Heidenreich, Land, Maleckar,
  Pathmanathan, Plank, Rodríguez, Roy, Sachse, Seemann, Skavhaug, and
  Smith}]{niederer11:_nversion_ep}
\bibinfo{author}{S.~A. Niederer}, \bibinfo{author}{E.~Kerfoot},
  \bibinfo{author}{A.~P. Benson}, \bibinfo{author}{M.~O. Bernabeu},
  \bibinfo{author}{O.~Bernus}, \bibinfo{author}{C.~Bradley},
  \bibinfo{author}{E.~M. Cherry}, \bibinfo{author}{R.~Clayton},
  \bibinfo{author}{F.~H. Fenton}, \bibinfo{author}{A.~Garny},
  \bibinfo{author}{E.~Heidenreich}, \bibinfo{author}{S.~Land},
  \bibinfo{author}{M.~Maleckar}, \bibinfo{author}{P.~Pathmanathan},
  \bibinfo{author}{G.~Plank}, \bibinfo{author}{J.~F. Rodríguez},
  \bibinfo{author}{I.~Roy}, \bibinfo{author}{F.~B. Sachse},
  \bibinfo{author}{G.~Seemann}, \bibinfo{author}{O.~Skavhaug},
  \bibinfo{author}{N.~P. Smith},
\newblock \bibinfo{title}{Verification of cardiac tissue electrophysiology
  simulators using an n-version benchmark.},
\newblock \bibinfo{journal}{Philosophical transactions. Series A, Mathematical,
  physical, and engineering sciences} \bibinfo{volume}{369}
  (\bibinfo{year}{2011}) \bibinfo{pages}{4331--4351}.
\bibitem[{ten Tusscher et~al.(2004)ten Tusscher, Noble, Noble, and
  Panfilov}]{tentusscher04:_TNNP}
\bibinfo{author}{K.~H. W.~J. ten Tusscher}, \bibinfo{author}{D.~Noble},
  \bibinfo{author}{P.~J. Noble}, \bibinfo{author}{A.~V. Panfilov},
\newblock \bibinfo{title}{A model for human ventricular tissue.},
\newblock \bibinfo{journal}{American Journal of Physiology. Heart and
  Circulatory Physiology} \bibinfo{volume}{286} (\bibinfo{year}{2004})
  \bibinfo{pages}{H1573--H1589}.
\bibitem[{Lumens et~al.(2009)Lumens, Delhaas, Kirn, and
  Arts}]{lumens2009triseg}
\bibinfo{author}{J.~Lumens}, \bibinfo{author}{T.~Delhaas},
  \bibinfo{author}{B.~Kirn}, \bibinfo{author}{T.~Arts},
\newblock \bibinfo{title}{{Three-Wall Segment (TriSeg) Model Describing
  Mechanics and Hemodynamics of Ventricular Interaction}},
\newblock \bibinfo{journal}{Annals of Biomedical Engineering}
  \bibinfo{volume}{37} (\bibinfo{year}{2009}) \bibinfo{pages}{2234--2255}.
\bibitem[{K{\"{u}}ttler et~al.(2006)K{\"{u}}ttler, F{\"{o}}rster, and
  Wall}]{Kuttler2006}
\bibinfo{author}{U.~K{\"{u}}ttler}, \bibinfo{author}{C.~F{\"{o}}rster},
  \bibinfo{author}{W.~A. Wall},
\newblock \bibinfo{title}{{A Solution for the Incompressibility Dilemma in
  Partitioned Fluid–Structure Interaction with Pure Dirichlet Fluid
  Domains}},
\newblock \bibinfo{journal}{Computational Mechanics} \bibinfo{volume}{38}
  (\bibinfo{year}{2006}) \bibinfo{pages}{417--429}.
\bibitem[{Gurev et~al.(2011)Gurev, Lee, Constantino, Arevalo, and
  Trayanova}]{gurev2011models}
\bibinfo{author}{V.~Gurev}, \bibinfo{author}{T.~Lee},
  \bibinfo{author}{J.~Constantino}, \bibinfo{author}{H.~Arevalo},
  \bibinfo{author}{N.~A. Trayanova},
\newblock \bibinfo{title}{Models of cardiac electromechanics based on
  individual hearts imaging data},
\newblock \bibinfo{journal}{Biomechanics and modeling in mechanobiology}
  \bibinfo{volume}{10} (\bibinfo{year}{2011}) \bibinfo{pages}{295--306}.
\bibitem[{Rumpel and Schweizerhof(2003)}]{rumpel2003volume}
\bibinfo{author}{T.~Rumpel}, \bibinfo{author}{K.~Schweizerhof},
\newblock \bibinfo{title}{Volume-dependent pressure loading and its influence
  on the stability of structures},
\newblock \bibinfo{journal}{International Journal for Numerical Methods in
  Engineering} \bibinfo{volume}{56} (\bibinfo{year}{2003})
  \bibinfo{pages}{211--238}.
\bibitem[{Sugiura et~al.(2012)Sugiura, Washio, Hatano, Okada, Watanabe, and
  Hisada}]{sugiura2012multiscale}
\bibinfo{author}{S.~Sugiura}, \bibinfo{author}{T.~Washio},
  \bibinfo{author}{A.~Hatano}, \bibinfo{author}{J.~Okada},
  \bibinfo{author}{H.~Watanabe}, \bibinfo{author}{T.~Hisada},
\newblock \bibinfo{title}{{Multi-scale simulations of cardiac electrophysiology
  and mechanics using the University of Tokyo heart simulator}},
\newblock \bibinfo{journal}{Progress in Biophysics and Molecular Biology}
  \bibinfo{volume}{110} (\bibinfo{year}{2012}) \bibinfo{pages}{380--9}.
\bibitem[{Sellier(2011)}]{Sellier2011}
\bibinfo{author}{M.~Sellier},
\newblock \bibinfo{title}{{An iterative method for the inverse elasto-static
  problem}},
\newblock \bibinfo{journal}{Journal of Fluids and Structures}
  \bibinfo{volume}{27} (\bibinfo{year}{2011}) \bibinfo{pages}{1461--1470}.
\bibitem[{Klotz et~al.(2007)Klotz, Dickstein, and
  Burkhoff}]{klotz2007computational}
\bibinfo{author}{S.~Klotz}, \bibinfo{author}{M.~L. Dickstein},
  \bibinfo{author}{D.~Burkhoff},
\newblock \bibinfo{title}{{A computational method of prediction of the
  end-diastolic pressure–volume relationship by single beat}},
\newblock \bibinfo{journal}{Nature Protocols} \bibinfo{volume}{2}
  (\bibinfo{year}{2007}) \bibinfo{pages}{2152--2158}.
\bibitem[{Willemen et~al.(2019)Willemen, Schreurs, Huntjens, Strik, Plank,
  Vigmond, Walmsley, Vernooy, Delhaas, Prinzen, and Lumens}]{willemen2019:_crt}
\bibinfo{author}{E.~Willemen}, \bibinfo{author}{R.~Schreurs},
  \bibinfo{author}{P.~R. Huntjens}, \bibinfo{author}{M.~Strik},
  \bibinfo{author}{G.~Plank}, \bibinfo{author}{E.~Vigmond},
  \bibinfo{author}{J.~Walmsley}, \bibinfo{author}{K.~Vernooy},
  \bibinfo{author}{T.~Delhaas}, \bibinfo{author}{F.~W. Prinzen},
  \bibinfo{author}{J.~Lumens},
\newblock \bibinfo{title}{{The Left and Right Ventricles Respond Differently to
  Variation of Pacing Delays in Cardiac Resynchronization Therapy: A Combined
  Experimental- Computational Approach}},
\newblock \bibinfo{journal}{Front. Physiol.} \bibinfo{volume}{10}
  (\bibinfo{year}{2019}) \bibinfo{pages}{1--13}.
\bibitem[{Burkhoff et~al.(2005)Burkhoff, Mirsky, and
  Suga}]{burkhoff2005assessment}
\bibinfo{author}{D.~Burkhoff}, \bibinfo{author}{I.~Mirsky},
  \bibinfo{author}{H.~Suga},
\newblock \bibinfo{title}{{Assessment of systolic and diastolic ventricular
  properties via pressure-volume analysis: a guide for clinical, translational,
  and basic researchers.}},
\newblock \bibinfo{journal}{Am. J. Physiol. Heart Circ. Physiol.}
  \bibinfo{volume}{289} (\bibinfo{year}{2005}) \bibinfo{pages}{H501--12}.
\bibitem[{Westerhof et~al.(2019)Westerhof, Stergiopulos, and
  Noble}]{westerhof2019:_snapshots}
\bibinfo{author}{N.~Westerhof}, \bibinfo{author}{N.~Stergiopulos},
  \bibinfo{author}{M.~I. Noble}, \bibinfo{title}{{Snapshots of Hemodynamics}},
  \bibinfo{year}{2019}.
\bibitem[{Balay et~al.(2018)Balay, Abhyankar, Adams, Brown, Brune, Buschelman,
  Dalcin, Dener, Eijkhout, Gropp, Kaushik, Knepley, Dave~A.~M, McInnes, Mills,
  Munson, Rupp, Sanan, Smith, Zampini, Zhang, and Zhang}]{petsc-user-ref}
\bibinfo{author}{S.~Balay}, \bibinfo{author}{S.~Abhyankar},
  \bibinfo{author}{M.~F. Adams}, \bibinfo{author}{J.~Brown},
  \bibinfo{author}{P.~Brune}, \bibinfo{author}{K.~Buschelman},
  \bibinfo{author}{L.~Dalcin}, \bibinfo{author}{A.~Dener},
  \bibinfo{author}{V.~Eijkhout}, \bibinfo{author}{W.~D. Gropp},
  \bibinfo{author}{D.~Kaushik}, \bibinfo{author}{M.~G. Knepley},
  \bibinfo{author}{a.~Dave~A.~M}, \bibinfo{author}{L.~C. McInnes},
  \bibinfo{author}{R.~T. Mills}, \bibinfo{author}{T.~Munson},
  \bibinfo{author}{K.~Rupp}, \bibinfo{author}{P.~Sanan}, \bibinfo{author}{B.~F.
  Smith}, \bibinfo{author}{S.~Zampini}, \bibinfo{author}{H.~Zhang},
  \bibinfo{author}{H.~Zhang}, \bibinfo{title}{{PETS}c Users Manual},
  \bibinfo{type}{Technical Report} \bibinfo{number}{ANL-95/11 - Revision 3.10},
  Argonne National Laboratory, \bibinfo{year}{2018}. \URLprefix
  \url{http://www.mcs.anl.gov/petsc}.
\bibitem[{Henson and Yang(2002)}]{henson2002boomeramg}
\bibinfo{author}{V.~E. Henson}, \bibinfo{author}{U.~M. Yang},
\newblock \bibinfo{title}{{BoomerAMG: A parallel algebraic multigrid solver and
  preconditioner}},
\newblock in: \bibinfo{booktitle}{Applied Numerical Mathematics},
  \bibinfo{year}{2002}, pp. \bibinfo{pages}{155--177}.
\bibitem[{Regazzoni and Quarteroni(2021)}]{Regazzoni2021}
\bibinfo{author}{F.~Regazzoni}, \bibinfo{author}{A.~Quarteroni},
\newblock \bibinfo{title}{{An oscillation-free fully staggered algorithm for
  velocity-dependent active models of cardiac mechanics}},
\newblock \bibinfo{journal}{Computer Methods in Applied Mechanics and
  Engineering} \bibinfo{volume}{373} (\bibinfo{year}{2021})
  \bibinfo{pages}{113506}.
\bibitem[{Pfaller et~al.(2019)Pfaller, H{\"{o}}rmann, Weigl, Nagler, Chabiniok,
  Bertoglio, and Wall}]{Pfaller2019}
\bibinfo{author}{M.~R. Pfaller}, \bibinfo{author}{J.~M. H{\"{o}}rmann},
  \bibinfo{author}{M.~Weigl}, \bibinfo{author}{A.~Nagler},
  \bibinfo{author}{R.~Chabiniok}, \bibinfo{author}{C.~Bertoglio},
  \bibinfo{author}{W.~A. Wall},
\newblock \bibinfo{title}{{The importance of the pericardium for cardiac
  biomechanics: from physiology to computational modeling}},
\newblock \bibinfo{journal}{Biomechanics and Modeling in Mechanobiology}
  \bibinfo{volume}{18} (\bibinfo{year}{2019}) \bibinfo{pages}{503--529}.
\bibitem[{Jacob and Ebecken(1994)}]{Jacob1994}
\bibinfo{author}{B.~Jacob}, \bibinfo{author}{N.~Ebecken},
\newblock \bibinfo{title}{{Towards an adaptive ‘semi-implicit' solution
  scheme for nonlinear structural dynamic problems}},
\newblock \bibinfo{journal}{Computers \& Structures} \bibinfo{volume}{52}
  (\bibinfo{year}{1994}) \bibinfo{pages}{495--504}.
\bibitem[{Vigmond et~al.(2008)Vigmond, Weber~dos Santos, Prassl, Deo, and
  Plank}]{vigmond2008solvers}
\bibinfo{author}{E.~Vigmond}, \bibinfo{author}{R.~Weber~dos Santos},
  \bibinfo{author}{A.~Prassl}, \bibinfo{author}{M.~Deo},
  \bibinfo{author}{G.~Plank},
\newblock \bibinfo{title}{Solvers for the cardiac bidomain equations},
\newblock \bibinfo{journal}{Prog Biophys Mol Biol} \bibinfo{volume}{96}
  (\bibinfo{year}{2008}) \bibinfo{pages}{3--18}.
\bibitem[{Deuflhard(2011)}]{deuflhard2011newton}
\bibinfo{author}{P.~Deuflhard}, \bibinfo{title}{{Newton Methods for Nonlinear
  Problems: Affine Invariance and Adaptive Algorithms}},
  volume~\bibinfo{volume}{35}, \bibinfo{publisher}{Springer Science \& Business
  Media}, \bibinfo{year}{2011}.
\bibitem[{Carruth et~al.(2016)Carruth, McCulloch, and Omens}]{carruth2016}
\bibinfo{author}{E.~D. Carruth}, \bibinfo{author}{A.~D. McCulloch},
  \bibinfo{author}{J.~H. Omens},
\newblock \bibinfo{title}{{Transmural gradients of myocardial structure and
  mechanics: Implications for fiber stress and strain in pressure overload}},
\newblock \bibinfo{journal}{Progress in Biophysics and Molecular Biology}
  \bibinfo{volume}{122} (\bibinfo{year}{2016}) \bibinfo{pages}{215--226}.
\bibitem[{Fung(2013)}]{fung2013biomechanics_circulation}
\bibinfo{author}{Y.-c. Fung}, \bibinfo{title}{Biomechanics: circulation},
  \bibinfo{publisher}{Springer Science \& Business Media},
  \bibinfo{year}{2013}.
\bibitem[{Omens and Fung(1990)}]{omens1990residual}
\bibinfo{author}{J.~H. Omens}, \bibinfo{author}{Y.-C. Fung},
\newblock \bibinfo{title}{Residual strain in rat left ventricle},
\newblock \bibinfo{journal}{Circulation Research} \bibinfo{volume}{66}
  (\bibinfo{year}{1990}) \bibinfo{pages}{37--45}.
\bibitem[{Rodriguez et~al.(1993)Rodriguez, Omens, Waldman, and
  McCulloch}]{Rodriguez1993}
\bibinfo{author}{E.~K. Rodriguez}, \bibinfo{author}{J.~H. Omens},
  \bibinfo{author}{L.~K. Waldman}, \bibinfo{author}{A.~D. McCulloch},
\newblock \bibinfo{title}{{Effect of residual stress on transmural sarcomere
  length distributions in rat left ventricle}},
\newblock \bibinfo{journal}{American Journal of Physiology - Heart and
  Circulatory Physiology}  (\bibinfo{year}{1993}).
\bibitem[{Karabelas et~al.(2018)Karabelas, Gsell, Augustin, Marx, Neic, Prassl,
  Goubergrits, Kuehne, and Plank}]{Karabelas2018towards}
\bibinfo{author}{E.~Karabelas}, \bibinfo{author}{M.~A.~F. Gsell},
  \bibinfo{author}{C.~M. Augustin}, \bibinfo{author}{L.~Marx},
  \bibinfo{author}{A.~Neic}, \bibinfo{author}{A.~J. Prassl},
  \bibinfo{author}{L.~Goubergrits}, \bibinfo{author}{T.~Kuehne},
  \bibinfo{author}{G.~Plank},
\newblock \bibinfo{title}{{Towards a Computational Framework for Modeling the
  Impact of Aortic Coarctations Upon Left Ventricular Load}},
\newblock \bibinfo{journal}{Frontiers in Physiology} \bibinfo{volume}{9}
  (\bibinfo{year}{2018}) \bibinfo{pages}{1--20}.
\bibitem[{Baillargeon et~al.(2014)Baillargeon, Rebelo, Fox, Taylor, and
  Kuhl}]{baillargeon2014:_lhp}
\bibinfo{author}{B.~Baillargeon}, \bibinfo{author}{N.~Rebelo},
  \bibinfo{author}{D.~D. Fox}, \bibinfo{author}{R.~L. Taylor},
  \bibinfo{author}{E.~Kuhl},
\newblock \bibinfo{title}{{The Living Heart Project: A robust and integrative
  simulator for human heart function}},
\newblock \bibinfo{journal}{Eur. J. Mech. - A/Solids}  (\bibinfo{year}{2014})
  \bibinfo{pages}{1--10}.
\bibitem[{Land et~al.(2015)Land, Gurev, Arens, Augustin, Baron, Blake, Bradley,
  Castro, Crozier, Favino, Fastl, Fritz, Gao, Gizzi, Griffith, Hurtado, Krause,
  Luo, Nash, Pezzuto, Plank, Rossi, Ruprecht, Seemann, Smith, Sundnes, Rice,
  Trayanova, Wang, {Jenny Wang}, and Niederer}]{land2015:_nversion_mech}
\bibinfo{author}{S.~Land}, \bibinfo{author}{V.~Gurev},
  \bibinfo{author}{S.~Arens}, \bibinfo{author}{C.~M. Augustin},
  \bibinfo{author}{L.~Baron}, \bibinfo{author}{R.~Blake},
  \bibinfo{author}{C.~Bradley}, \bibinfo{author}{S.~Castro},
  \bibinfo{author}{A.~Crozier}, \bibinfo{author}{M.~Favino},
  \bibinfo{author}{T.~E. Fastl}, \bibinfo{author}{T.~Fritz},
  \bibinfo{author}{H.~Gao}, \bibinfo{author}{A.~Gizzi}, \bibinfo{author}{B.~E.
  Griffith}, \bibinfo{author}{D.~E. Hurtado}, \bibinfo{author}{R.~Krause},
  \bibinfo{author}{X.~Luo}, \bibinfo{author}{M.~P. Nash},
  \bibinfo{author}{S.~Pezzuto}, \bibinfo{author}{G.~Plank},
  \bibinfo{author}{S.~Rossi}, \bibinfo{author}{D.~Ruprecht},
  \bibinfo{author}{G.~Seemann}, \bibinfo{author}{N.~P. Smith},
  \bibinfo{author}{J.~Sundnes}, \bibinfo{author}{J.~J. Rice},
  \bibinfo{author}{N.~Trayanova}, \bibinfo{author}{D.~Wang},
  \bibinfo{author}{Z.~{Jenny Wang}}, \bibinfo{author}{S.~A. Niederer},
\newblock \bibinfo{title}{{Verification of cardiac mechanics software:
  benchmark problems and solutions for testing active and passive material
  behaviour.}},
\newblock \bibinfo{journal}{Proc. Math. Phys. Eng. Sci.} \bibinfo{volume}{471}
  (\bibinfo{year}{2015}) \bibinfo{pages}{20150641}.
\bibitem[{Kerckhoffs et~al.(2009)Kerckhoffs, McCulloch, Omens, and
  Mulligan}]{Kerckhoffs2009bivscar}
\bibinfo{author}{R.~C.~P. Kerckhoffs}, \bibinfo{author}{A.~D. McCulloch},
  \bibinfo{author}{J.~H. Omens}, \bibinfo{author}{L.~J. Mulligan},
\newblock \bibinfo{title}{{Effects of biventricular pacing and scar size in a
  computational model of the failing heart with left bundle branch block.}},
\newblock \bibinfo{journal}{Medical Image Analysis} \bibinfo{volume}{13}
  (\bibinfo{year}{2009}) \bibinfo{pages}{362--9}.
\bibitem[{Xi et~al.(2011)Xi, Lamata, Lee, Moireau, Chapelle, and
  Smith}]{Xi2011}
\bibinfo{author}{J.~Xi}, \bibinfo{author}{P.~Lamata}, \bibinfo{author}{J.~Lee},
  \bibinfo{author}{P.~Moireau}, \bibinfo{author}{D.~Chapelle},
  \bibinfo{author}{N.~P. Smith},
\newblock \bibinfo{title}{{Myocardial transversely isotropic material parameter
  estimation from in-silico measurements based on a reduced-order unscented
  Kalman filter}},
\newblock \bibinfo{journal}{Journal of the Mechanical Behavior of Biomedical
  Materials} \bibinfo{volume}{4} (\bibinfo{year}{2011})
  \bibinfo{pages}{1090--1102}.
\bibitem[{Lamata et~al.(2014)Lamata, Sinclair, Kerfoot, Lee, Crozier, Blazevic,
  Land, Lewandowski, Barber, Niederer, and Smith}]{Lamata2014}
\bibinfo{author}{P.~Lamata}, \bibinfo{author}{M.~Sinclair},
  \bibinfo{author}{E.~Kerfoot}, \bibinfo{author}{A.~Lee},
  \bibinfo{author}{A.~Crozier}, \bibinfo{author}{B.~Blazevic},
  \bibinfo{author}{S.~Land}, \bibinfo{author}{A.~J. Lewandowski},
  \bibinfo{author}{D.~Barber}, \bibinfo{author}{S.~Niederer},
  \bibinfo{author}{N.~Smith},
\newblock \bibinfo{title}{{An automatic service for the personalization of
  ventricular cardiac meshes}},
\newblock \bibinfo{journal}{Journal of the Royal Society Interface}
  \bibinfo{volume}{11} (\bibinfo{year}{2014}).
\bibitem[{Pathmanathan et~al.(2009)Pathmanathan, Gavaghan, and
  Whiteley}]{Pathmanathan2009}
\bibinfo{author}{P.~Pathmanathan}, \bibinfo{author}{D.~Gavaghan},
  \bibinfo{author}{J.~Whiteley},
\newblock \bibinfo{title}{{A comparison of numerical methods used for finite
  element modelling of soft tissue deformation}},
\newblock \bibinfo{journal}{Journal of Strain Analysis for Engineering Design}
  \bibinfo{volume}{44} (\bibinfo{year}{2009}) \bibinfo{pages}{391--406}.
\bibitem[{Wang et~al.(2021)Wang, Santiago, Zhou, Wang, Margara,
  Levrero-Florencio, Das, Kelly, Dallarmellina, Vazquez, and
  Rodriguez}]{Wang2021}
\bibinfo{author}{Z.~J. Wang}, \bibinfo{author}{A.~Santiago},
  \bibinfo{author}{X.~Zhou}, \bibinfo{author}{L.~Wang},
  \bibinfo{author}{F.~Margara}, \bibinfo{author}{F.~Levrero-Florencio},
  \bibinfo{author}{A.~Das}, \bibinfo{author}{C.~Kelly},
  \bibinfo{author}{E.~Dallarmellina}, \bibinfo{author}{M.~Vazquez},
  \bibinfo{author}{B.~Rodriguez},
\newblock \bibinfo{title}{{Human biventricular electromechanical simulations on
  the progression of electrocardiographic and mechanical abnormalities in
  post-myocardial infarction}},
\newblock \bibinfo{journal}{Europace} \bibinfo{volume}{23}
  (\bibinfo{year}{2021}) \bibinfo{pages}{I143--I152}.
\bibitem[{Kariya et~al.(2020)Kariya, Washio, ichi Okada, Nakagawa, Watanabe,
  Kadooka, Sano, Nagai, Sugiura, and Hisada}]{Kariya2020}
\bibinfo{author}{T.~Kariya}, \bibinfo{author}{T.~Washio},
  \bibinfo{author}{J.~ichi Okada}, \bibinfo{author}{M.~Nakagawa},
  \bibinfo{author}{M.~Watanabe}, \bibinfo{author}{Y.~Kadooka},
  \bibinfo{author}{S.~Sano}, \bibinfo{author}{R.~Nagai},
  \bibinfo{author}{S.~Sugiura}, \bibinfo{author}{T.~Hisada},
\newblock \bibinfo{title}{{Personalized Perioperative Multi-scale,
  Multi-physics Heart Simulation of Double Outlet Right Ventricle}},
\newblock \bibinfo{journal}{Annals of Biomedical Engineering}
  \bibinfo{volume}{48} (\bibinfo{year}{2020}) \bibinfo{pages}{1740--1750}.
\bibitem[{Gillette et~al.(2021)Gillette, Gsell, Prassl, Karabelas, Reiter,
  Reiter, Grandits, Payer, {\v{S}}tern, Urschler, Bayer, Augustin, Neic, Pock,
  Vigmond, and Plank}]{gillette2021:_ep_twin}
\bibinfo{author}{K.~Gillette}, \bibinfo{author}{M.~A. Gsell},
  \bibinfo{author}{A.~J. Prassl}, \bibinfo{author}{E.~Karabelas},
  \bibinfo{author}{U.~Reiter}, \bibinfo{author}{G.~Reiter},
  \bibinfo{author}{T.~Grandits}, \bibinfo{author}{C.~Payer},
  \bibinfo{author}{D.~{\v{S}}tern}, \bibinfo{author}{M.~Urschler},
  \bibinfo{author}{J.~D. Bayer}, \bibinfo{author}{C.~M. Augustin},
  \bibinfo{author}{A.~Neic}, \bibinfo{author}{T.~Pock}, \bibinfo{author}{E.~J.
  Vigmond}, \bibinfo{author}{G.~Plank},
\newblock \bibinfo{title}{{A Framework for the generation of digital twins of
  cardiac electrophysiology from clinical 12-leads ECGs}},
\newblock \bibinfo{journal}{Med. Image Anal.} \bibinfo{volume}{71}
  (\bibinfo{year}{2021}) \bibinfo{pages}{102080}.
\bibitem[{van Osta et~al.(2020)van Osta, Lyon, Kirkels, Koopsen, van Loon,
  Cramer, Teske, Delhaas, Huberts, and Lumens}]{VanOsta2020a}
\bibinfo{author}{N.~van Osta}, \bibinfo{author}{A.~Lyon},
  \bibinfo{author}{F.~Kirkels}, \bibinfo{author}{T.~Koopsen},
  \bibinfo{author}{T.~van Loon}, \bibinfo{author}{M.~J. Cramer},
  \bibinfo{author}{A.~J. Teske}, \bibinfo{author}{T.~Delhaas},
  \bibinfo{author}{W.~Huberts}, \bibinfo{author}{J.~Lumens},
\newblock \bibinfo{title}{{Parameter subset reduction for patient-specific
  modelling of arrhythmogenic cardiomyopathy-related mutation carriers in the
  CircAdapt model}},
\newblock \bibinfo{journal}{Philosophical Transactions of the Royal Society A:
  Mathematical, Physical and Engineering Sciences} \bibinfo{volume}{378}
  (\bibinfo{year}{2020}) \bibinfo{pages}{20190347}.
\bibitem[{van Osta et~al.(2021)van Osta, Kirkels, Lyon, Koopsen, van Loon,
  Cramer, Teske, Delhaas, and Lumens}]{VanOsta2021}
\bibinfo{author}{N.~van Osta}, \bibinfo{author}{F.~Kirkels},
  \bibinfo{author}{A.~Lyon}, \bibinfo{author}{T.~Koopsen},
  \bibinfo{author}{T.~van Loon}, \bibinfo{author}{M.~J. Cramer},
  \bibinfo{author}{A.~J. Teske}, \bibinfo{author}{T.~Delhaas},
  \bibinfo{author}{J.~Lumens},
\newblock \bibinfo{title}{{Electromechanical substrate characterization in
  arrhythmogenic cardiomyopathy using imaging-based patient-specific computer
  simulations}},
\newblock \bibinfo{journal}{Europace} \bibinfo{volume}{23}
  (\bibinfo{year}{2021}) \bibinfo{pages}{I153--I160}.
\bibitem[{Levine et~al.(1991)Levine, Lane, Buckey, Friedman, and {Gunnar
  Blomqvist}}]{Levine1991:_fsr}
\bibinfo{author}{B.~D. Levine}, \bibinfo{author}{L.~D. Lane},
  \bibinfo{author}{J.~C. Buckey}, \bibinfo{author}{D.~B. Friedman},
  \bibinfo{author}{C.~{Gunnar Blomqvist}},
\newblock \bibinfo{title}{{Left ventricular pressure - Volume and
  Frank-Starling relations in endurance athletes. Implications for orthostatic
  tolerance and exercise performance}},
\newblock \bibinfo{journal}{Circulation} \bibinfo{volume}{84}
  (\bibinfo{year}{1991}) \bibinfo{pages}{1016--1023}.
\bibitem[{Ponnaluri et~al.(2019)Ponnaluri, Verzhbinsky, Eldredge, Garfinkel,
  Ennis, and Perotti}]{Ponnaluri2019}
\bibinfo{author}{A.~V.~S. Ponnaluri}, \bibinfo{author}{I.~A. Verzhbinsky},
  \bibinfo{author}{J.~D. Eldredge}, \bibinfo{author}{A.~Garfinkel},
  \bibinfo{author}{D.~B. Ennis}, \bibinfo{author}{L.~E. Perotti},
\newblock \bibinfo{title}{{Model of Left Ventricular Contraction: Validation
  Criteria and Boundary Conditions}},
\newblock in: \bibinfo{booktitle}{Lecture Notes in Computer Science}, volume
  \bibinfo{volume}{11504 LNCS}, \bibinfo{publisher}{Springer International
  Publishing}, \bibinfo{year}{2019}, pp. \bibinfo{pages}{294--303}.
\bibitem[{Sermesant et~al.(2012)Sermesant, Chabiniok, Chinchapatnam, Mansi,
  Billet, Moireau, Peyrat, Wong, Relan, Rhode, Ginks, Lambiase, Delingette,
  Sorine, Rinaldi, Chapelle, Razavi, and Ayache}]{Sermesant2012}
\bibinfo{author}{M.~Sermesant}, \bibinfo{author}{R.~Chabiniok},
  \bibinfo{author}{P.~Chinchapatnam}, \bibinfo{author}{T.~Mansi},
  \bibinfo{author}{F.~Billet}, \bibinfo{author}{P.~Moireau},
  \bibinfo{author}{J.~Peyrat}, \bibinfo{author}{K.~Wong},
  \bibinfo{author}{J.~Relan}, \bibinfo{author}{K.~Rhode},
  \bibinfo{author}{M.~Ginks}, \bibinfo{author}{P.~Lambiase},
  \bibinfo{author}{H.~Delingette}, \bibinfo{author}{M.~Sorine},
  \bibinfo{author}{C.~Rinaldi}, \bibinfo{author}{D.~Chapelle},
  \bibinfo{author}{R.~Razavi}, \bibinfo{author}{N.~Ayache},
\newblock \bibinfo{title}{{Patient-specific electromechanical models of the
  heart for the prediction of pacing acute effects in CRT: A preliminary
  clinical validation}},
\newblock \bibinfo{journal}{Medical Image Analysis} \bibinfo{volume}{16}
  (\bibinfo{year}{2012}) \bibinfo{pages}{201--215}.
\bibitem[{Finsberg et~al.(2018)Finsberg, Xi, Tan, Zhong, Genet, Sundnes, Lee,
  and Wall}]{Finsberg2018}
\bibinfo{author}{H.~Finsberg}, \bibinfo{author}{C.~Xi}, \bibinfo{author}{J.~L.
  Tan}, \bibinfo{author}{L.~Zhong}, \bibinfo{author}{M.~Genet},
  \bibinfo{author}{J.~Sundnes}, \bibinfo{author}{L.~C. Lee},
  \bibinfo{author}{S.~T. Wall},
\newblock \bibinfo{title}{{Efficient estimation of personalized biventricular
  mechanical function employing gradient-based optimization}},
\newblock \bibinfo{journal}{International Journal for Numerical Methods in
  Biomedical Engineering} \bibinfo{volume}{34} (\bibinfo{year}{2018})
  \bibinfo{pages}{1--20}.
\bibitem[{Amzulescu et~al.(2019)Amzulescu, {De Craene}, Langet, Pasquet,
  Vancraeynest, Pouleur, Vanoverschelde, and Gerber}]{Amzulescu2019}
\bibinfo{author}{M.~S. Amzulescu}, \bibinfo{author}{M.~{De Craene}},
  \bibinfo{author}{H.~Langet}, \bibinfo{author}{A.~Pasquet},
  \bibinfo{author}{D.~Vancraeynest}, \bibinfo{author}{A.~C. Pouleur},
  \bibinfo{author}{J.~L. Vanoverschelde}, \bibinfo{author}{B.~L. Gerber},
\newblock \bibinfo{title}{{Myocardial strain imaging: Review of general
  principles, validation, and sources of discrepancies}},
\newblock \bibinfo{journal}{European Heart Journal Cardiovascular Imaging}
  \bibinfo{volume}{20} (\bibinfo{year}{2019}) \bibinfo{pages}{605--619}.
\bibitem[{Reichek(2017)}]{Reichek2017}
\bibinfo{author}{N.~Reichek},
\newblock \bibinfo{title}{{Myocardial Strain}},
\newblock \bibinfo{journal}{Circulation: Cardiovascular Imaging}
  \bibinfo{volume}{10} (\bibinfo{year}{2017}) \bibinfo{pages}{1--3}.
\bibitem[{Niederer et~al.(2020)Niederer, Aboelkassem, Cantwell, Corrado,
  Coveney, Cherry, Delhaas, Fenton, Panfilov, Pathmanathan, Plank, Riabiz,
  Roney, {Dos Santos}, and Wang}]{Niederer2020creation}
\bibinfo{author}{S.~A. Niederer}, \bibinfo{author}{Y.~Aboelkassem},
  \bibinfo{author}{C.~D. Cantwell}, \bibinfo{author}{C.~Corrado},
  \bibinfo{author}{S.~Coveney}, \bibinfo{author}{E.~M. Cherry},
  \bibinfo{author}{T.~Delhaas}, \bibinfo{author}{F.~H. Fenton},
  \bibinfo{author}{A.~V. Panfilov}, \bibinfo{author}{P.~Pathmanathan},
  \bibinfo{author}{G.~Plank}, \bibinfo{author}{M.~Riabiz},
  \bibinfo{author}{C.~H. Roney}, \bibinfo{author}{R.~W. {Dos Santos}},
  \bibinfo{author}{L.~Wang},
\newblock \bibinfo{title}{{Creation and application of virtual patient cohorts
  of heart models: Virtual Cohorts of Heart Models}},
\newblock \bibinfo{journal}{Philosophical Transactions of the Royal Society A:
  Mathematical, Physical and Engineering Sciences} \bibinfo{volume}{378}
  (\bibinfo{year}{2020}).
\bibitem[{Arts et~al.(1991)Arts, Bovendeerd, Prinzen, and
  Reneman}]{arts1991relation}
\bibinfo{author}{T.~J. Arts}, \bibinfo{author}{P.~H.~M. Bovendeerd},
  \bibinfo{author}{F.~W. Prinzen}, \bibinfo{author}{R.~S. Reneman},
\newblock \bibinfo{title}{{Relation between left ventricular cavity pressure
  and volume and systolic fibre stress and strain in the wall}},
\newblock \bibinfo{journal}{Biophys. J.} \bibinfo{volume}{59}
  (\bibinfo{year}{1991}) \bibinfo{pages}{93--102}.
\bibitem[{Firstenberg et~al.(2001)Firstenberg, Greenberg, Smedira, McCarthy,
  Garcia, and Thomas}]{firstenberg2001noninvasive}
\bibinfo{author}{M.~S. Firstenberg}, \bibinfo{author}{N.~Greenberg},
  \bibinfo{author}{N.~Smedira}, \bibinfo{author}{P.~McCarthy},
  \bibinfo{author}{M.~J. Garcia}, \bibinfo{author}{J.~D. Thomas},
\newblock \bibinfo{title}{{Noninvasive assessment of mitral inertness: clinical
  results with numerical model validation}},
\newblock \bibinfo{journal}{Comput. Cardiol.}  (\bibinfo{year}{2001})
  \bibinfo{pages}{613--616}.
\bibitem[{Otto(2006)}]{otto2006valvular}
\bibinfo{author}{C.~M. Otto},
\newblock \bibinfo{title}{{Valvular aortic stenosis: disease severity and
  timing of intervention.}},
\newblock \bibinfo{journal}{Journal of the American College of Cardiology}
  \bibinfo{volume}{47} (\bibinfo{year}{2006}) \bibinfo{pages}{2141--51}.
\bibitem[{Hairer et~al.(1993)Hairer, N{\o}rsett, and
  Wanner}]{hairer1993solving}
\bibinfo{author}{E.~Hairer}, \bibinfo{author}{S.~P. N{\o}rsett},
  \bibinfo{author}{G.~Wanner}, \bibinfo{title}{Solving ordinary differential
  equations 1}, \bibinfo{publisher}{Springer}, \bibinfo{year}{1993}.
\bibitem[{Holzapfel(2000)}]{holzapfel2000nonlinear}
\bibinfo{author}{G.~A. Holzapfel}, \bibinfo{title}{{Nonlinear Solid Mechanics.
  A Continuum Approach for Engineering}}, \bibinfo{publisher}{John Wiley \&
  Sons Ltd}, \bibinfo{address}{Chichester}, \bibinfo{year}{2000}.
\bibitem[{Press et~al.(2007)Press, Teukolsky, Vettering, and
  Flannery}]{Press2007}
\bibinfo{author}{W.~H. Press}, \bibinfo{author}{S.~A. Teukolsky},
  \bibinfo{author}{W.~T. Vettering}, \bibinfo{author}{B.~P. Flannery},
  \bibinfo{title}{{Numerical Recipies: The Art of Scientific Computing}},
  \bibinfo{edition}{3rd} ed., \bibinfo{publisher}{Cambridge University Press},
  \bibinfo{year}{2007}.
\bibitem[{Kadapa et~al.(2017)Kadapa, Dettmer, and
  Peri{\'{c}}}]{kadapa2017advantages}
\bibinfo{author}{C.~Kadapa}, \bibinfo{author}{W.~G. Dettmer},
  \bibinfo{author}{D.~Peri{\'{c}}},
\newblock \bibinfo{title}{{On the advantages of using the first-order
  generalised-alpha scheme for structural dynamic problems}},
\newblock \bibinfo{journal}{Computers and Structures} \bibinfo{volume}{193}
  (\bibinfo{year}{2017}) \bibinfo{pages}{226--238}.
\bibitem[{Chung and Hulbert(1993)}]{chung1993time}
\bibinfo{author}{J.~Chung}, \bibinfo{author}{G.~M. Hulbert},
\newblock \bibinfo{title}{{A Time Integration Algorithm for Structural Dynamics
  With Improved Numerical Dissipation: The Generalized-$\alpha$ Method}},
\newblock \bibinfo{journal}{Journal of Applied Mechanics} \bibinfo{volume}{60}
  (\bibinfo{year}{1993}) \bibinfo{pages}{371}.
\bibitem[{Newmark(1959)}]{newmark1959method}
\bibinfo{author}{N.~M. Newmark},
\newblock \bibinfo{title}{A method of computation for structural dynamics},
\newblock \bibinfo{journal}{Journal of the engineering mechanics division}
  \bibinfo{volume}{85} (\bibinfo{year}{1959}) \bibinfo{pages}{67--94}.

\end{thebibliography}

\end{document}